\documentclass[a4paper,11pt]{article}
\pdfoutput=1

\usepackage{jheppub}

\usepackage{graphicx}
\usepackage{amsmath}
\usepackage{xcolor}
\usepackage{amssymb}
\usepackage{microtype}
\usepackage[utf8]{inputenc} 

\newcommand{\D}[1]{\mathrm{D^{#1}}}
\newcommand{\pT}{p_{\mathrm{T}}}
\newcommand{\PT}{P_{\mathrm{T}}}

\title{A QCD analysis of LHCb D-meson data in p+Pb collisions}

\author[a,b]{Kari J.~Eskola,}
\author[a,b]{Ilkka Helenius,}
\author[a,b,c]{Petja Paakkinen}
\author[a,b]{and Hannu Paukkunen}

\affiliation[a]{University of Jyvaskyla, Department of Physics, P.O. Box 35, FI-40014 University of Jyvaskyla, Finland}
\affiliation[b]{Helsinki Institute of Physics, P.O. Box 64, FI-00014 University of Helsinki, Finland}
\affiliation[c]{Instituto Galego de F\'\i sica de Altas Enerx\'\i as IGFAE, Universidade de Santiago de Compostela, E-15782 Galicia-Spain}

\emailAdd{kari.eskola@jyu.fi}
\emailAdd{ilkka.m.helenius@jyu.fi}
\emailAdd{petja.paakkinen@usc.es}
\emailAdd{hannu.t.paukkunen@jyu.fi}

\abstract{%
We scrutinize the recent LHCb data for $\D0$-meson production in p+Pb collisions within a next-to-leading order QCD framework. Our calculations are performed in the SACOT-$m_{\rm T}$ variant of the general-mass variable-flavour-number scheme (GM-VFNS), which has previously been shown to provide a realistic description of the LHC p+p data. Using the EPPS16 and nCTEQ15 nuclear parton distribution functions (PDFs) we show that a very good agreement is obtained also in the p+Pb case both for cross sections and nuclear modification ratios in the wide rapidity range covered by the LHCb data. Encouraged by the good correspondence, we quantify the impact of these data on the nuclear PDFs by the Hessian reweighting technique. We find compelling direct evidence of gluon shadowing at small momentum fractions $x$, with no signs of parton dynamics beyond the collinear factorization. We also compare our theoretical framework to a fixed-order calculation supplemented with a parton shower. While the two frameworks differ in the absolute cross sections, these differences largely cancel in the nuclear modification ratios. Thus, the constraints for nuclear PDFs appear solid.
}

\begin{document}

\maketitle
\flushbottom

\section{Introduction}

In the collinear-factorization approach to describe scattering of protons and heavier nuclei in Quantum Chromodynamics (QCD), the non-perturbative structure of the hadrons --- parton distribution functions (PDFs) --- is factorized from the perturbatively calculable coefficient functions \cite{Collins:1989gx, Kovarik:2019xvh}. The PDFs are typically extracted from experimental data via global analysis and their accurate determination has been a long-standing effort in the community \cite{Gao:2017yyd, Kovarik:2019xvh}. For the free proton PDF fits there are plenty of accurate data available and the most recent global analyses \cite{Harland-Lang:2014zoa, Dulat:2015mca, Abramowicz:2015mha, Ball:2017nwa, Alekhin:2018pai} result with PDFs that are reasonably well constrained within the typical kinematics probed at the Large Hadron Collider (LHC). 

For PDFs in heavier nuclei, nuclear PDFs (nPDFs), the available data have been rather sparse until very lately \cite{Paukkunen:2018kmm}. Indeed, even some recent analyses still rely only on older fixed-target deep inelastic scattering (DIS) and Drell-Yan (DY) data \cite{AbdulKhalek:2019mzd,Khanpour:2016pph}. Due to the relatively low center-of-mass (c.m.) energy $\sqrt{s}$, these data provide constraints only for momentum fractions $x \gtrsim 0.01$, and the gluons are constrained only indirectly via scale-evolution effects and momentum sum rule \cite{Eskola:1998iy}. To obtain better gluon constraints, the potential of inclusive pion production in d+Au collisions at RHIC \cite{Adler:2006wg,Adams:2006nd,Abelev:2009hx,Adare:2013esx} was first discussed in ref.~\cite{deFlorian:2003qf} and eventually the data were incorporated into the global fits \cite{Eskola:2008ca,Eskola:2009uj,deFlorian:2011fp,Kovarik:2015cma}. The $x$ reach was still, however, rather similar to the available DIS data. The currently most comprehensive nPDF analysis, EPPS16 \cite{Eskola:2016oht}, includes also LHC Run-I data for electroweak-boson ($\mathrm{W}^{\pm}$ and $\mathrm{Z}^0$) \cite{Khachatryan:2015pzs,Aad:2015gta,Khachatryan:2015hha} and dijet production \cite{Chatrchyan:2014hqa} in p+Pb collisions. Because of the large masses of the $\mathrm{W}^{\pm}$ and $\mathrm{Z}^0$ bosons, the interaction scale is high and a significant sensitivity to gluons via evolution effects will eventually set constraints on gluons, as has been shown in ref.~\cite{Citron:2018lsq} (sect. 10.4.2). However, the Run-I $\mathrm{W}^{\pm}$ and $\mathrm{Z}^0$ data have still a rather limited impact due to the low statistics. The dijet production, on the other hand, probes the gluon density much more directly and already the Run-I data clearly helps to narrow down the gluons in the $x \gtrsim 0.002$ region \cite{Eskola:2019dui}. All this still leaves the small-$x$ region only weakly constrained. To probe gluons at small $x$, almost any conceivable observable at lowish interaction scales and forward rapidity $y \gg 0$ would do. Good candidates at hadron colliders include e.g. low-mass Drell-Yan dilepton and isolated-photon production at low transverse momentum $p_{\mathrm{T}}$ \cite{Arleo:2007js, Stavreva:2010mw, Arleo:2011gc, Brandt:2014vva, Helenius:2014qla, Goharipour:2018sip, Helenius:2019lop}. Isolated photons in p+Pb collisions have already been measured at central rapidities \cite{Aaboud:2019tab}, and the large-$y$ measurements appear to be within the capabilities of the LHCb collaboration \cite{Boettcher:2019kxa}. In further future, measurements of isolated-photon production would be a central goal of the ALICE FoCal upgrade \cite{Peitzmann:2014isa}. 

Another promising observable for gluon constraints is the inclusive D- and B-meson production where the heavy-quark mass provides the hard scale even at zero $p_{\mathrm{T}}$. In fact, the LHCb collaboration has published low-$p_{\mathrm{T}}$ data on $\D{}$-meson production at forward kinematics in p+p collisions at different $\sqrt{s}$ \cite{Aaij:2013mga, Aaij:2015bpa, Aaij:2016jht}, and recently also in the p+Pb case at $\sqrt{s}=5~\text{TeV}$ \cite{Aaij:2017gcy}. The use of these D-meson data as a free proton and nuclear PDF constraint has been advocated e.g. in refs.~\cite{Cacciari:2015fta, Gauld:2015yia, Gauld:2016kpd, Zenaiev:2015rfa,Gauld:2015lxa, Kusina:2017gkz} and studied otherwise \cite{Kramer:2017gct}, but for the moment the default sets of globally fitted general-purpose PDFs \cite{Harland-Lang:2014zoa, Dulat:2015mca, Abramowicz:2015mha, Ball:2017nwa, Alekhin:2018pai,Eskola:2016oht,Kovarik:2015cma} do not include any D-meson data. Here, our purpose is to provide a first estimate of the impact the recent LHCb p+Pb data have on globally fitted nPDFs within a rigorous next-to-leading order (NLO) perturbative-QCD framework. We will focus only on the LHCb measurements \cite{Aaij:2017gcy}, as the central-rapidity ALICE \cite{Adam:2016ich} data are not as precise and as the ATLAS central-rapidity data \cite{Acharya:2019zup} are only preliminary.

As the nPDF sets we consider in this work, EPPS16 \cite{Eskola:2016oht} and nCTEQ15 \cite{Kovarik:2015cma}, are of a variable-flavour type, where the charm and bottom quarks are “active” partons above their mass thresholds, our default setup for the heavy-meson cross section calculations is based on the general-mass variable-flavour-number scheme (GM-VFNS) approach. The concept of this formalism is to match the fixed-flavour-number scheme (FFNS) valid at very low $p_{\mathrm{T}}$ with a massless variable-flavour calculation valid at high $p_{\mathrm{T}}$. Such an approach was first developed for leptoproduction of heavy-quarks \cite{Aivazis:1993pi, Collins:1998rz, Tung:2001mv, Thorne:2008xf, Forte:2010ta, Guzzi:2011ew} and has also been applied to heavy-quark hadroproduction \cite{Olness:1997yc, Kniehl:2004fy, Kniehl:2005mk, Kniehl:2012ti}. In this framework the mass-dependent logarithms arising from collinear emissions are resummed into scale-dependent PDFs and fragmentation functions (FFs). Similar parton-level resummation of collinear emissions is also achieved within the FONLL formalism \cite{Cacciari:1998it, Cacciari:2012ny}, which essentially constitutes a particular GM-VFNS scheme. Also parton showers in general-purpose Monte Carlo event generators, such as \textsc{Pythia} \cite{Sjostrand:2014zea}, provide an effective (leading-logarithm) resummation. In this work, we will use the SACOT-$m_{\rm T}$ variant of the GM-VFNS formalism, introduced in ref.~\cite{Helenius:2018uul}. This framework takes fully into account the $\D{}$ mesons produced by gluon fragmentation --- something that is neglected in the FFNS approach. However, when the partonic $p_{\mathrm{T}}$ scale is less or close to the heavy-quark mass, $p_{\mathrm{T}} \lesssim m$, the inherent uncertainties of the GM-VFNS approach grow and somewhere close to this region the pure FFNS approach becomes arguably more reliable. For this and other reasons discussed later on, in our main results we restrict to region with minimum $p_{\mathrm{T}} = 3\,{\rm GeV}$ for the produced D mesons, though also the lower $p_{\mathrm{T}}$ regime is explored. 
To decide with confidence which $p_{\mathrm{T}}$ scale sets the borderline between the two approaches is a question that would probably require calculations at next-to-NLO level which are not yet available. Thus, in parallel to the GM-VFNS calculations, we perform the cross-section calculations also in the FFNS-based approach to further chart the uncertainties. To quantify the impact on the nPDFs, we will use the Hessian reweighting technique \cite{Paukkunen:2013grz, Paukkunen:2014zia, Schmidt:2018hvu, Eskola:2019dui} that facilitates an estimate of the data impact without re-doing the complete global analysis.

The paper will now continue as follows: In section \ref{sec:framework}, we introduce our theoretical setup, including the GM-VFNS framework and the applied reweighting machinery. Then, in section \ref{sec:results}, we compare the resulting cross sections and nuclear modification ratios with the LHCb data, demonstrate the impact these data have on nPDFs, and discuss their sensitivity to small-$x$ gluons. We summarize our findings in section \ref{sec:summary}.

\section{Theoretical framework}
\label{sec:framework}

\subsection{SACOT-$m_{\mathrm{T}}$ scheme for heavy-quark production}

The general idea of D-meson hadroproduction in the GM-VFNS approach \cite{Kniehl:2004fy,Helenius:2018uul} is to reproduce the results of (3-flavour) fixed flavour-number scheme (FFNS) at the small $p_{\mathrm{T}}$ limit and match to the massless calculation at high values of $p_{\mathrm{T}}$.
Let us first discuss the FFNS limit, in which the cross section for inclusive production of a heavy-flavoured hadron $h_3$ at a given transverse momentum $P_{\mathrm{T}}$ and rapidity $Y$ in a collision of two hadrons, $h_1$ and $h_2$, can be written as
\begin{equation}
\begin{split}
\frac{\mathrm{d}\sigma^{h_1 + h_2 \rightarrow h_3 + X}}{\mathrm{d}P_{\rm T}\mathrm{d}Y}&\Big|_{\rm FFNS} =
 \sum _{ij} \int_{z^{\rm min}}^1 \frac{\mathrm{d}z}{z} \int_{x_1^{\rm min}}^1 \mathrm{d}x_1 \int _{x_2^{\rm min}}^1 \mathrm{d}x_2 \\
& \hspace{-3em} \times D_{Q \rightarrow h_3}(z) \, 
f_i^{h_1}(x_1,\mu^2_{\rm fact}) f_j^{h_2}(x_2,\mu^2_{\rm fact}) \frac{\mathrm{d}\hat{\sigma}^{ij\rightarrow Q+X}}{\mathrm{d}p_{\rm T} \mathrm{d}y}(\tau_1, \tau_2, m, \mu^2_{\rm ren}, \mu^2_{\rm fact})\,.
\label{eq:FFNS} 
\end{split}
\end{equation}
In this expression, $f_{i,j}^{h_{1,2}}$ are the PDFs (in 3-flavour scheme) for partons $i$ and $j$ in hadrons $h_1$ and $h_2$ with momentum fractions $x_1$ and $x_2$, and $\mathrm{d}\hat{\sigma}^{ij\rightarrow Q+X}/\mathrm{d}p_{\rm T}\mathrm{d}y$ denote the perturbatively calculable coefficient functions for inclusive heavy-quark $Q$ (here charm) production \cite{Nason:1989zy} with fixed rapidity $y$ and transverse momentum $p_{\rm T}$ of $Q$. The renormalization and factorization scales are denoted by $\mu^2_{\rm ren}, \mu^2_{\rm fact}$ and $m$ is the heavy-quark (here charm) mass. The fragmentation of a heavy-quark to hadron $h_3$ is described by a scale-independent fragmentation function (FF) $D_{Q \rightarrow h_3}$ (such as in ref.~\cite{Peterson:1982ak}). The invariants $\tau_i$ can be calculated from the partonic transverse mass $m_{\mathrm{T}} = \sqrt{p^2_{\mathrm{T}} + m^2}$ and rapidity $y$ as
\begin{equation}
\tau_1 \equiv \frac{p_1 \cdot p_3}{p_1 \cdot p_2} = \frac{m_{\mathrm{T}}\mathrm{e}^{-y}}{x_2\sqrt{s}} \quad \text{and} \quad 
\tau_2 \equiv \frac{p_2 \cdot p_3}{p_1 \cdot p_2}= \frac{m_{\mathrm{T}}\mathrm{e}^{y}}{x_1\sqrt{s}}.
\end{equation}
where $p_1$ and $p_2$ are the momenta of the incoming massless partons, and $p_3$ is the final-state heavy-quark momentum. When masses are neglected, the relation between partonic and hadronic variables is simply $y = Y$ and $P_{\mathrm{T}} = z p_{\mathrm{T}}$. However, when the masses of the heavy quark and the final-state hadron are taken into account, the definition of $z$ becomes ambiguous \cite{Albino:2008fy}. Adopting the choice made in \cite{Helenius:2018uul},
\begin{equation}
z \equiv \frac{P_3 \cdot (P_1 + P_2) }{p_3 \cdot (P_1 + P_2)},
\end{equation}
where $P_{i}$ is the momentum of hadron $h_i$, the $z$ variable can be interpreted as the fraction of partonic energy carried by the outgoing hadron in the c.m. frame of the initial-state hadrons $h_1$ and $h_2$. The relations between partonic and hadronic variables become somewhat more involved, but eq.~(\ref{eq:FFNS}) stays intact.

When the transverse momentum of the produced hadron $h_3$ is large, $P_{\mathrm{T}} \gg m$, the heavy-quark mass can be neglected and thus the zero-mass description becomes the most relevant. In this limit, the cross section can be written as \cite{Aversa:1988vb},
\begin{equation}
\begin{split}
\frac{\mathrm{d}\sigma^{h_1 + h_2 \rightarrow h_3 + X}}{\mathrm{d}P_{\rm T}\mathrm{d}Y}&\Big|_{\rm ZM} =
 \sum _{ijk} \int_{z^{\rm min}}^1 \frac{\mathrm{d}z}{z} \int_{x_1^{\rm min}}^1 \mathrm{d}x_1 \int _{x_2^{\rm min}}^1 \mathrm{d}x_2 \label{eq:masterZM} \\
& \hspace{-5em} \times D_{k \rightarrow h_3}(z,\mu^2_{\rm frag}) \, 
f_i^{h_1}(x_1,\mu^2_{\rm fact}) f_j^{h_2}(x_2,\mu^2_{\rm fact}) \frac{\mathrm{d}\hat{\sigma}^{ij\rightarrow k+X}}{\mathrm{d}p_{\rm T} \mathrm{d}y}(\tau_1^0, \tau_2^0, \mu^2_{\rm ren}, \mu^2_{\rm fact}, \mu^2_{\rm frag})\,.
\end{split}
\end{equation}
The formal difference with respect to eq.~(\ref{eq:FFNS}) is that now the FFs are fragmentation-scale $\mu^2_{\rm frag}$ dependent, and a summation over all partonic channels is included. For massless partons the invariants $\tau_i^0$ are obtained as
\begin{equation}
\tau_1^0 = \lim_{m\rightarrow 0} \tau_1 = \frac{p_{\mathrm{T}}\mathrm{e}^{-y}}{x_2\sqrt{s}} \quad \text{and} \quad 
\tau_2^0 = \lim_{m\rightarrow 0} \tau_2 = \frac{p_{\mathrm{T}}\mathrm{e}^{y}}{x_1\sqrt{s}}.
\end{equation}

The GM-VFNS technique \cite{Kniehl:2004fy,Helenius:2018uul} provides a general framework to match the two extremes of eq.~(\ref{eq:FFNS}) and eq.~(\ref{eq:masterZM}) in a way that is consistent with collinear factorization. If we start from the FFNS description and increase $P_{\rm T}$, the cross sections will quickly be dominated by $\log(p_{\mathrm{T}}/m)$ terms whose origin is in the initial- and final-state partons' collinear splittings into $Q\overline{Q}$ pairs. In GM-VFNS these logarithms are resummed to the scale-dependent heavy-quark PDFs and scale-dependent FFs. Because the FFNS expressions already contain the first of the resummed logarithmic terms, subtraction terms are needed to avoid double counting and ensure the correct zero-mass limit of eq.~(\ref{eq:masterZM}). For example, the inclusion of the gluon production channel $gg \rightarrow gg$, 
\begin{equation}
\begin{split}
& \frac{\mathrm{d}\sigma^{h_1 + h_2 \rightarrow h_3 + X}}{\mathrm{d}P_{\rm T}\mathrm{d}Y}\Big|_{gg \rightarrow gg} =
 \int_{z^{\rm min}}^1 \frac{\mathrm{d}z}{z} \int_{x_1^{\rm min}}^1 \mathrm{d}x_1 \int _{x_2^{\rm min}}^1 \mathrm{d}x_2 \\
& \hspace{2em} \times D_{g \rightarrow h_3}(z,\mu^2_{\rm frag}) \, 
f_g^{h_1}(x_1,\mu^2_{\rm fact}) f_g^{h_2}(x_2,\mu^2_{\rm fact}) \frac{\mathrm{d}\hat{\sigma}^{gg\rightarrow g+X}}{\mathrm{d}p_{\rm T} \mathrm{d}y}(\tilde\tau_1, \tilde\tau_2, \mu^2_{\rm ren}, \mu^2_{\rm fact}, \mu^2_{\rm frag})
\label{eq:add} 
\end{split}
\end{equation}
on top of eq.~(\ref{eq:FFNS}), must be accompanied
by a subtraction term which has otherwise the same expression as eq. (\ref{eq:add}) but where the gluon-to-$h_3$ FF is replaced by
\begin{equation}
\begin{split}
D_{g\rightarrow h_3}(x,\mu_{\mathrm{frag}}^2) & = \frac{\alpha_{\mathrm{s}}}{2 \pi} \log \left( \frac{\mu_{\mathrm{frag}}^2}{m^2} \right) \int_x^1\frac{\mathrm{d}z}{z}P_{qg}(x/z) D_{Q \rightarrow h_3}(z) \label{eq:subt} \\ 
& = \frac{\alpha_{\mathrm{s}}}{2 \pi} \log \left( \frac{\mu_{\mathrm{frag}}^2}{m^2} \right) \int_x^1\frac{\mathrm{d}z}{z}P_{qg}(x/z) D_{Q \rightarrow h_3}(z, \mu^2_{\rm frag}) + \mathcal{O}(\alpha_s^2) \,,
\end{split}
\end{equation}
which is the first term in the definition of scale-dependent FFs with massive quarks. In an NLO-accurate $\mathcal{O}(\alpha_s^3)$ calculation, only the leading-order part of $\mathrm{d}\hat{\sigma}^{gg\rightarrow g+X}$ is included in the subtraction term. However, the exact form of $\mathrm{d}\hat{\sigma}^{gg\rightarrow g+X}$ in the equation above is not fixed by this construction. The only condition is that we recover the standard zero-mass $\overline{\rm MS}$ expression at $p_{\mathrm{T}} \rightarrow \infty$ to meet eq.~(\ref{eq:masterZM}). This means that we can include mass-dependent terms in $\mathrm{d}\hat{\sigma}^{gg\rightarrow g+X}$ as we like, and a specific choice defines a scheme. The difference between the added and subtracted contributions discussed above is formally of order $\mathcal{O}(\alpha_s^4)$, so that different schemes are formally equivalent up to $\mathcal{O}(\alpha_s^3)$. Here we adopt the so-called SACOT-$m_{\mathrm{T}}$ scheme \cite{Helenius:2018uul}. It is rooted in a simple observation that in order to make a heavy-flavoured hadron in QCD, a $Q\bar{Q}$ pair must be first produced. That is, the relevant invariants to describe the process are the massive ones, $\tilde \tau_{1,2} = \tau_{1,2}$, even for seemingly massless partonic contribution (like the $gg \rightarrow gg$ channel). Importantly, the mass then prevents the partonic cross sections from diverging towards small $p_{\mathrm{T}}$ exactly in the same way as the FFNS cross section are finite at $p_{\mathrm{T}}=0$. In the previous GM-VFNS approach \cite{Kniehl:2004fy} such a physical behaviour is obtained only by a particular choice of QCD scales \cite{Kniehl:2012ti, Kniehl:2015fla}. However, we still stress that when the partonic $p_{\mathrm{T}}$ scale is less or similar to the heavy-quark mass $m$, the arbitrariness related to the GM-VFNS scheme choice reduces the reliability of the predictions. The arbitrariness related to the choice of the fragmentation variable $z$ is also most prominent at low $p_{\mathrm{T}}$. For these two reasons in our main results we will concentrate on the region $\PT > 3 \, {\rm GeV}$ where the associated uncertainties are smaller.

The final differential cross sections are then calculated by using the FFNS expressions for the explicit $Q\overline{Q}$ production, and for all other channels zero-mass expressions with the mentioned massive kinematics. The subtraction terms discussed above are included to avoid double counting and to ensure proper matching between $\alpha_s$ and PDFs in 3- and 4-flavour schemes. The switch from 3- to 4-flavour scheme is done at the charm-mass threshold. The bottom decays to $\D{0}$ are an order of magnitude smaller \cite{Acharya:2019mgn} than the ``direct'' charm fragmentation to $\D{0}$. Thus, the treatment of the bottom mass is not as critical, and in our present setup we switch from 4- to 5-flavour scheme at the bottom-mass threshold with no matching conditions and ignoring the bottom mass. For the numerical implementation of the described SACOT-$m_{\mathrm{T}}$ scheme the massless NLO matrix elements are obtained from the \textsc{incnlo} \cite{Aversa:1988vb} code and the FFNS part with explicit heavy-quark production is obtained from the \textsc{mnr} code \cite{Mangano:1991jk}. As presented in refs.~\cite{Helenius:2018uul, Acharya:2019mgn}, this framework is in a very good agreement with the ALICE \cite{Acharya:2019mgn, Adam:2016ich} and LHCb \cite{Aaij:2013mga, Aaij:2015bpa, Aaij:2016jht} data for inclusive $\D{}$-meson production in p+p collisions in a broad rapidity range. The GM-VFNS approach also broadly reproduces the LHCb data on double D-meson production \cite{Helenius:2019uge}.

\subsection{\textsc{Powheg}+\textsc{Pythia} approach}
\label{sec:Powheg}

We will also contrast our results in the SACOT-$m_{\mathrm{T}}$ framework with a Monte-Carlo based NLO computation that is often applied to heavy-meson phenomenology at the LHC in the context of PDFs \cite{Gauld:2015yia, Gauld:2016kpd, Garzelli:2016xmx}. This approach is based on the \textsc{Powheg} method \cite{Frixione:2007vw} to combine NLO matrix elements with a parton shower and hadronization from a general-purpose Monte-Carlo event generator. The underlying idea is to generate the partonic $2 \rightarrow 2$ and $2 \rightarrow 3$ events with the NLO-correct matrix elements. These events are then passed to any parton shower generator that provides the rest of the partonic branchings, accounting for the fact that the first one may already have occurred. The parton shower can be considered as being analogous to the scale evolution of FFs and PDFs as the splitting probabilities are based on the DGLAP evolution equations in both cases. 

We generate the partonic events with the heavy-quark pair production (\textsc{hvq}) scenario \cite{Frixione:2007nw} of the \textsc{Powheg Box} framework \cite{Alioli:2010xd} which we pass on to \textsc{Pythia~8} \cite{Sjostrand:2014zea} for showering and hadronization. As \textsc{Powheg} generates only events where the heavy-quark pair is produced in the Born-level process or in the first (hardest) splitting, it ignores the component where the $Q\overline{Q}$ would be created only later on in the shower e.g. starting from a hard $gg \rightarrow gg$ process. Such contributions are, however, effectively included in any GM-VFNS framework via the scale-dependent PDFs and FFs. Since charm quarks are abundantly produced in parton showers at the LHC energies \cite{Norrbin:2000zc}, truncating the resummation of the splittings to the first one may miss a significant source of heavy quarks, as was pointed out in ref.~\cite{Helenius:2018uul}. This interpretation is supported by noting that within the GM-VFNS framework, the fixed-order production channels (the ones included in the \textsc{hvq} scenario of \textsc{Powheg}) were observed to constitute less than 10\% of the full cross section at $\PT \gtrsim 3~\text{GeV}$ \cite{Helenius:2018uul} once the subtraction terms were included. In addition, as demonstrated in Refs.~\cite{Aaij:2015bpa, Aaij:2016jht, Helenius:2018uul}, the uncertainties arising from scale variations within the \textsc{Powheg}+\textsc{Pythia} setup become considerably larger at $\PT \gtrsim 3~\text{GeV}$ than what they are in GM-VFNS implementations. It is thus conceivable that the logarithmic terms resummed in GM-VFNS are significant already at $\PT \sim 3~\text{GeV}$. However, as mentioned before, the uncertainties related to the choice of the GM-VFNS scheme are large at low $\PT$ and it is thus impossible to draw a decisive conclusion.

At high enough $\PT$, the truncation of the chain of partonic splittings the \textsc{Powheg+Pythia} method potentially overestimates the sensitivity to low-$x$ PDFs as the neglected contributions with several emissions would always require a higher value of $x$ to produce a heavy meson at a fixed $P_{\mathrm{T}}$ and $Y$. Within its large scale uncertainties the \textsc{Powheg+Pythia} method nevertheless agrees with the $\D{}$-meson data measured by LHCb even at $\PT \gg m_{\rm charm}$, though the central predictions are generally below the data \cite{Gauld:2015yia}. 

\subsection{Reweighting machinery}
\label{sec:reweight}

We will quantify the impact of the single inclusive $\D0$-meson production data in p+Pb collisions on nuclear PDFs by the Hessian reweighting method \cite{Paukkunen:2013grz, Paukkunen:2014zia, Schmidt:2018hvu,Eskola:2019dui}. The method has recently been discussed at length e.g. in ref.~\cite{Eskola:2019dui} so here we only outline the basic underlying idea. Let us consider a global PDF analysis whose fit parameters $a_i$ are tuned to minimize a global $\chi^2$ function, $\chi^2_0 = \min \chi^2 = \chi^2\{a_i=a_i^0\}$. The $\chi^2$ function is expanded around the best fit as
\begin{equation}
 \chi^2\{a\} \approx \chi^2_0 + \sum_{ij} (a_i-a_i^0) H_{ij} (a_j-a_j^0) = \chi^2_0  + \sum_i  z_i^2 \label{eq:chi2orig} \,,
\end{equation}
where $H_{ij}$ is the Hessian matrix, $H_{ij} =  \frac{1}{2}\partial^2\chi^2/(\partial a_i \partial a_j)$. Denoting by $O$ the orthogonal matrix that diagonalizes the Hessian matrix, $OHO^{\rm T} = I$, the $z_i$ variables are linear combinations $z_i \equiv O_{ij}(a_j-a_j^0)$. We refer to the best-fit as $S_0$, and it corresponds to the point $z=0$. The Hessian error sets $S^\pm_k$ can then be defined by $z_i({S^\pm_k}) = \pm \sqrt{\Delta \chi^2} \delta_{ik}$, where $\Delta \chi^2$ is the estimated tolerance. It follows \cite{Pumplin:2001ct} that for any PDF-dependent quantity $X$ there are unique points in the $z$ space that extremize its positive and negative deviations from the central value $X(S_0)$. These deviations, $\Delta X^\pm$, are given by
\begin{equation}
 \Delta X^\pm = \pm \frac{1}{2} \sqrt{\sum_k \left[X({S^+_k}) - X({S^-_k})\right]^2} \,.
\end{equation}
This, or its asymmetric version (see later), is normally quoted as the uncertainty in Hessian PDF fits. In a global analysis, the $\chi^2$ contributions of individual data sets are simply summed in the overall $\chi^2$. Thus, if we wish to include a new set of data into our global fit, we just add its contribution to eq.~(\ref{eq:chi2orig}), 
\begin{equation}
\chi^2_{\rm new} \equiv \chi^2_0  +   \sum_k z_k^2  + 
\sum_{i,j} \left(y_i\{z\}-y_i^{\rm data}\right) C_{ij}^{-1} \left(y_j\{z\}-y_j^{\rm data}\right), \label{eq:newchi2}
\end{equation}
where $y_i^{\rm data}$ denote the new data points with a covariance matrix $C_{ij}$. The PDF-dependent values $y_i\{z\}$ can now be approximated linearly as
\begin{equation}
 y_i \{z\}  \approx y_i \left[{S_0} \right] + \sum_{k} \frac{\partial y_i [{S}]}{\partial z_k}{\Big|_{S=S_0}} z_k \approx y_i \left[S_0 \right] + \sum_{k} \frac{y_i[S_k^+] - y_i[S_k^-]}{2} \frac{z_k}{\sqrt{\Delta \chi^2}} \,,
\label{eq:XS}
\end{equation}
and by substituting this into eq.~(\ref{eq:newchi2}), we see that $\chi^2_{\rm new}$ is still quadratic in variables $z_k$ and has therefore a unique minimum which we denote by $z_k=z^{\rm min}_k$. Note that we do not need to know the value of $\chi^2_0$. The PDFs $f_i^{\rm new}(x,Q^2)$ that correspond to this new minimum are obtained by replacing $y_i$ in eq.~(\ref{eq:XS}) by PDFs,
\begin{equation}
 f_i^{\rm new}(x,Q^2) \approx f_i^{S_0}(x,Q^2) + \sum_{k}  \frac{f^{S^+_k}_i(x,Q^2)-f_i^{S^-_k}(x,Q^2)}{2}  \frac{z^{\rm min}_k}{\sqrt{\Delta \chi^2}}.
\label{eq:newPDF}
\end{equation}
Since we now know $\chi^2_{\rm new}$ analytically, we can repeat the original treatment by computing the new Hessian matrix and diagonalizing it exactly the same way as outlined above. As a result, we have an approximation of how a new set of data has affected a set of PDFs and its errors. In comparison to a full global analysis, the advantage of the reweighting technique is that it avoids the time-consuming fitting procedure which, in practice, is only available to the people that performed the PDF analysis itself. In addition, and also importantly, there is no need to implement a potentially CPU-expensive cross-section computation as a part of the fitting framework or to compute partial cross sections to form three dimensional ($x_1$,$x_2$,$\mu^2_{\rm fact}$) grids to facilitate a rapid cross-section evaluation. The downside is that since the reweighting method relies completely on the assumptions made in the prior PDF analysis, including e.g. a specific parametrization which may artificially overestimate the impact in a kinematic region beyond the reach of a given observable.

The Hessian reweighting method sketched above relied on a linear approximation for the PDFs and observables in the $z$ space, and on a quadratic expansion of the original $\chi^2$ function. These are not always good approximations and, as described in ref.~\cite{Eskola:2019dui}, the results can be refined by taking into account higher order terms in $z$. The results presented in this paper (section~\ref{sec:impact}) have been obtained using a quadratic extension of the approximation made in eq.~(\ref{eq:XS}). In the case of EPPS16 we also take into account cubic terms in the original $\chi^2$ profile, eq.~(\ref{eq:chi2orig}). See ref.~\cite{Eskola:2019dui} for further technical details.

\section{Results}
\label{sec:results}

Throughout this section, we will use two recent globally-fitted nPDF sets, EPPS16 \cite{Eskola:2016oht} and nCTEQ15 \cite{Kovarik:2015cma}, in our calculations. In the case of EPPS16 we use CT14NLO \cite{Dulat:2015mca} as the free proton PDF set and with nCTEQ15 we use its own proton PDF (with no uncertainties on it). As a default setup for the GM-VFNS calculation we adopt the KKKS08 \cite{Kneesch:2007ey} parton-to-hadron FFs and set the renormalization and factorization scales as $\mu_{\mathrm{ren}} = \mu_{\mathrm{fact}} = \sqrt{\PT^2+m_\mathrm{c}^2}$ with $m_{\mathrm{c}} = 1.3~\text{GeV}$ for the charm quark mass. For the fragmentation scale we set $\mu_{\mathrm{frag}} = \sqrt{\PT^2+(1.5~\text{GeV})^2}$ as the KKKS08 analysis assumed this slightly higher value for the charm-quark mass. In the matrix elements we always use $m_{\mathrm{c}} = 1.3~\text{GeV}$.  For the $\D0$ mass, relevant for transforming the partonic kinematics to hadronic ones, we adopt the value $M_{\D0} = 1.87~\text{GeV}$ \cite{Tanabashi:2018oca}. With the \textsc{Powheg} approach, we use the same nuclear and proton PDFs and the same value for the charm mass but the renormalization and factorization scales are fixed to transverse mass of the produced charm quark, $\sqrt{\pT^2+m_\mathrm{c}^2}$. At the time of generating the partonic events with \textsc{Powheg} it is not yet known which $\PT$ the D meson will have (if formed at all), so relating the scales to the partonic variables is the only reasonable option. The parton shower and hadronization for the \textsc{Powheg} events are generated with the \textsc{Pythia} version 8.235 \cite{Sjostrand:2014zea} using parameters from the default \textsc{Monash} tune \cite{Skands:2014pea}.

\subsection{Double-differential cross section for $\D0$ production in p+Pb collisions}

To benchmark our GM-VFNS framework in p+Pb collisions we first compare our calculations with the double-differential single-inclusive $\D0$ production cross section measured by LHCb \cite{Aaij:2017gcy}. This comparison is important since a good agreement with the measured cross sections would indicate that the framework includes e.g. all the relevant partonic processes. In this way we ensure that the framework is realistic.  

\begin{figure}[ptbh]
\begin{center}
\includegraphics[width=0.45\textwidth]{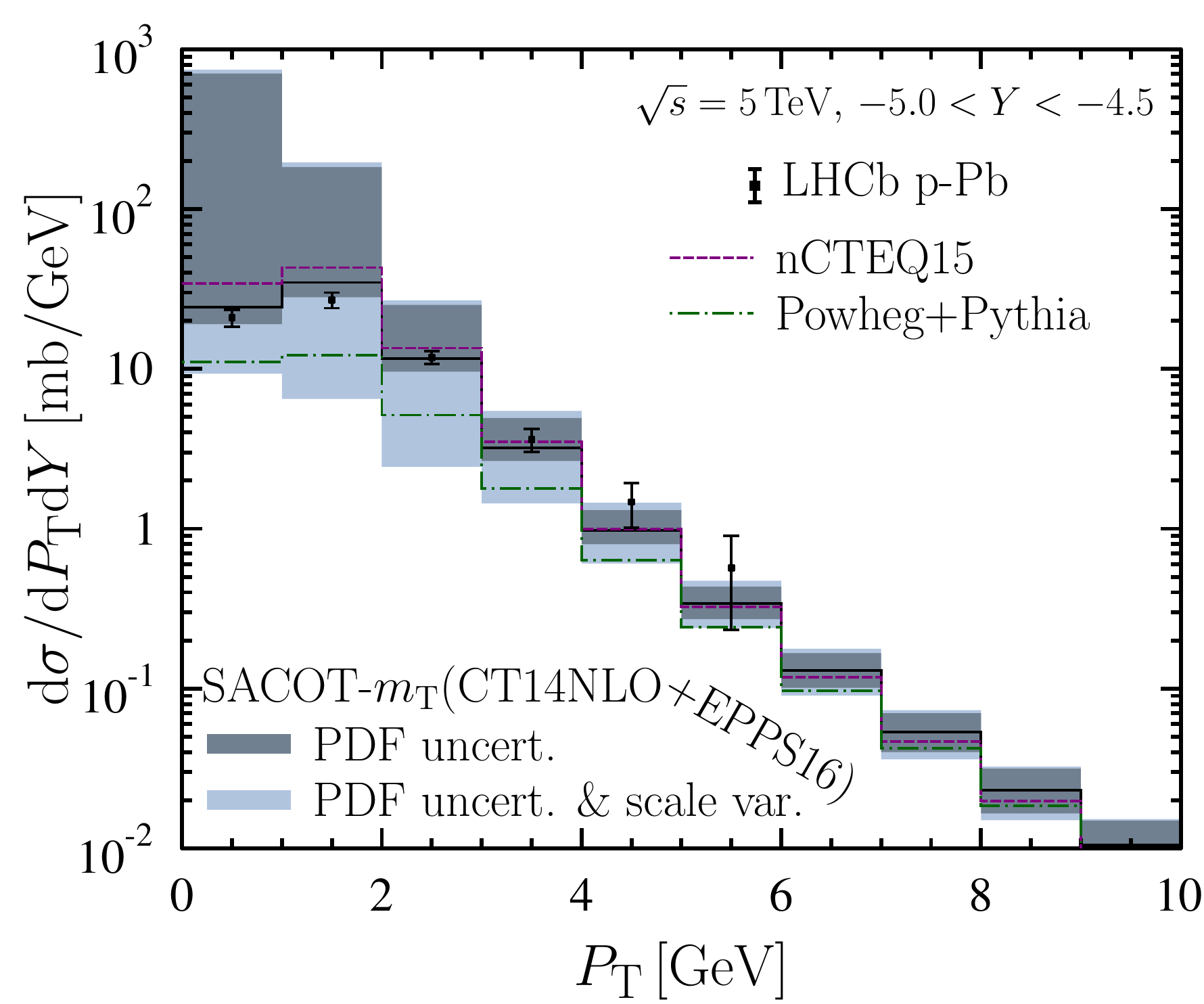}
\includegraphics[width=0.45\textwidth]{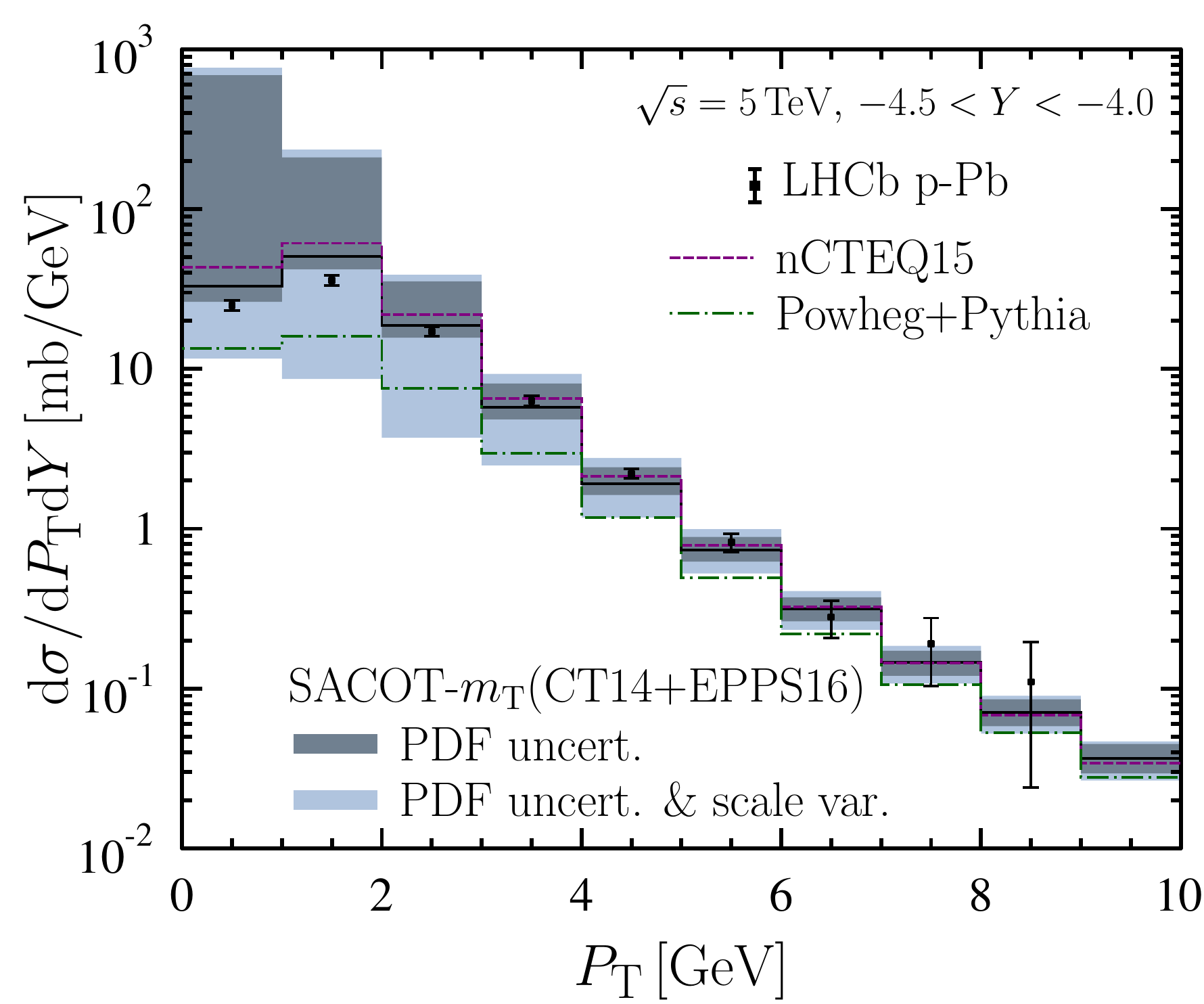}
\includegraphics[width=0.45\textwidth]{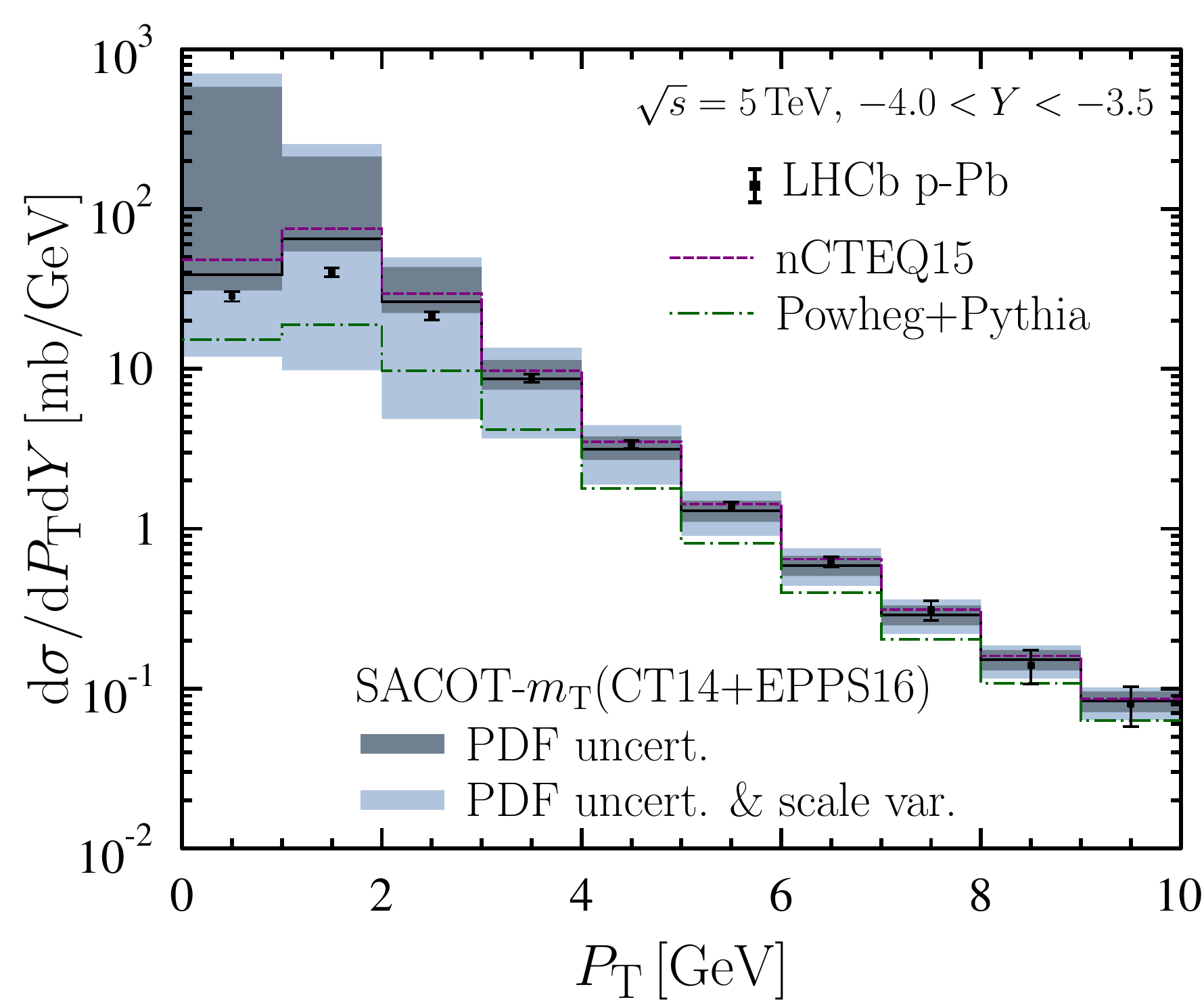}
\includegraphics[width=0.45\textwidth]{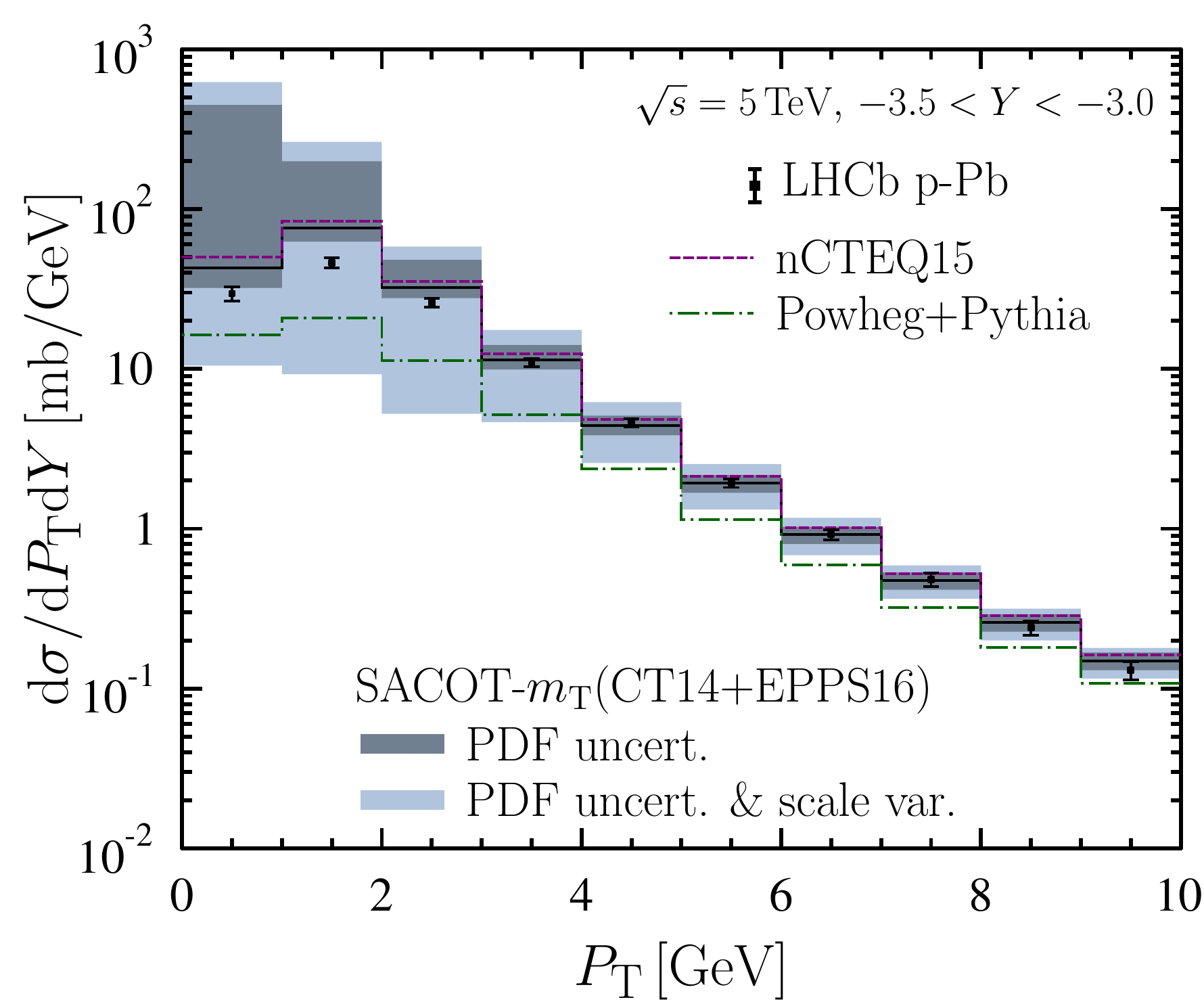}
\includegraphics[width=0.45\textwidth]{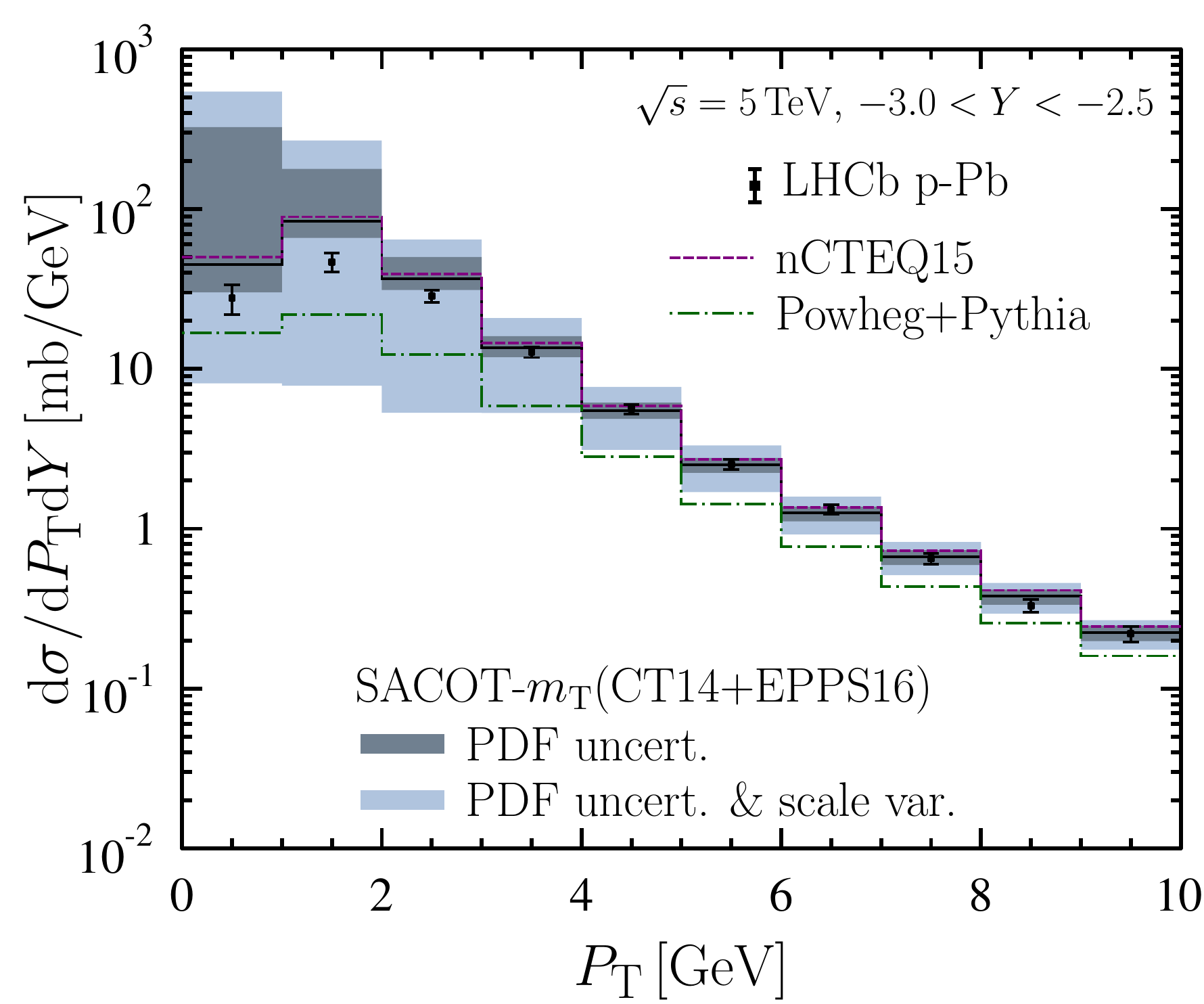}
\caption{Double-differential cross section for $\D0$ production in p+Pb collisions at $\sqrt{s_{\mathrm{NN}}}=5.0~\text{TeV}$ in five different $Y$ bins at backward rapidities. Data from LHCb \cite{Aaij:2017gcy} are compared to the GM-VFNS calculations with EPPS16 (solid black) and nCTEQ15 (dashed purple) nPDFs, and to a \textsc{Powheg}+\textsc{Pythia} setup with EPPS16 nPDFs (dot-dashed green). The theoretical uncertainties related to the PDFs are shown with dark grey and the combination of the scale variations and PDF uncertainties with light blue.}
\label{fig:dsigma_pPb_backward}
\end{center}
\end{figure}

In figure \ref{fig:dsigma_pPb_backward} we compare the calculated cross sections with the LHCb data at backward rapidities (Pb-going direction) in five different rapidity bins spanning $-5.0 < Y < -2.5$ in the nucleon-nucleon (NN) c.m. frame. The resulting cross sections with the default setup are shown for both the EPPS16 and nCTEQ15 nPDFs, whereas the theoretical uncertainties are quantified with EPPS16 only. These include now scale variations and PDF uncertainties. The former are calculated by varying the three QCD scales independently by a factor of two around the default choice. In addition, ratios $\mu_{\mathrm{fact}}/\mu_{\mathrm{ren}}$ and $\mu_{\mathrm{frag}}/\mu_{\mathrm{ren}}$ are required to stay within $[0.5,2]$ and the mass of the charm quark is used as a lower limit for all scales. For the PDF uncertainties the error bands from proton and nuclear PDFs are added in quadrature as they are approximately independent in the EPPS16 global analysis. Here, we use the asymmetric error prescription
\begin{align}
\Delta X^+ & = \sqrt{\sum_k {\max}\left[X({S^+_k}) - X({S^0}), X({S^-_k}) - X({S^0_k}),0\right]^2} \,, \\
\Delta X^- & = \sqrt{\sum_k {\min}\left[X({S^+_k}) - X({S^0}), X({S^-_k}) - X({S^0_k}),0\right]^2}  \,,
\end{align}
where the sum now runs over both the EPPS16 and CT14NLO error sets. Uncertainties due to the mentioned ambiguity in defining the fragmentation variable $z$, FFs, or e.g. variation in charm-quark mass are not considered. In addition to the GM-VFNS results, comparison with the \textsc{Powheg}+\textsc{Pythia} setup is shown.
The correspondence between the data and the GM-VFNS calculation with both EPPS16 and nCTEQ15 is found to be very good, though the theoretical uncertainties become large at $\PT<3~\text{GeV}$. Interestingly the PDF uncertainty at small $\PT$ is large above the central result but small below it. This can be traced back to the parametrization applied in the CT14 analysis where the requirement for positive-definite PDFs limits the small-$x$ behaviour as already the central set for gluons near the initial scale $Q^2_0$ at small $x$ is close to zero. Since similar positivity restriction was not applied in NNPDF3.1 \cite{Ball:2017nwa}, the PDF uncertainties shown in ref.~\cite{Helenius:2018uul} behave in a different manner at small values of $\PT$. As in the p+p case \cite{Helenius:2018uul}, the cross sections obtained with the \textsc{Powheg}+\textsc{Pythia} approach fall below the GM-VFNS results, albeit the spread is of the same order as the theoretical uncertainties in the applied GM-VFNS formalism.

\begin{figure}[ptbh]
\begin{center}
\includegraphics[width=0.45\textwidth]{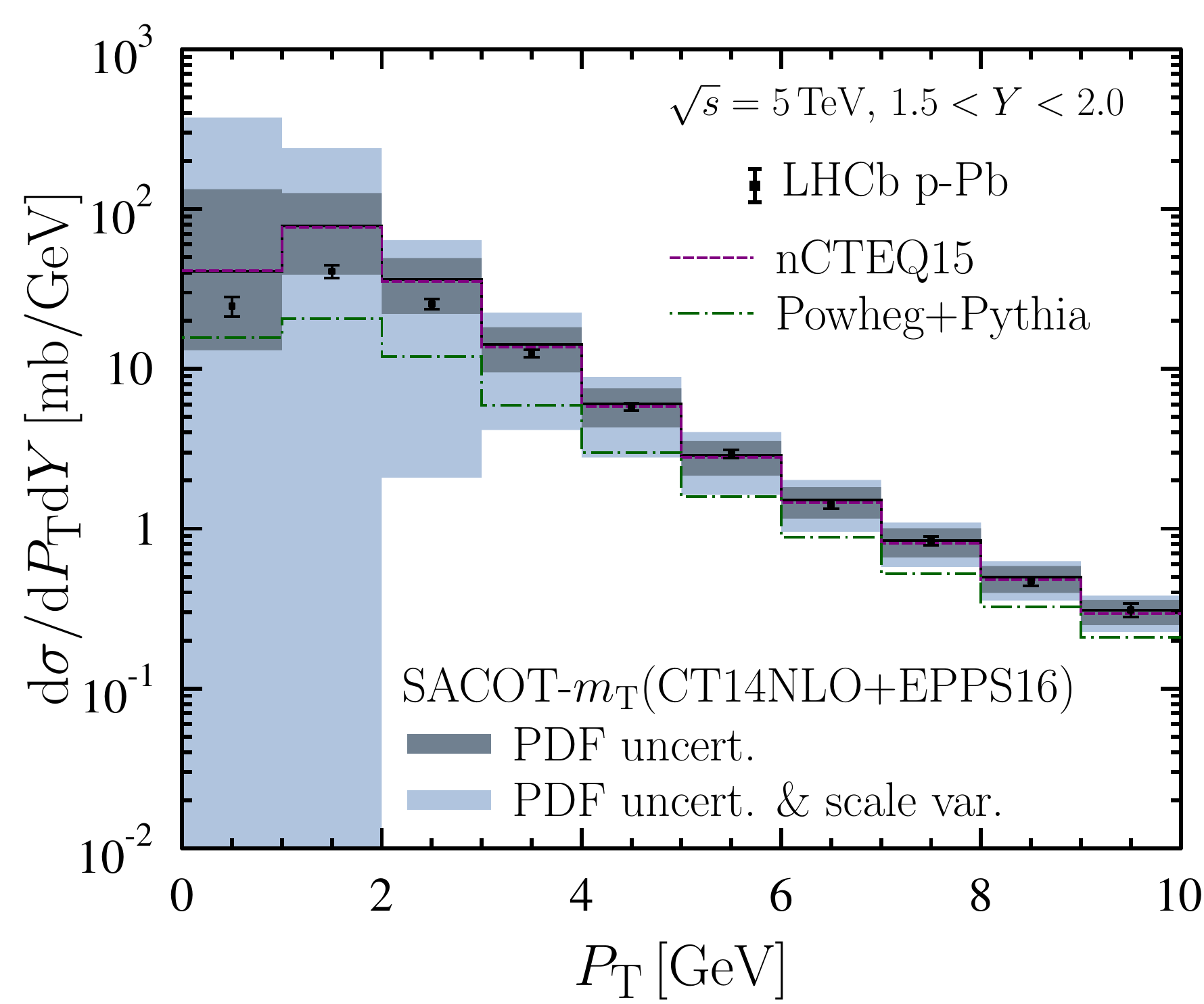}
\includegraphics[width=0.45\textwidth]{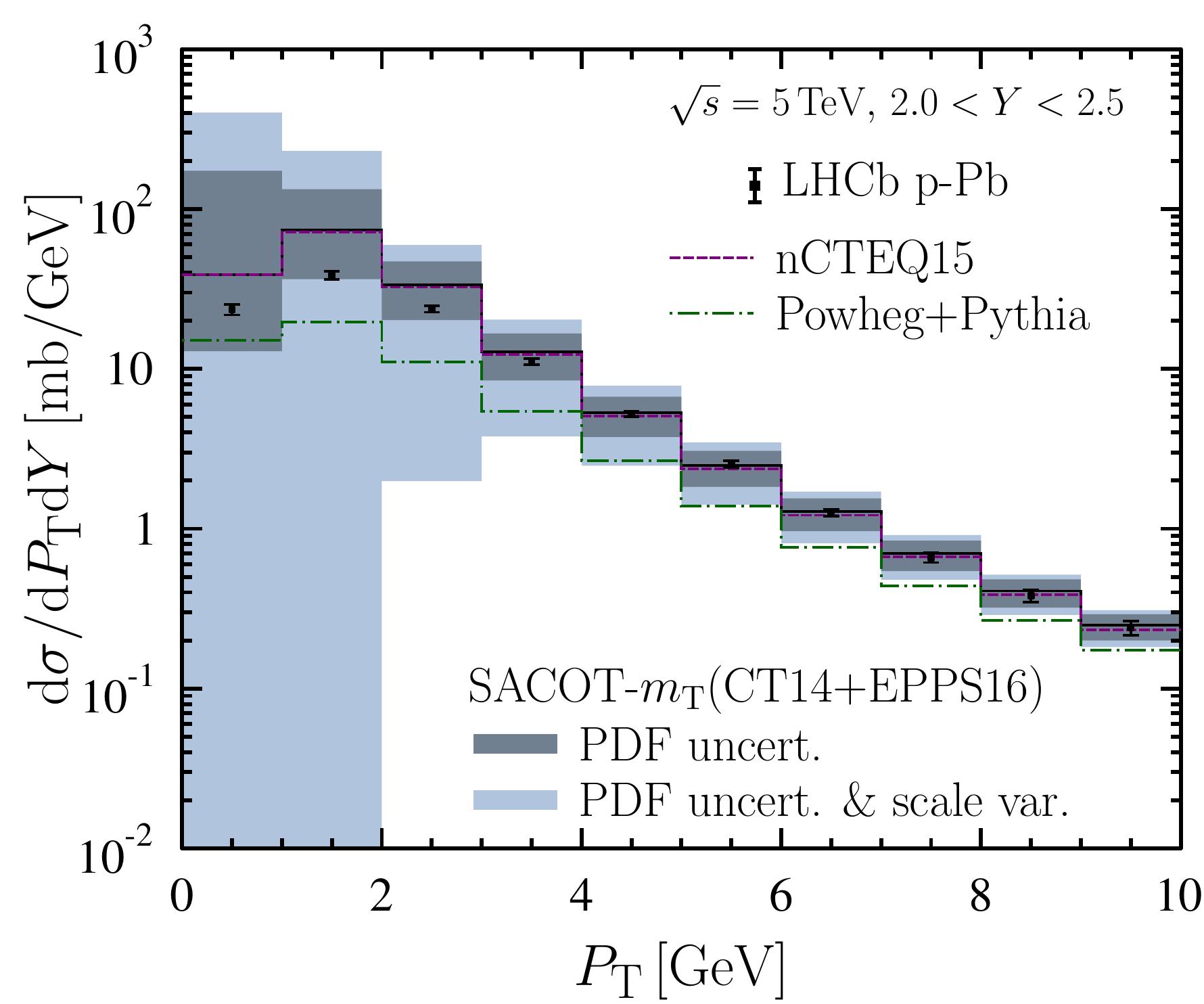}
\includegraphics[width=0.45\textwidth]{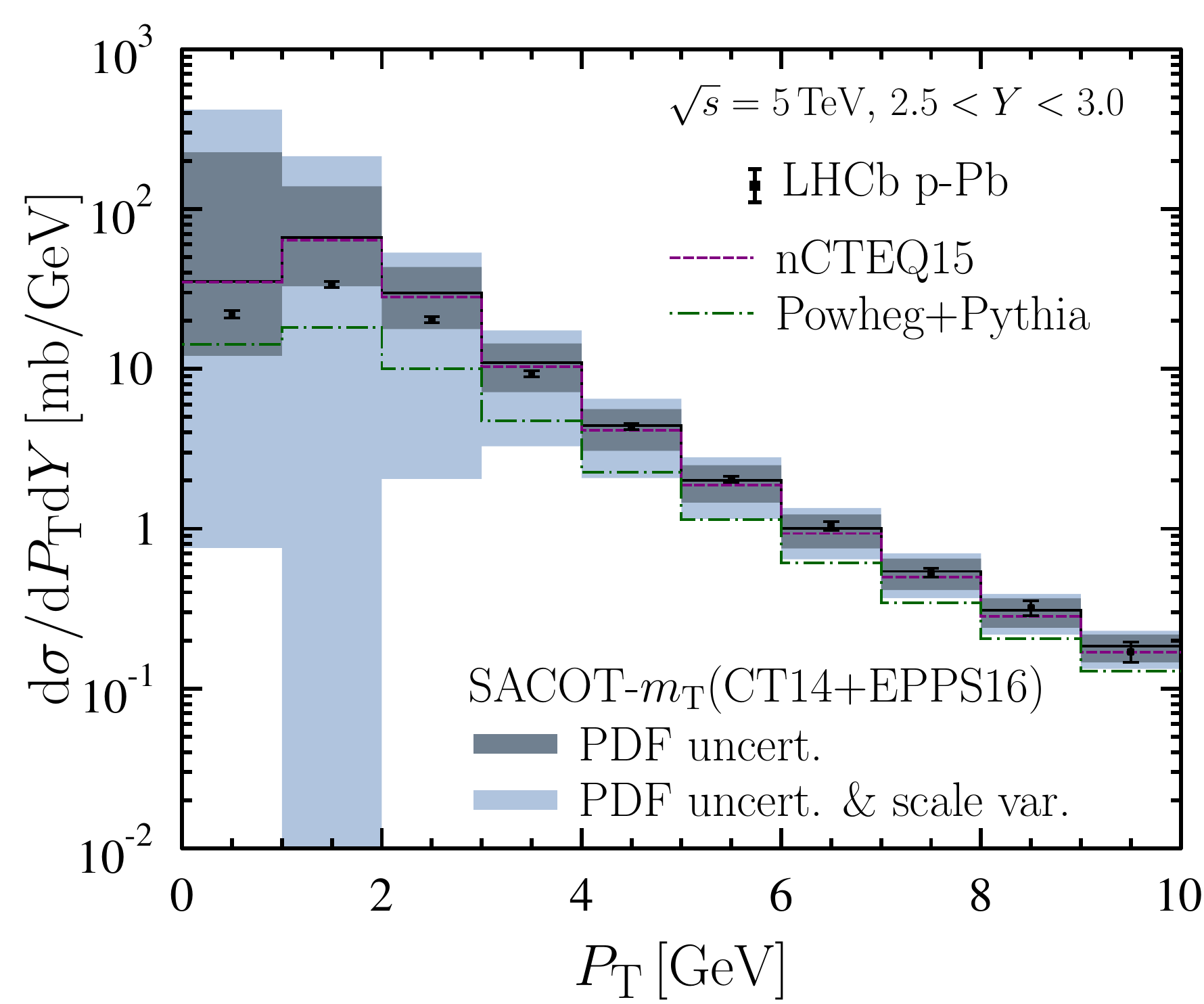}
\includegraphics[width=0.45\textwidth]{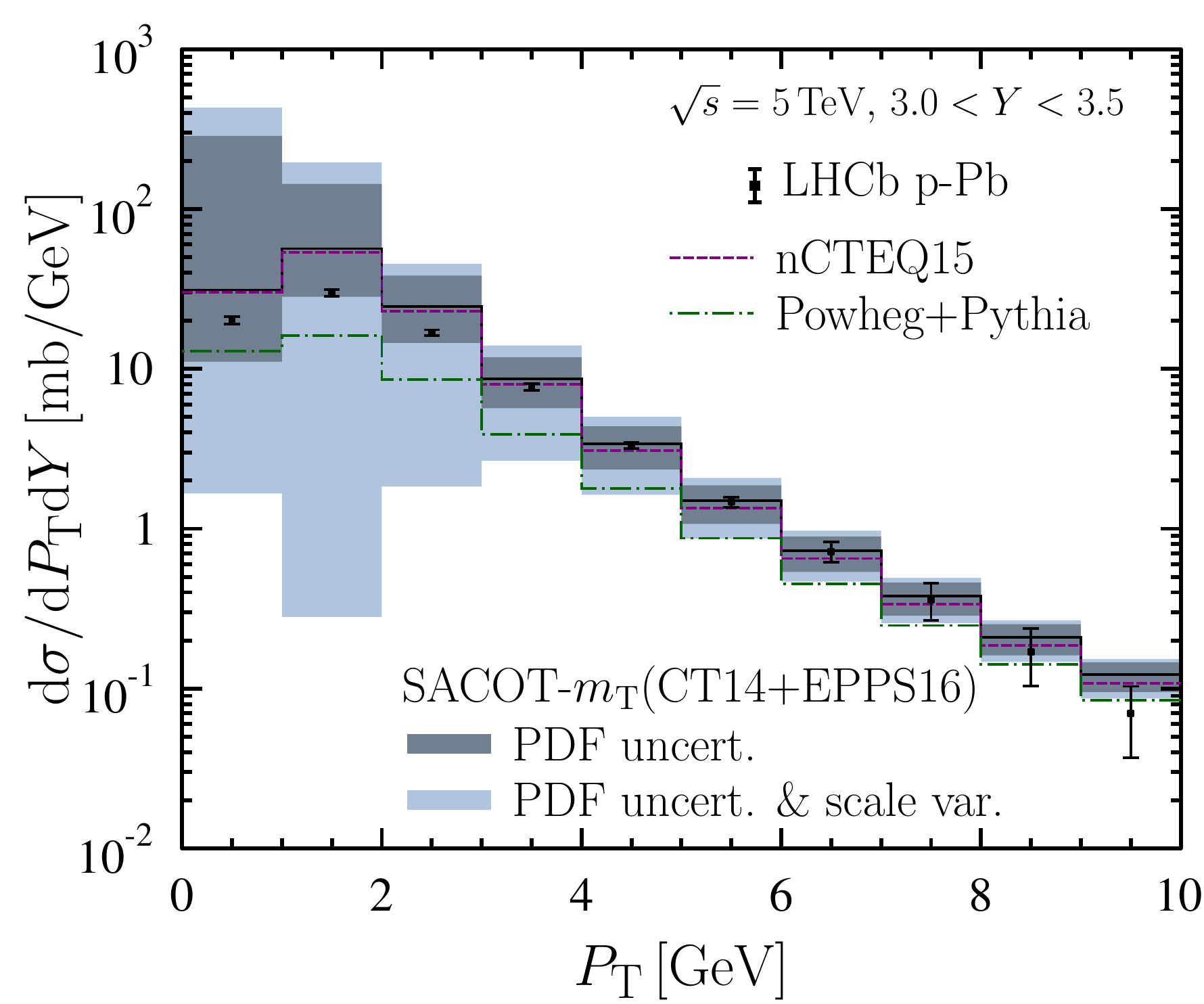}
\includegraphics[width=0.45\textwidth]{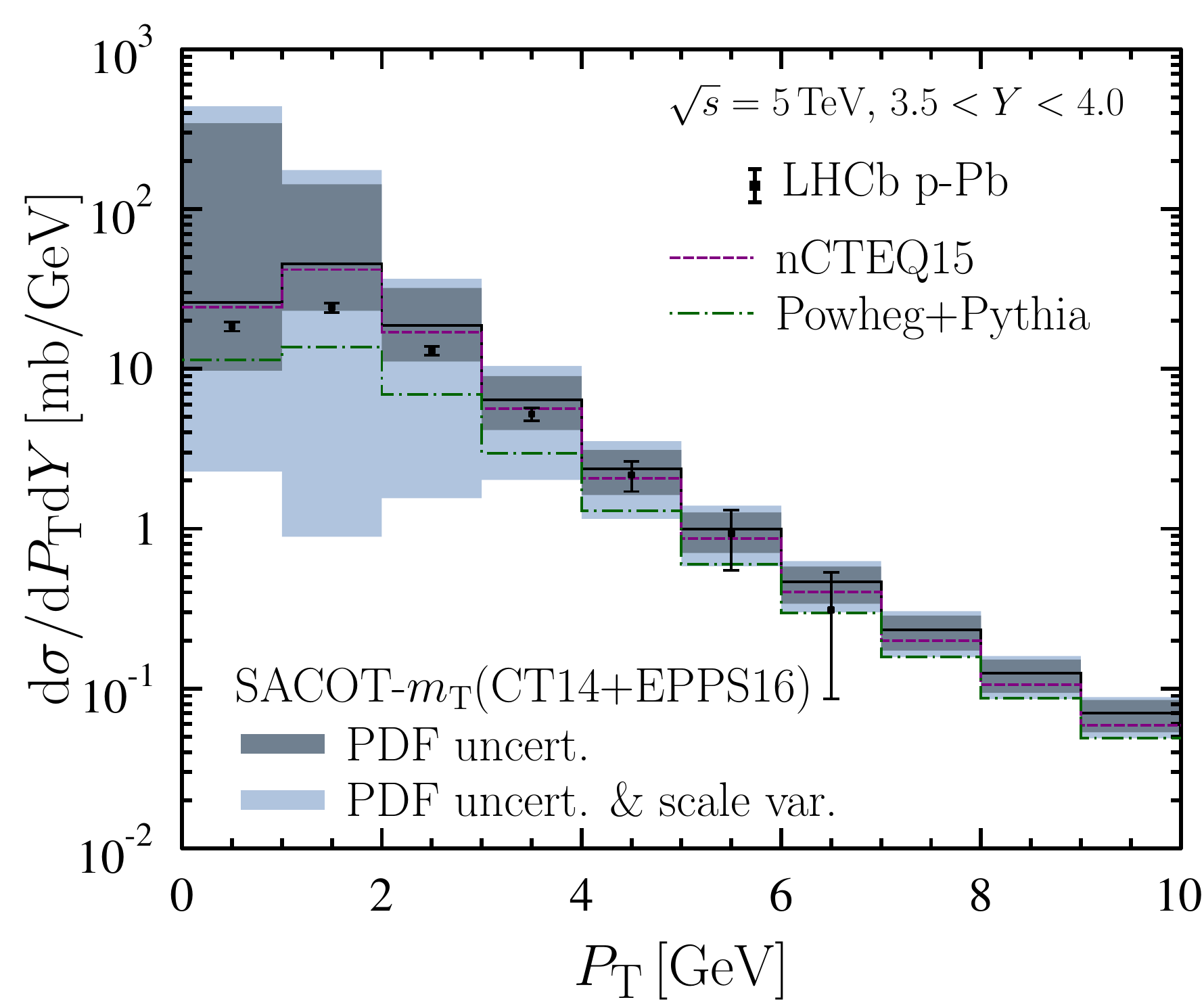}
\caption{Same as figure \ref{fig:dsigma_pPb_backward} but at forward rapidities.}
\label{fig:dsigma_pPb_forward}
\end{center}
\end{figure}

The corresponding cross sections at forward rapidities (p-going direction) are shown in figure \ref{fig:dsigma_pPb_forward}. Here the five rapidity bins cover the range $1.5<Y<4.0$. The conclusions are very similar as at backwards rapidities, the agreement between the GM-VFNS calculation and the data being very good, particularly at $\PT \gtrsim 3~\text{GeV}$ where the theoretical uncertainties are in control. The comparisons with the absolute cross sections lead us to conclude that the SACOT-$m_{\rm T}$ framework \cite{Helenius:2018uul} works very well also for p+Pb collisions and can be faithfully applied to study the nPDF constraints --- at least for $\PT \gtrsim 3~\text{GeV}$.

\subsection{Nuclear modification ratio for $\D0$ production in p+Pb collisions}

\begin{figure}[ptbh]
\begin{center}
\includegraphics[width=0.4\textwidth]{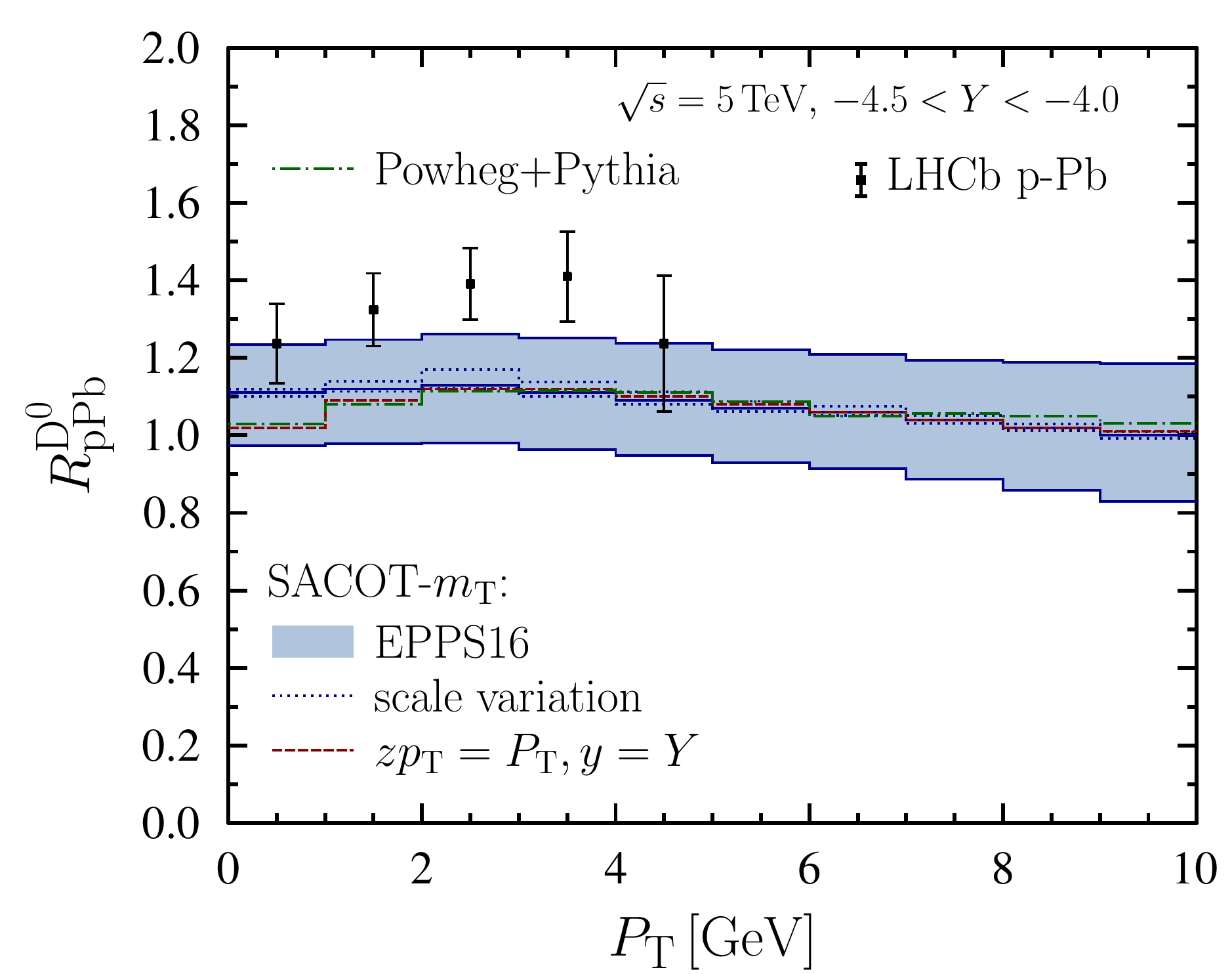}
\includegraphics[width=0.4\textwidth]{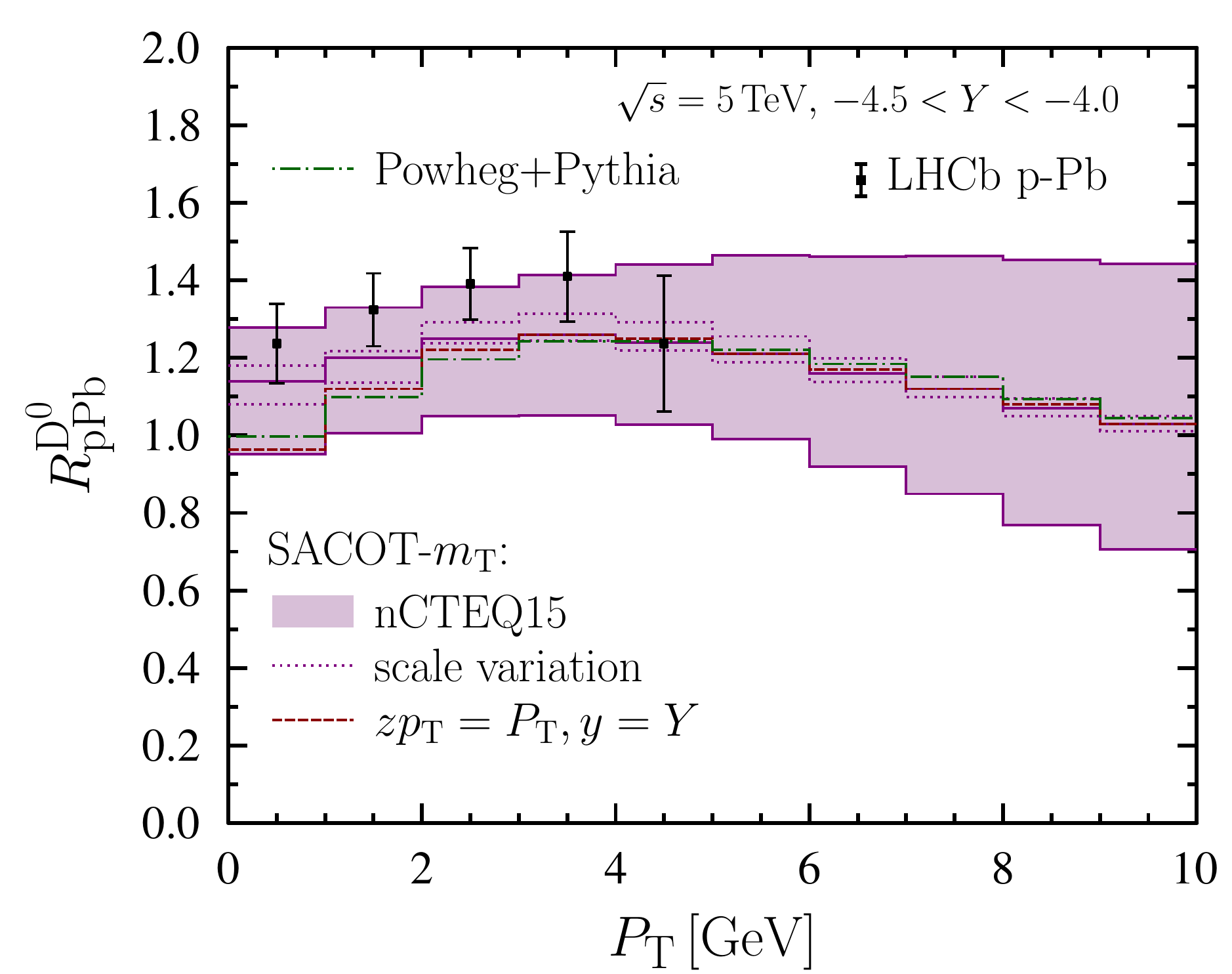}
\includegraphics[width=0.4\textwidth]{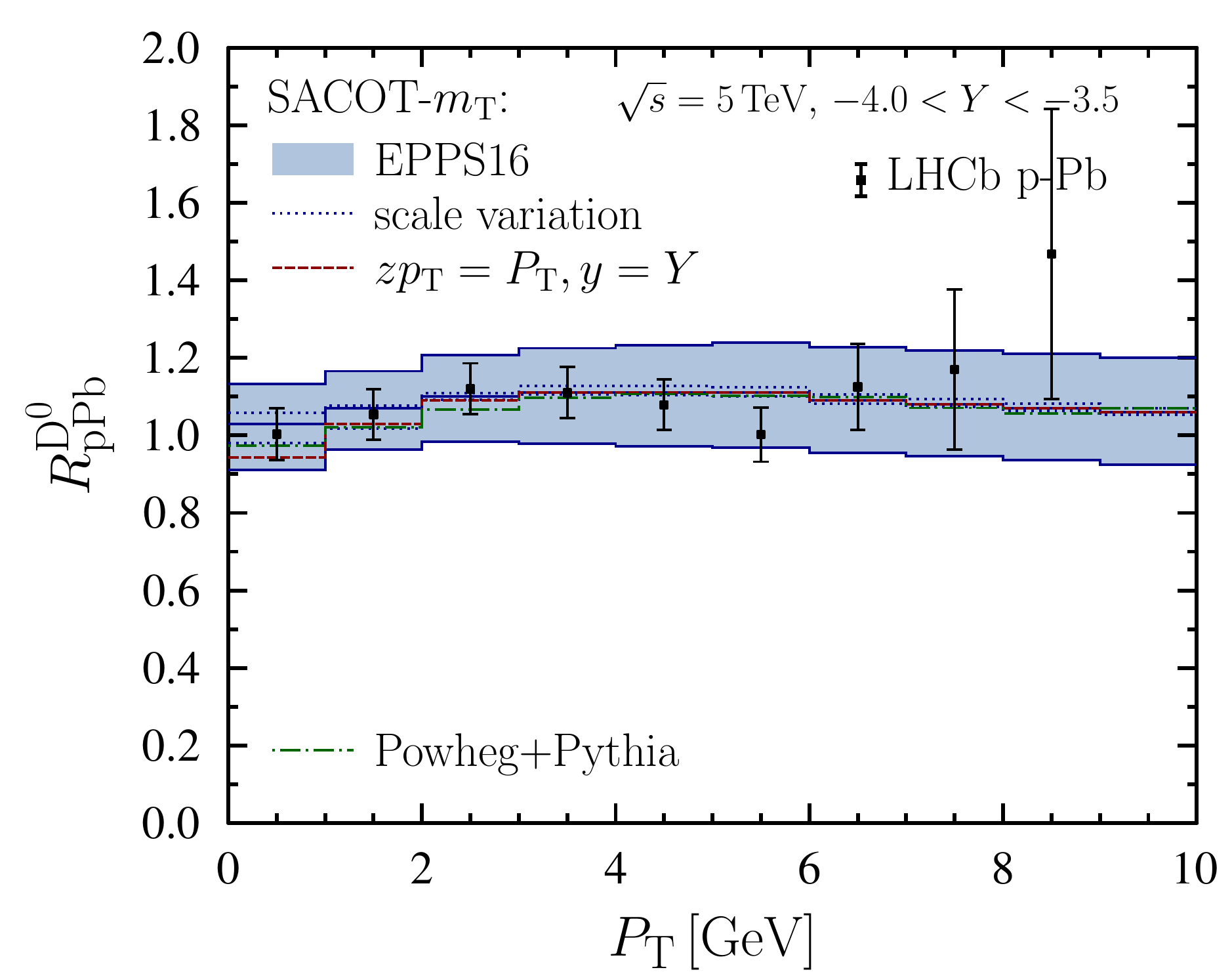}
\includegraphics[width=0.4\textwidth]{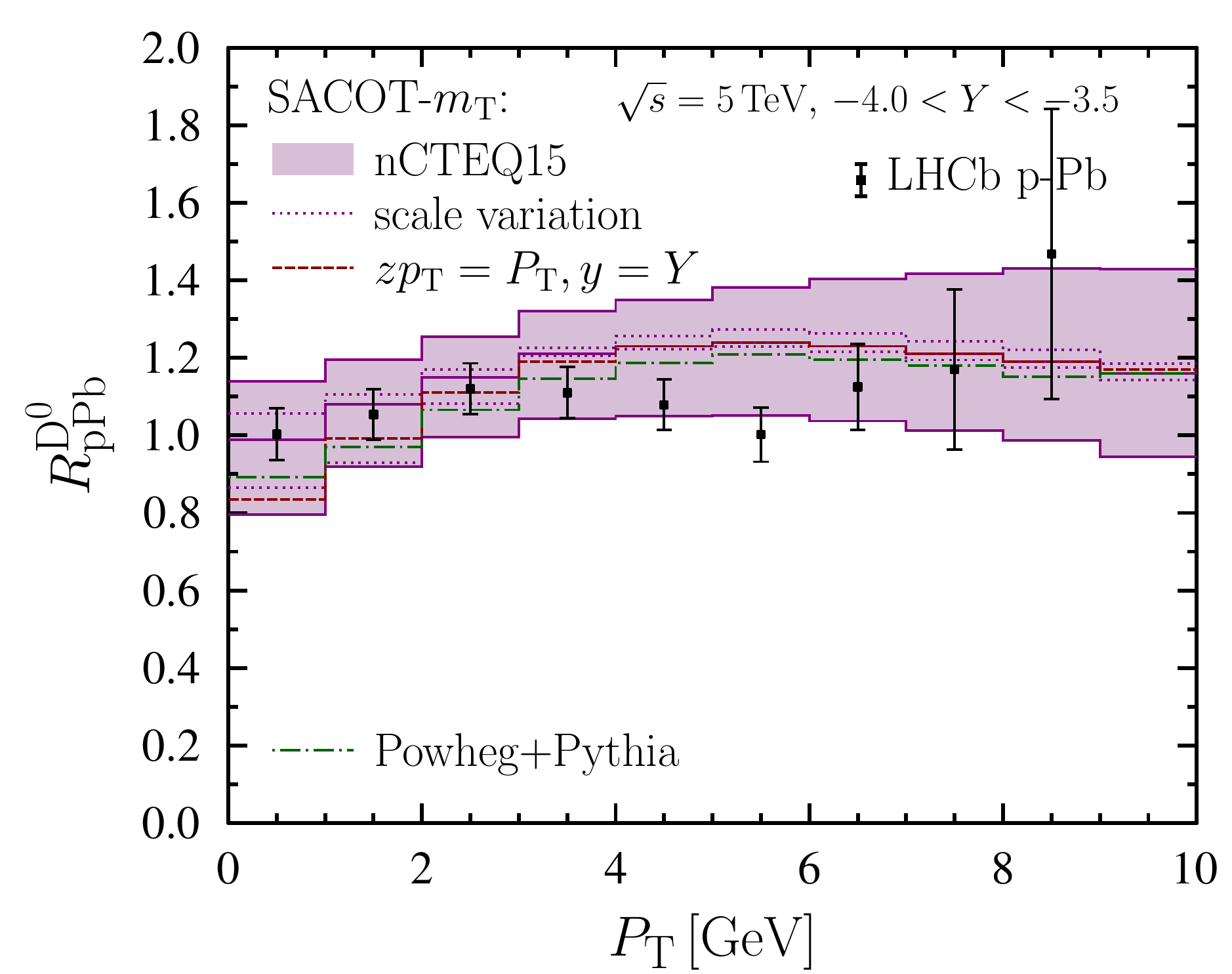}
\includegraphics[width=0.4\textwidth]{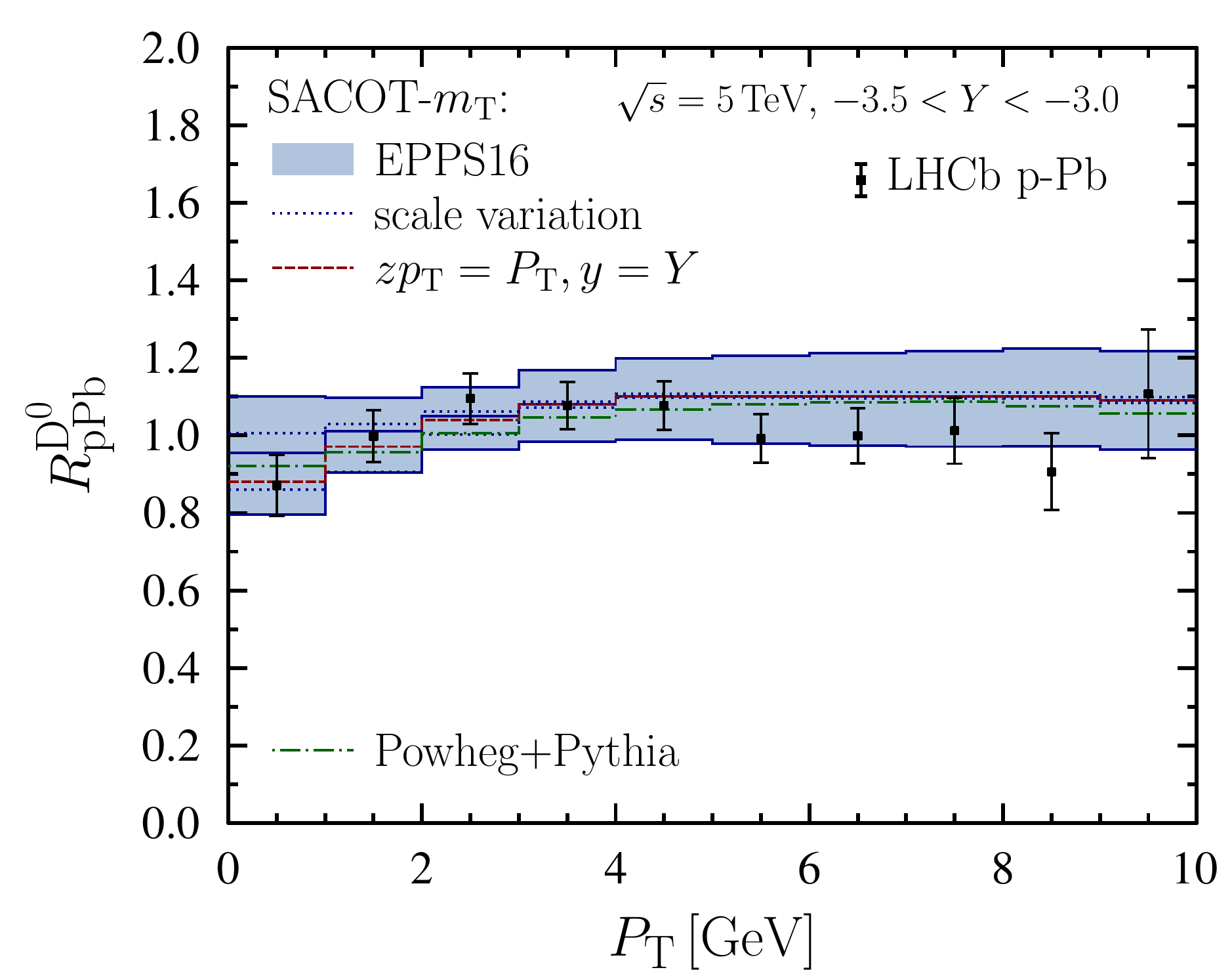}
\includegraphics[width=0.4\textwidth]{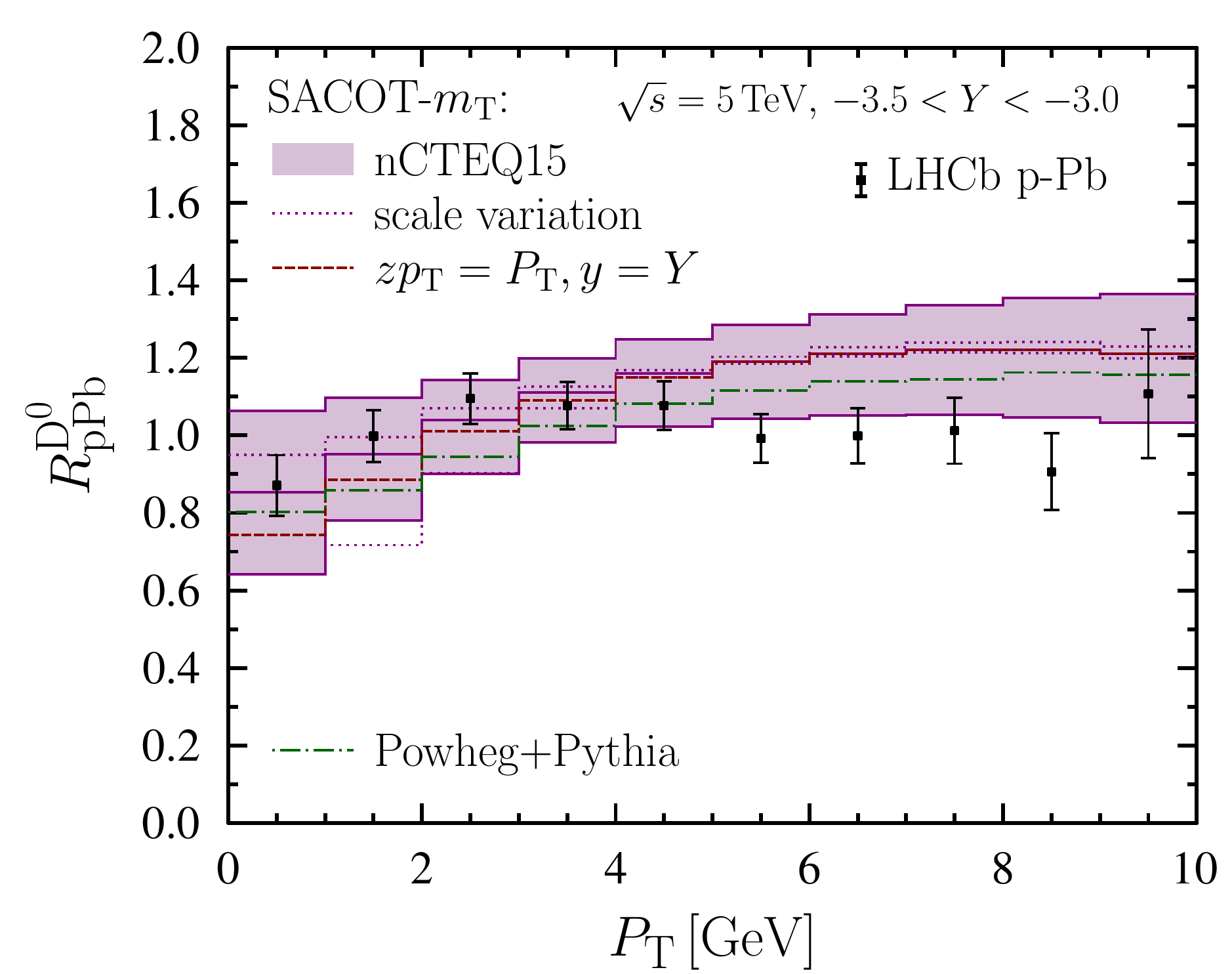}
\includegraphics[width=0.4\textwidth]{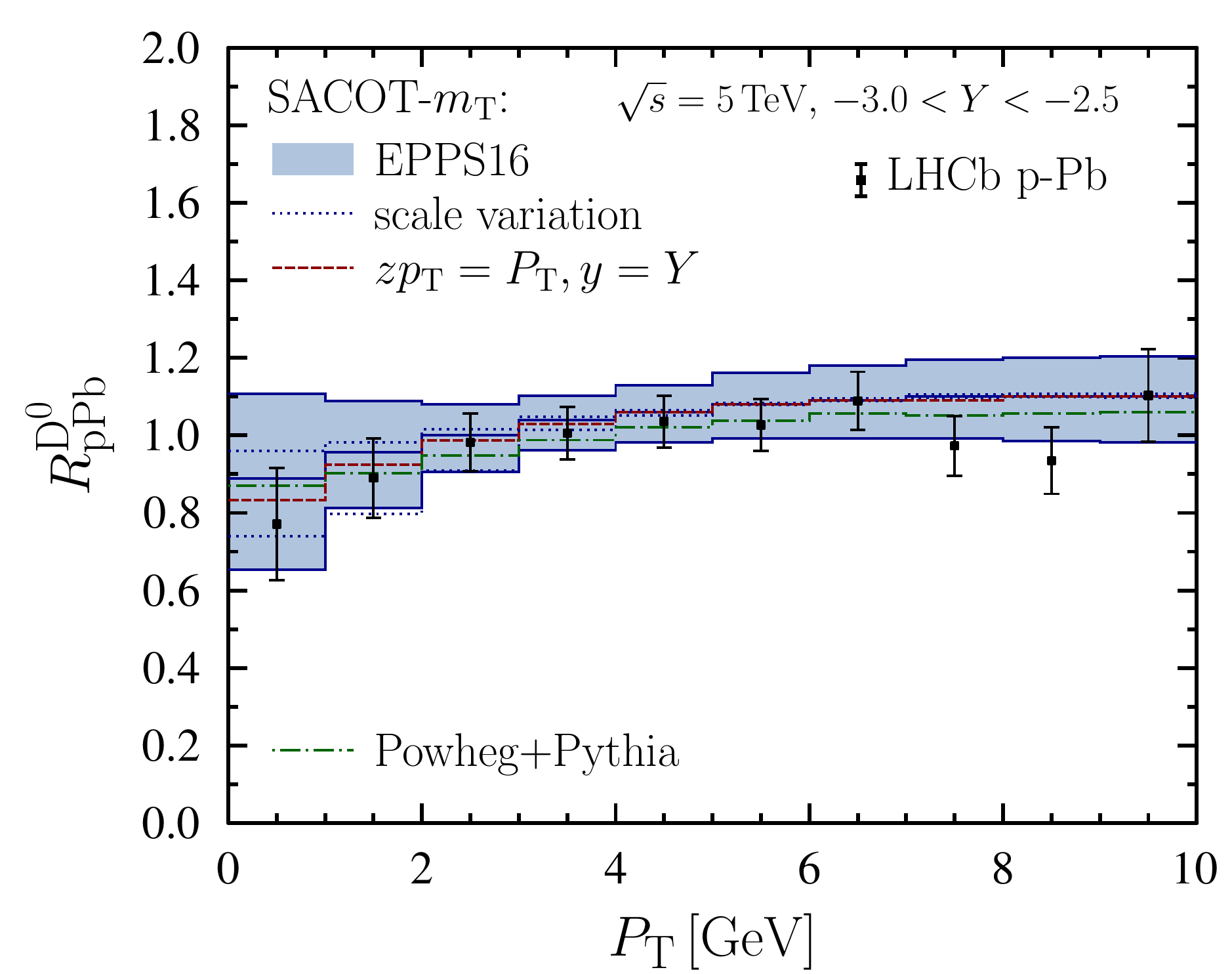}
\includegraphics[width=0.4\textwidth]{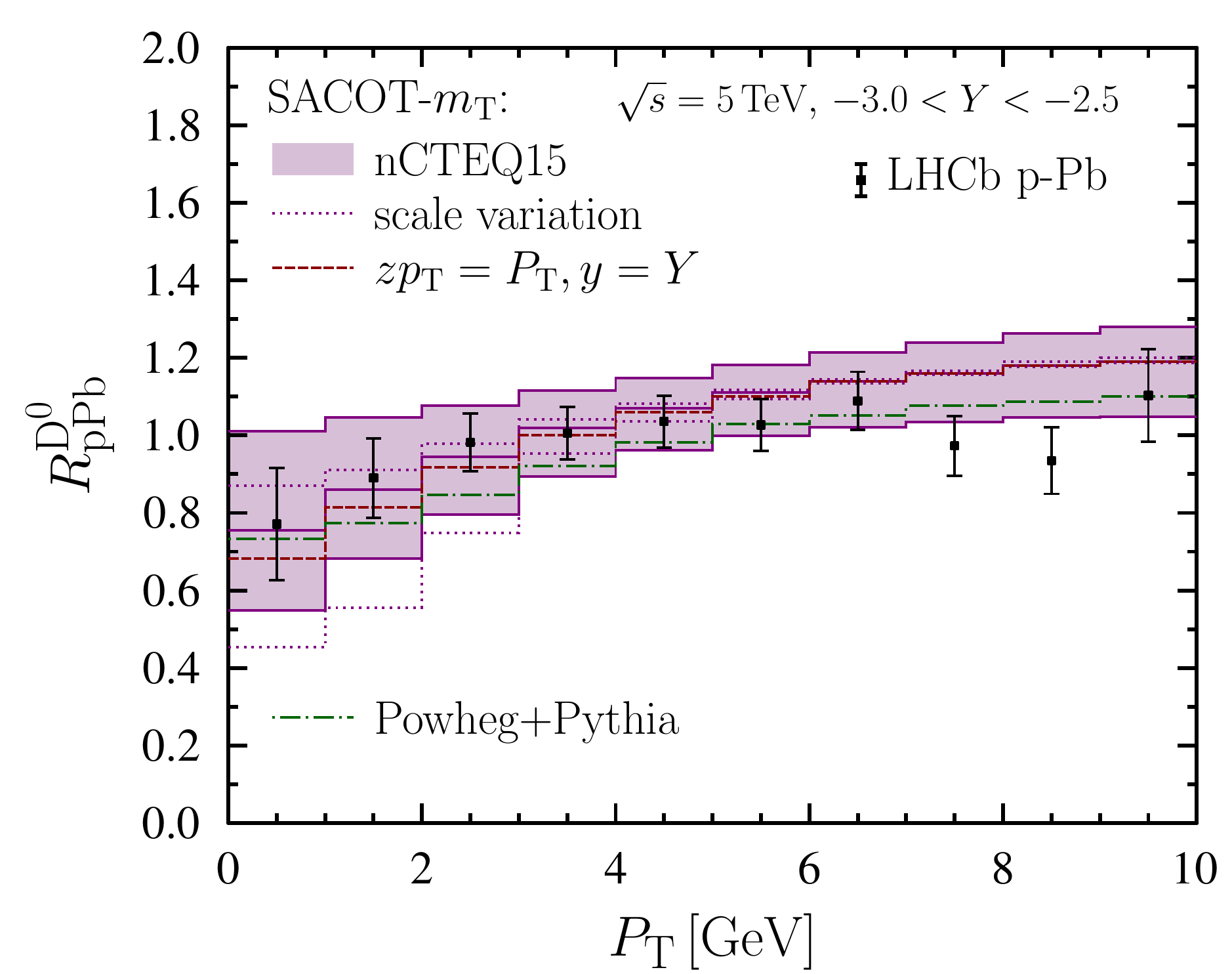}
\caption{Nuclear modification ratio for $\D0$ production in p+Pb collisions in different backward-rapidity bins from the LHCb measurement \cite{Aaij:2017gcy} (black points with error bars) and the SACOT-$m_{\mathrm{T}}$ calculation with the EPPS16 (left) and nCTEQ15 (right) nPDFs. In addition to the central result (solid) and the nPDF-originating uncertainties (coloured bands), the scale variations (dotted band) and the result with massless definition of the fragmentation variable (dashed) are shown, as well as the \textsc{Powheg}+\textsc{Pythia} predictions (dot-dashed).}
\label{fig:R_pPb_backward}
\end{center}
\end{figure}

\begin{figure}[ptbh]
\begin{center}
\includegraphics[width=0.4\textwidth]{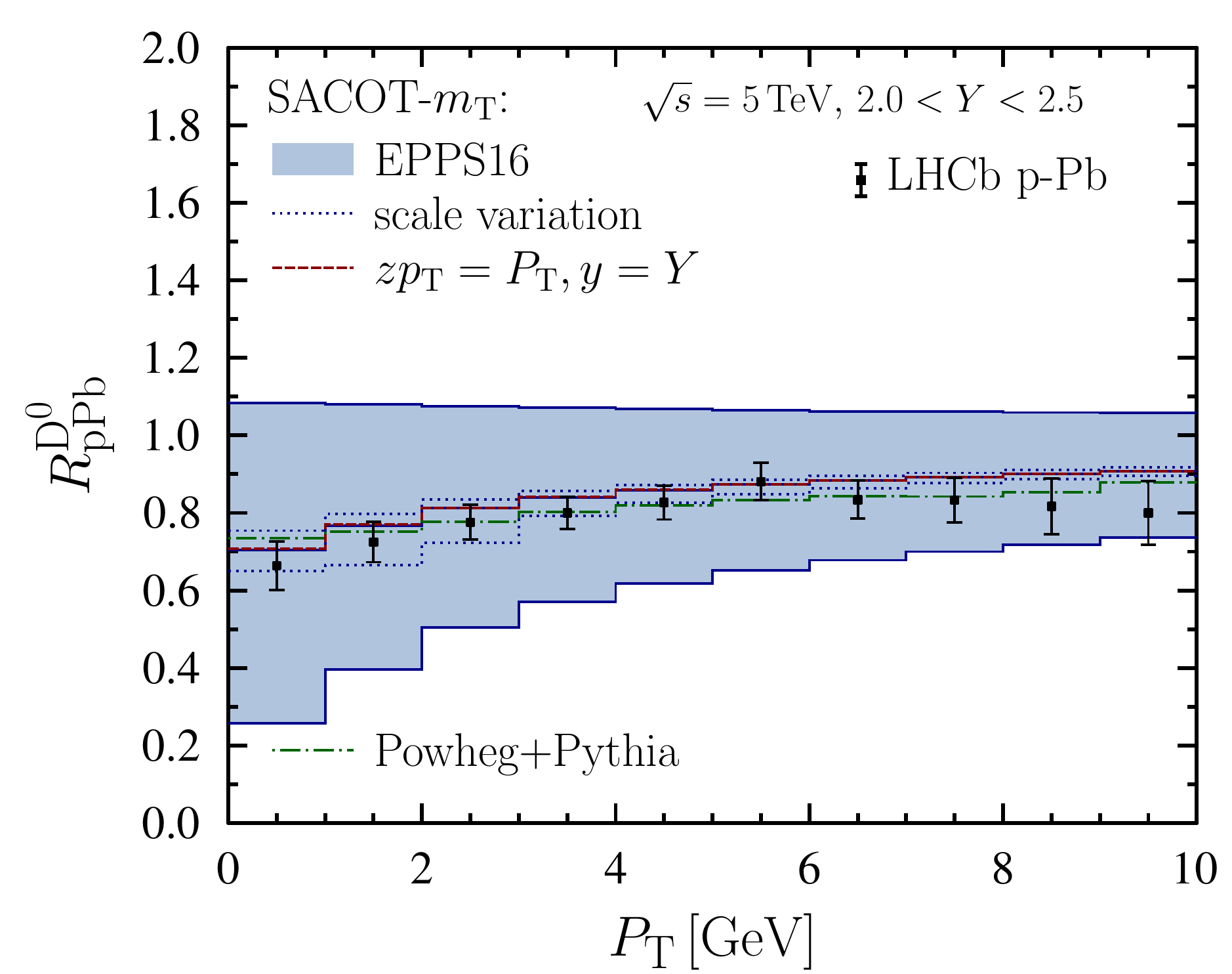}
\includegraphics[width=0.4\textwidth]{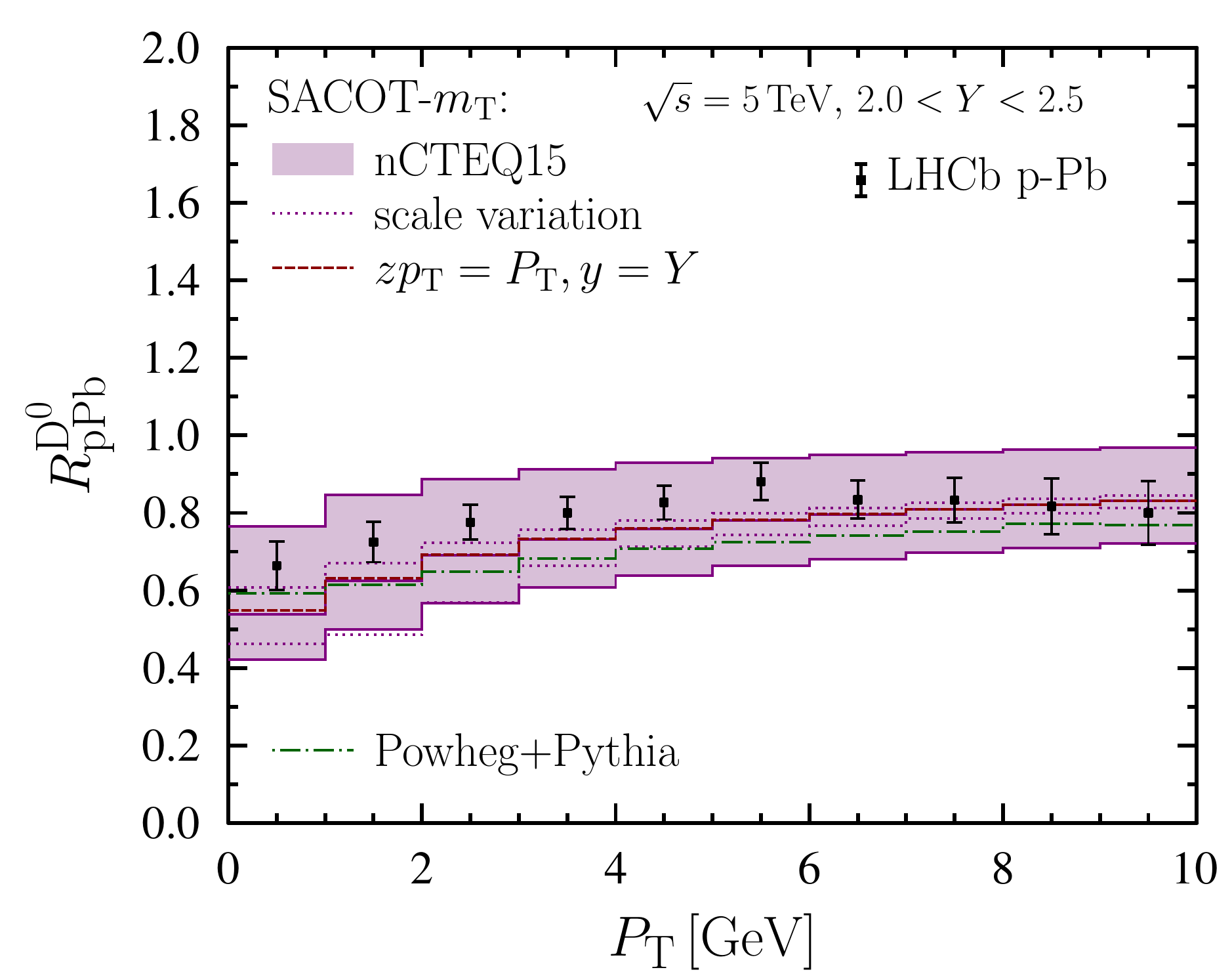}
\includegraphics[width=0.4\textwidth]{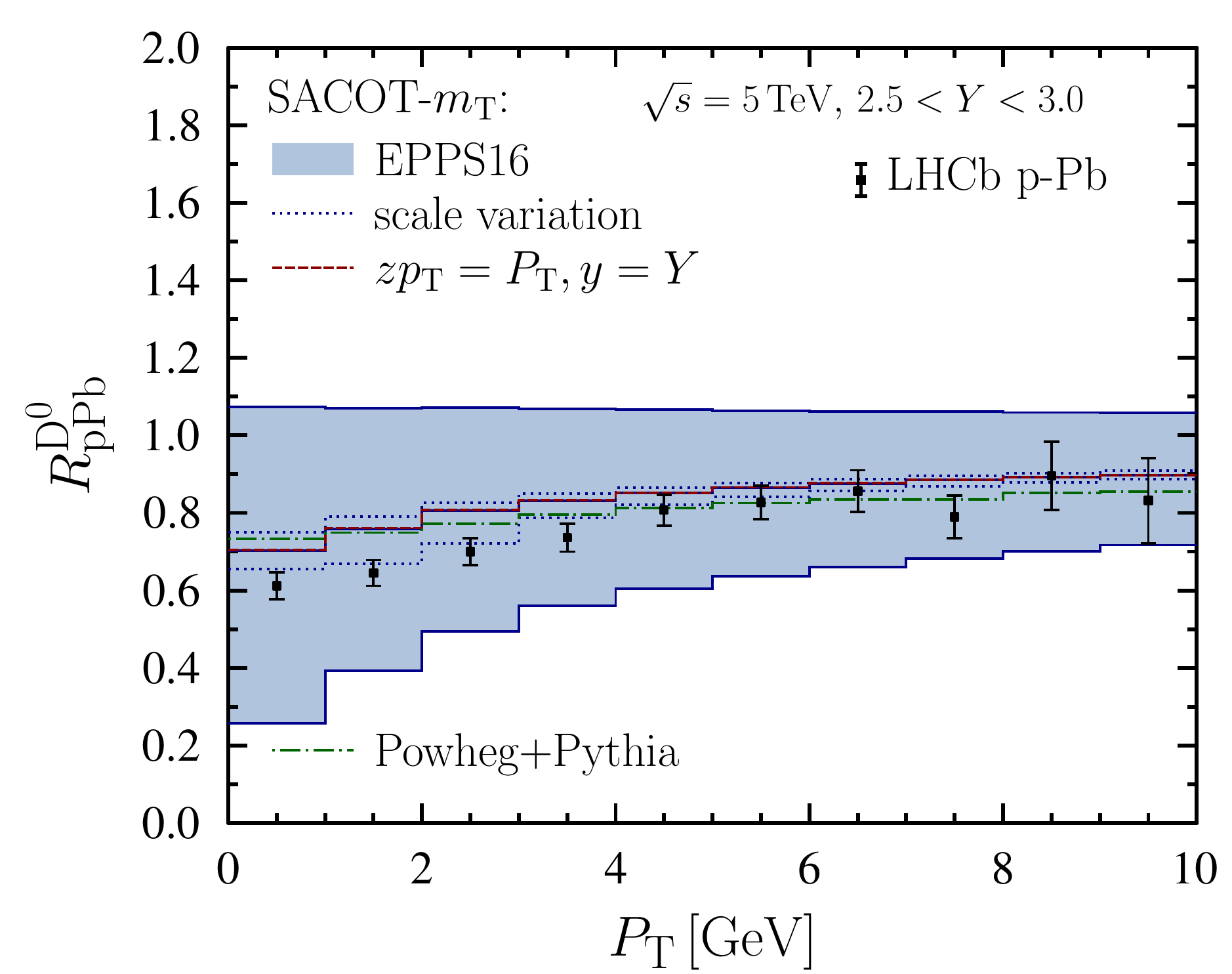}
\includegraphics[width=0.4\textwidth]{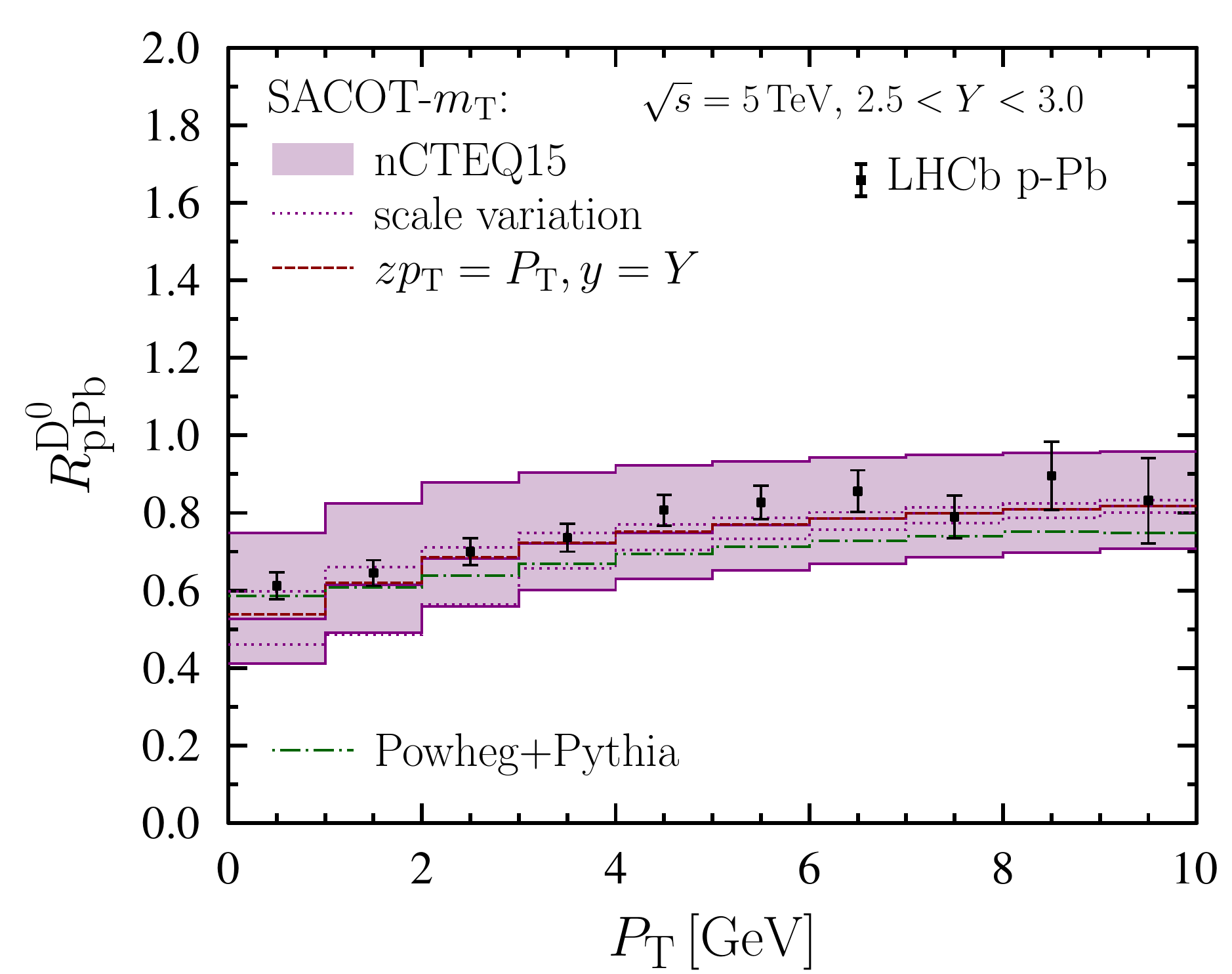}
\includegraphics[width=0.4\textwidth]{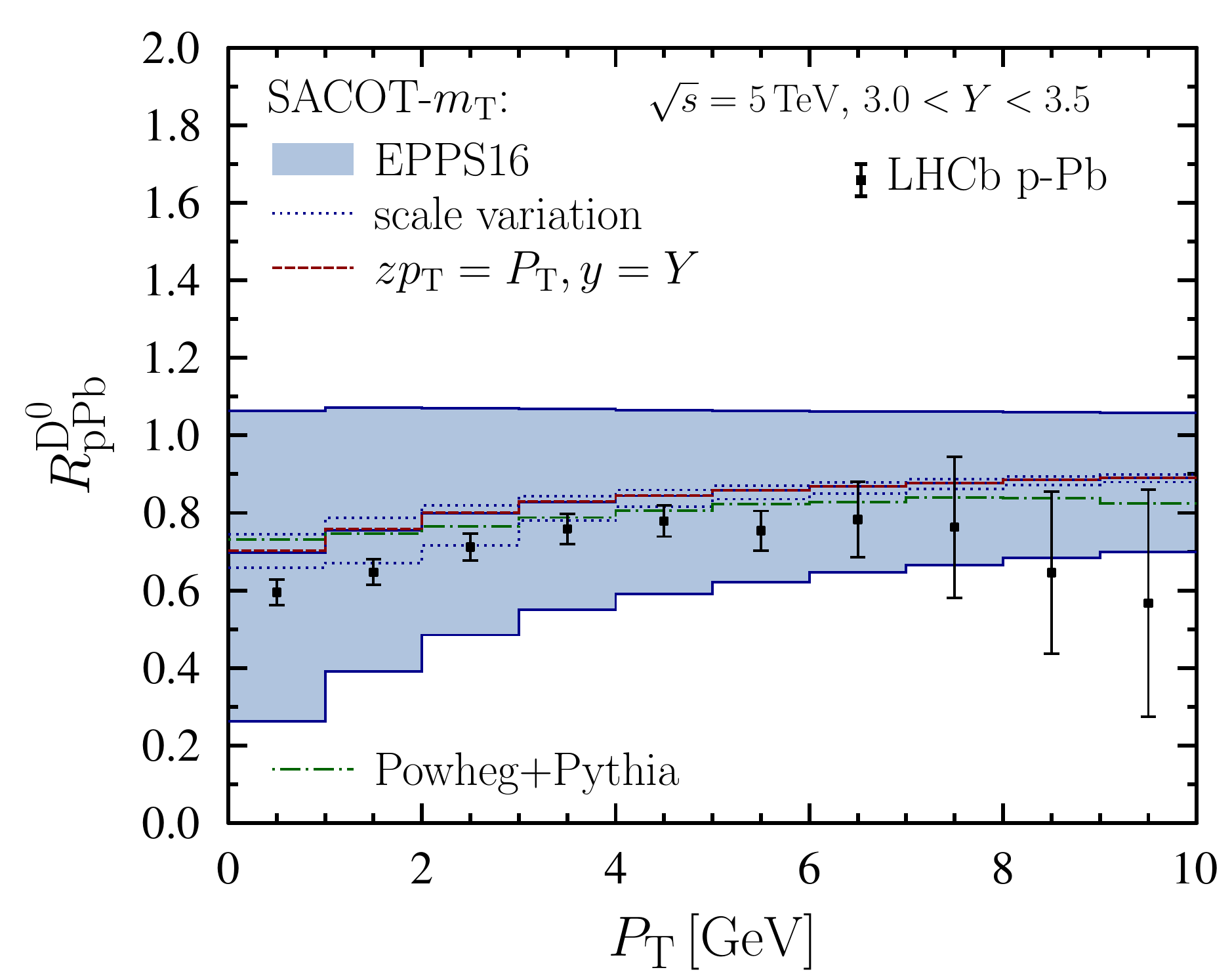}
\includegraphics[width=0.4\textwidth]{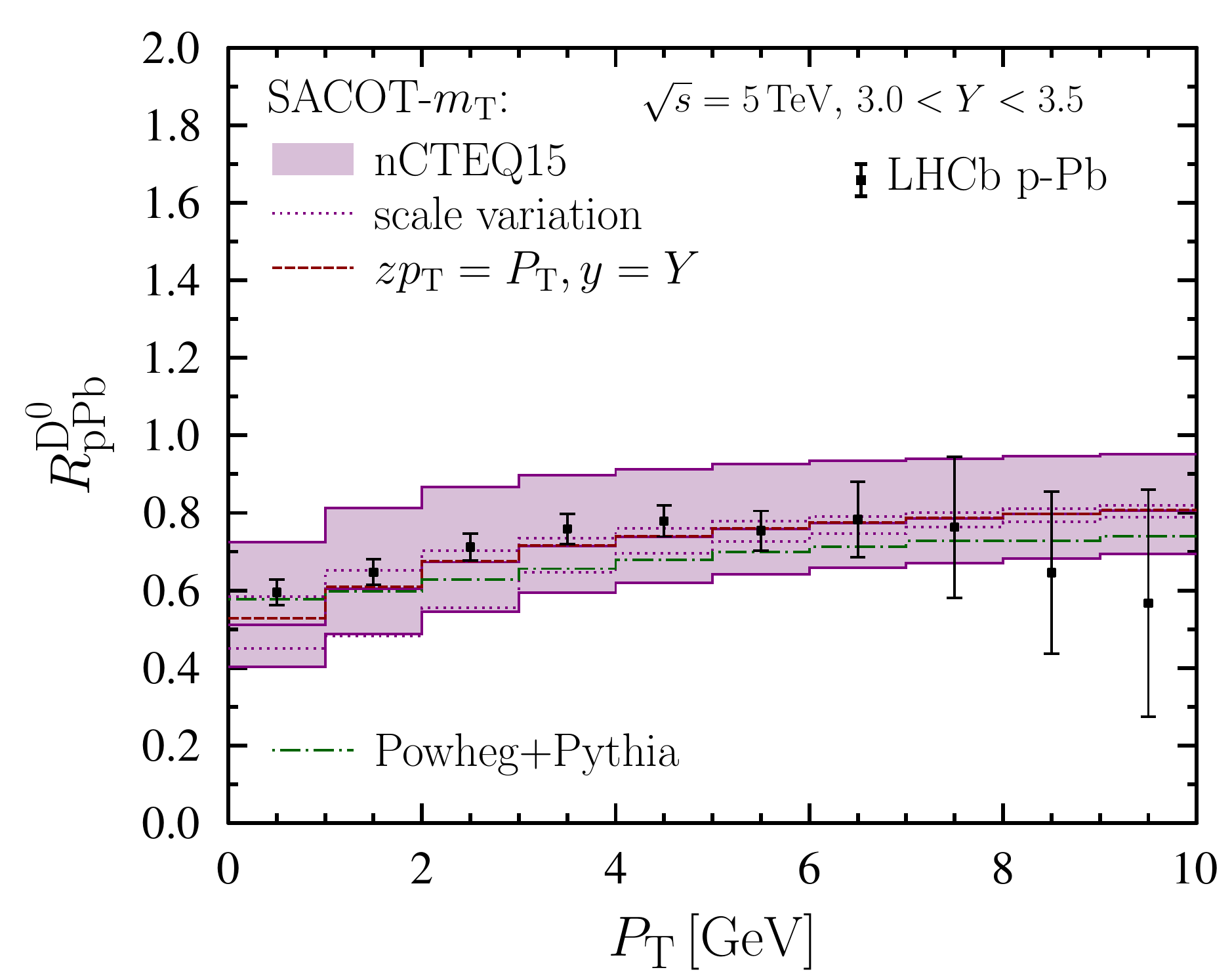}
\includegraphics[width=0.4\textwidth]{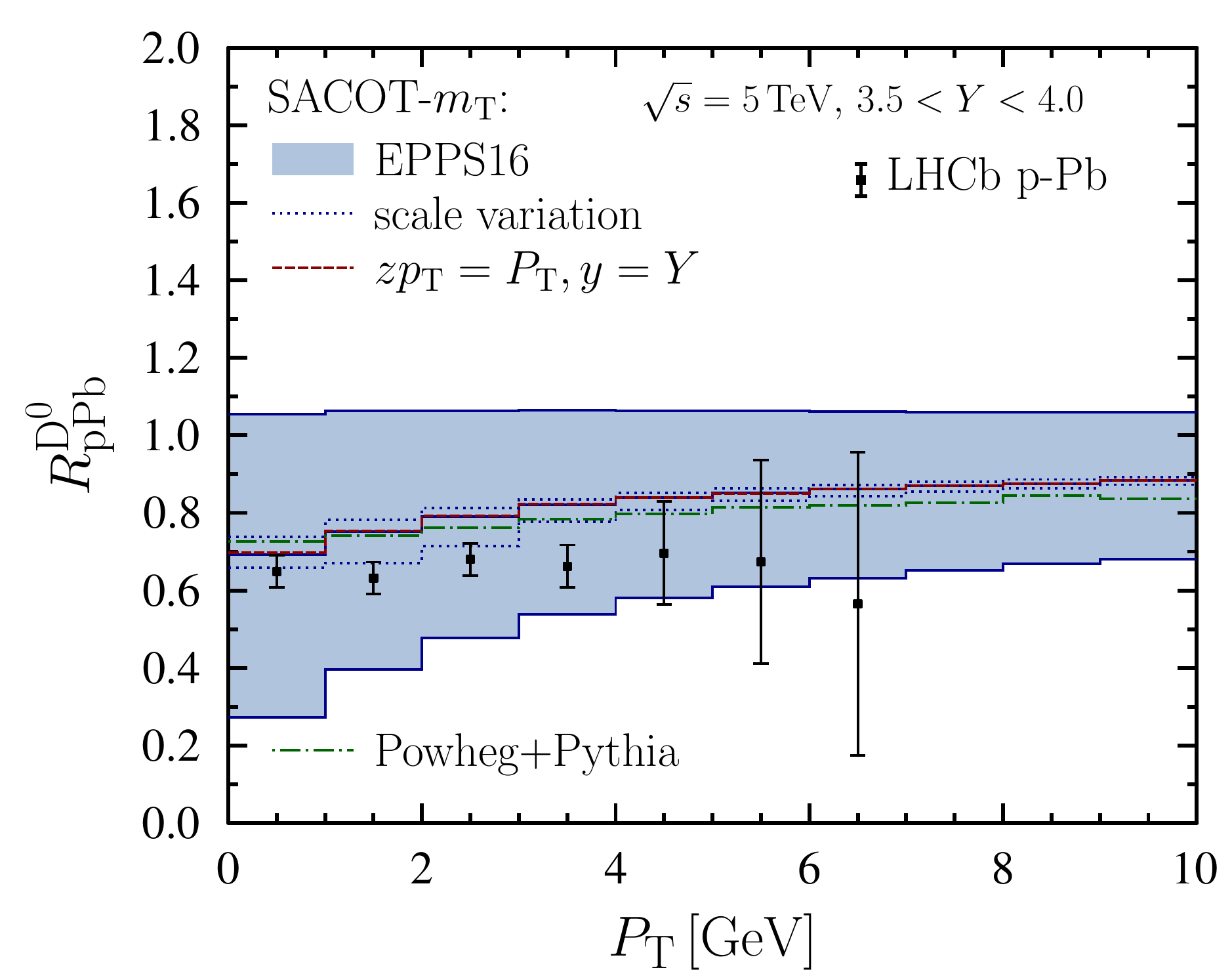}
\includegraphics[width=0.4\textwidth]{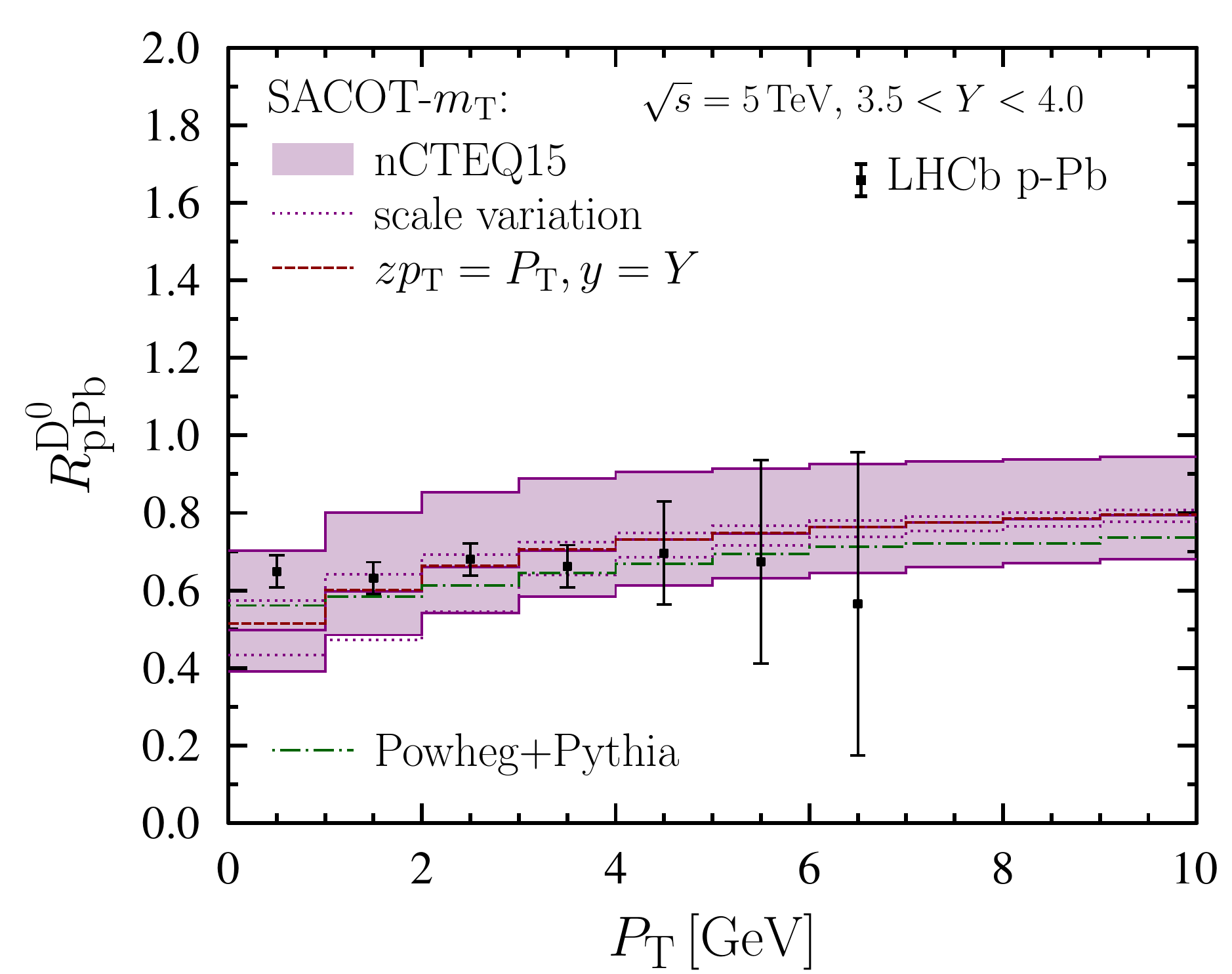}
\caption{Same as figure \ref{fig:R_pPb_backward} but at forward rapidities.}
\label{fig:R_pPb_forward}
\end{center}
\end{figure}

To constrain nPDFs with D mesons, it is useful to consider an observable in which theoretical uncertainties related to scale variations, free proton PDFs, and FFs  cancel out to a large extent. In the case of single-inclusive hadron production a suitable observable is the nuclear modification factor $R_{AB}^{h_3}$, defined for inclusive $\D0$ meson production in p+Pb collisions at the LHC as
\begin{equation} 
R_{\mathrm{pPb}}^{\D0}(\PT,Y) \equiv \frac{1}{208}\frac{\mathrm{d}\sigma^{\mathrm{p+Pb}\rightarrow \D0 + X}}{\mathrm{d}\PT \mathrm{d}Y} \bigg/ \frac{\mathrm{d}\sigma^{\mathrm{p+p}\rightarrow \D0 + X}}{\mathrm{d}\PT \mathrm{d}Y}.
\label{eq:RpA}
\end{equation}
We compare our calculations with the measured $R_{\mathrm{pPb}}^{\rm D^0}$ in figures \ref{fig:R_pPb_backward} and \ref{fig:R_pPb_forward} at backward and forward rapidities, respectively. The LHCb data span over four $Y$ bins in a range $-4.5 < Y < -2.5$ at backward rapidities and $2.0 < Y < 4.0$ at forward rapidities. Comparisons with the EPPS16 and nCTEQ15 nPDFs using the GM-VFNS framework and \textsc{Powheg}+\textsc{Pythia} setup are separately shown in each panel, and the uncertainty bands correspond to the nPDF errors calculated in the GM-VFNS approach. Furthermore, also the GM-VFNS result using the zero-mass definition for the fragmentation variable, and the scale variation band, are shown in each kinematic bin.

First observation is that the data uncertainties are in most of the cases smaller than the nPDF-originating ones with both nPDF sets considered. Especially at forward rapidities the EPPS16 nPDF uncertainty bands are much larger than the experimental uncertainties due to the poorly-constrained small-$x$ nuclear gluon distributions. This demonstrates the potential of these data to significantly constrain the current nPDFs at small-$x$ where no other data currently exist. Also, the good overall agreement with the calculated and measured $R_{\mathrm{pPb}}^{\rm D^0}$ over the wide rapidity range provides a strong indication of the applicability of factorization-based approach in this previously unconstrained kinematic region. The large uncertainties from scale variations observed for the differential cross sections largely cancel out in the nuclear modification ratio. However, at $\PT < 3~\text{GeV}$ they start to grow and the downward uncertainty is limited by the minimum scale $Q=1.3~\text{GeV}$ of EPPS16 and nCTEQ15. If the PDF parametrizations would extend to lower values, the downward uncertainty would probably be much larger. Similarly, the use of massless definition for the fragmentation variable $z$ --- taken here as an indicator of the associated uncertainty --- can lead to a significant variation in the calculated $R_{\mathrm{pPb}}^{\rm D^0}$ at small values of $\PT$ at backward rapidities. The reason is that the definition of $z$ provides the link between hadronic and partonic kinematics and therefore the probed $x$ regions are slightly different from one definition to another. In backward direction we are sensitive to the mid-$x$ region where the slope in both EPPS16 and nCTEQ15 nuclear gluon modifications is somewhat steepish (see figures \ref{fig:RPb_EPPS16_Q10} and \ref{fig:RPb_nCTEQ15_Q10} ahead), and changes in the probed $x$ regions matter. To make sure that we stay in a region where these theoretical uncertainties are in control, it seems sufficient to discard the data points below $\PT = 3~\text{GeV}$. 

Since many theoretical uncertainties get suppressed in $R_{\mathrm{pPb}}^{\rm D^0}$, we might expect that the \textsc{Powheg}+\textsc{Pythia} results would be very close to GM-VFNS ones. While the two are indeed very similar, we find that the \textsc{Powheg}+\textsc{Pythia} results tend to lie systematically below the GM-VFNS calculations. In part, the differences can be explained by the different scale choices ($\pT$ instead of $\PT$) but since the differences persist even at the largest $\PT$ bins, this cannot be the full explanation. Indeed, the main factor seems to be, as argued also in ref.~\cite{Helenius:2018uul}, that \textsc{Powheg}+\textsc{Pythia} framework misses the contributions in which the $\mathrm{c\bar{c}}$ pair would be produced only at later stages of the shower and therefore biases the kinematics to lower values of $x_2$ compared to the GM-VFNS calculation. Thus, the nuclear effects in the \textsc{Powheg}+\textsc{Pythia} predictions at a given $\PT$ come from smaller $x_2$ than in GM-VFNS. This explains why, when compared to the GM-VFNS results, the nuclear effects in \textsc{Powheg}+\textsc{Pythia} predictions are seemingly shifted towards higher values of $\PT$ in all rapidity bins, apart from the very lowest $\PT$ bins where the impact of the scale choice becomes important. 


\subsection{Impact of the LHCb data on nPDFs}
\label{sec:impact}

\begin{figure}[ptbh]
\begin{center}
\includegraphics[width=0.47\textwidth]{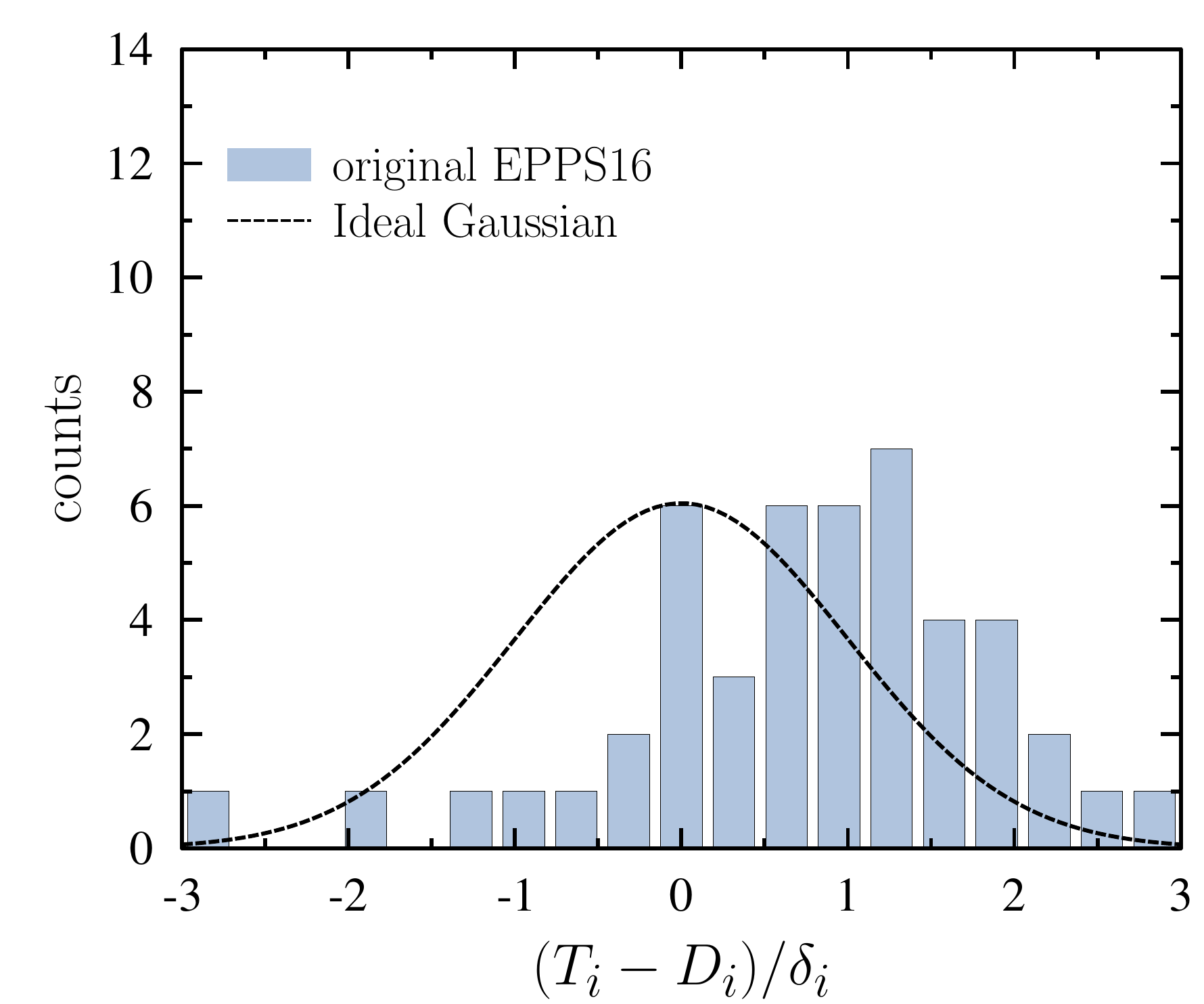}
\includegraphics[width=0.47\textwidth]{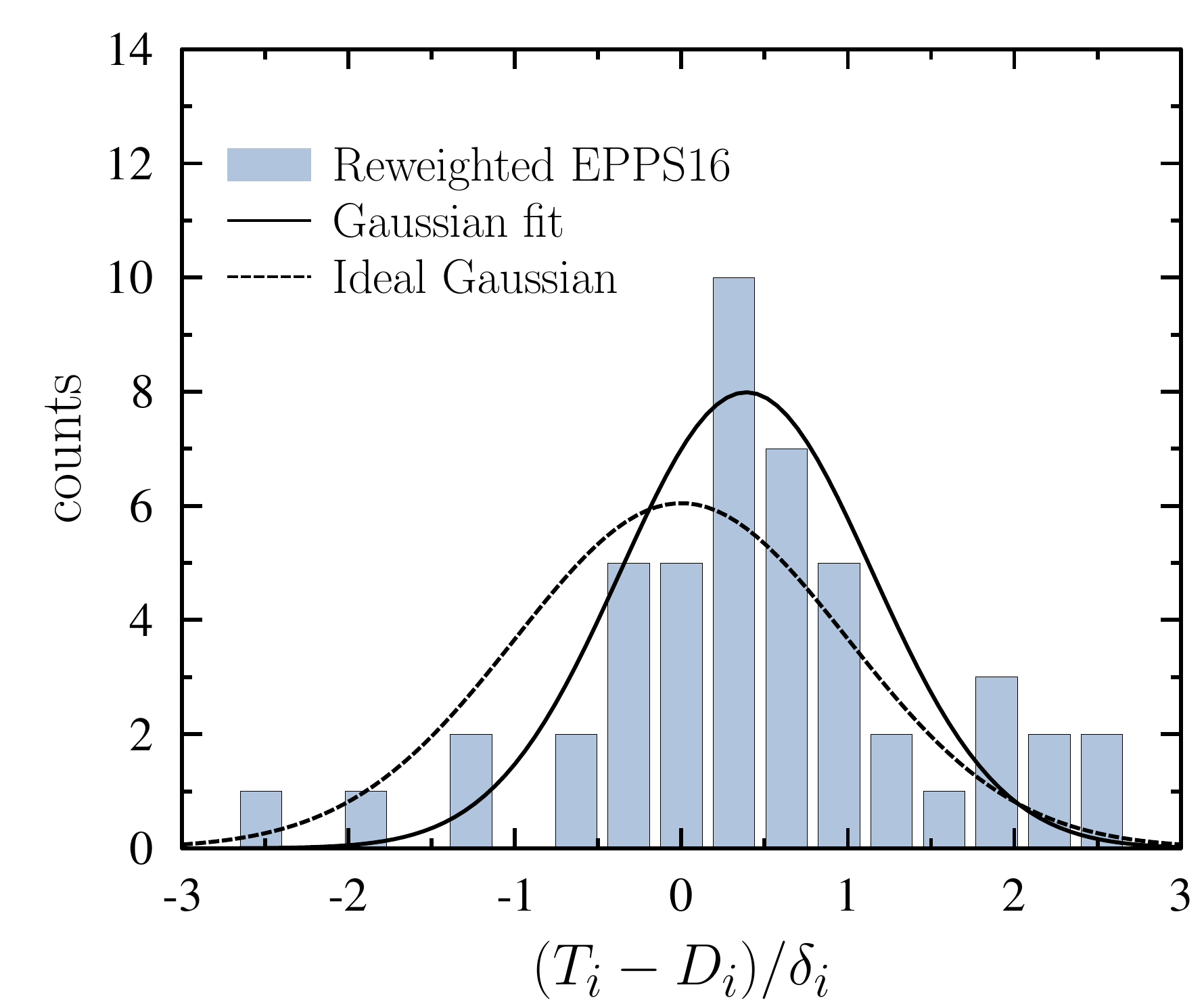}
\includegraphics[width=0.47\textwidth]{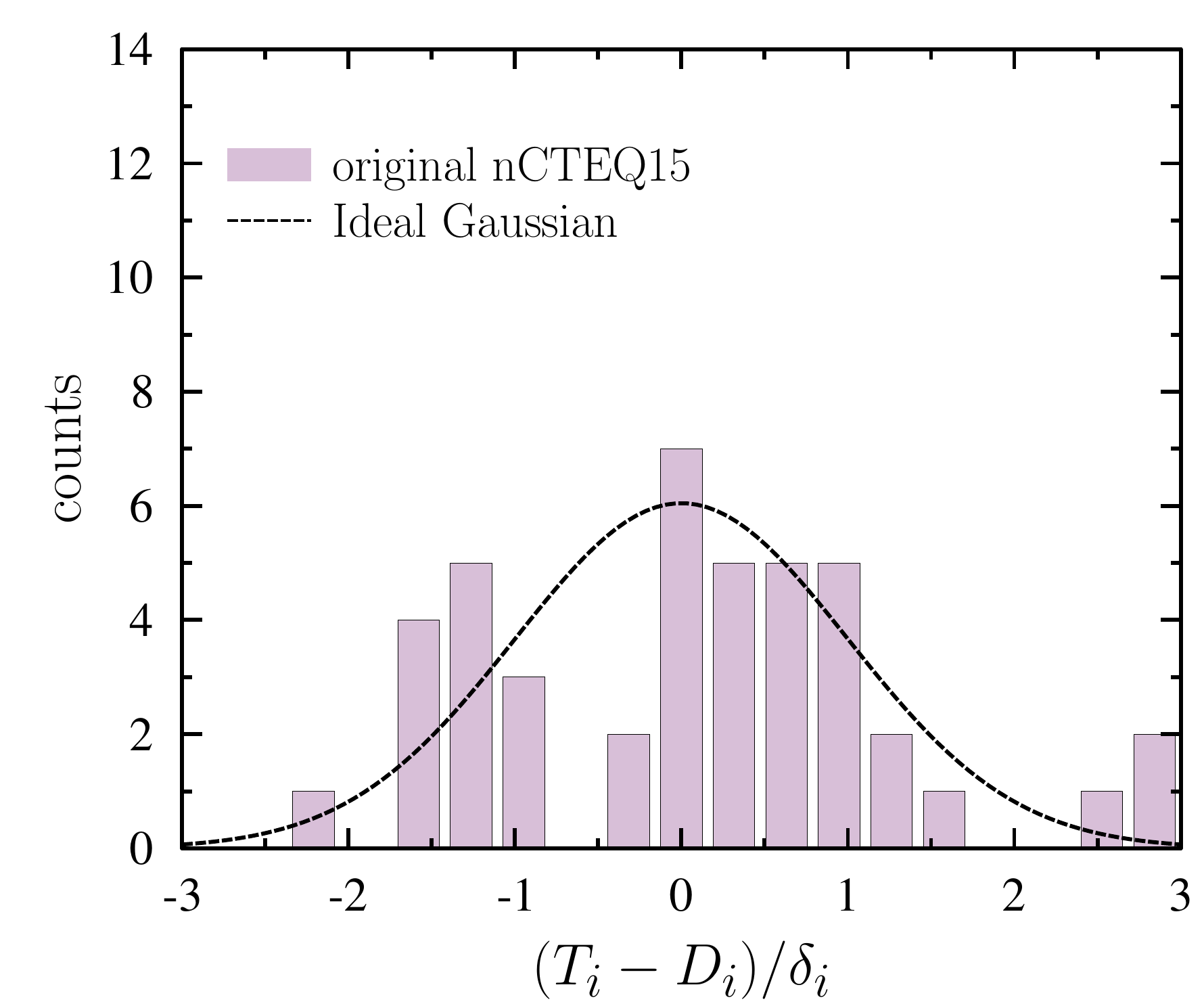}
\includegraphics[width=0.47\textwidth]{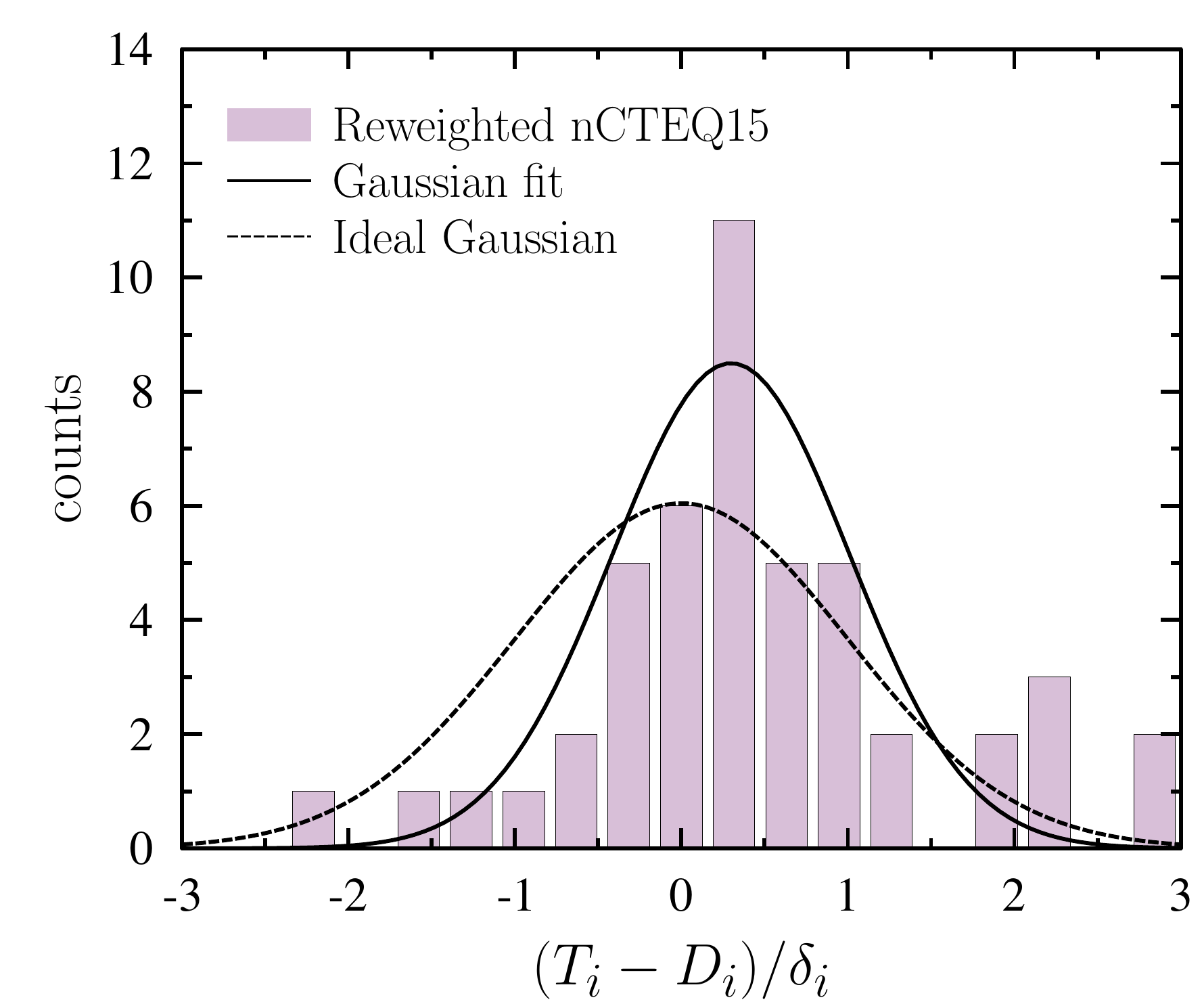}
\caption{The histograms of the $R_{\rm pPb}^{\rm D^0}$-data residuals obtained before (left) and after (right) the reweighting with EPPS16 (top) and nCTEQ15 (bottom). Ideal Gaussian distributions (dashed) are compared to fitted ones (solid) in the reweighted case.}
\label{fig:residuals}
\end{center}
\end{figure}

The observed consistency between the measured and calculated $R_{\mathrm{pPb}}^{\rm D^0}$ indicates that these data could be used in a global nPDF analysis. As a preparation for this, we now estimate the impact of the LHCb data for $R_{\mathrm{pPb}}^{\rm D^0}$ on the EPPS16 and nCTEQ15 nPDFs by applying the reweighting method outlined in section \ref{sec:reweight}. By excluding the data points at $\PT < 3~\text{GeV}$ we are left with $N_{\mathrm{data}}=48$ data points. The level of agreement is quantified by calculating the standard figure-of-merit $\chi^2$ before and after reweighting. The numbers are presented in table~\ref{tab:chi2}. Before the reweighting, the central nCTEQ15 value is somewhat high, but upon performing the reweighting both the EPPS16 and nCTEQ15 values are close to unity, indicating a good agreement with the data. 
\begin{table}[htb]
\caption{Values of $\chi^2/N_{\mathrm{data}}$ for the EPPS16 and nCTEQ15 nPDFs before and after reweighting.}
\begin{center}
\begin{tabular}{ccc}
$\chi^2/N_{\mathrm{data}} $ & EPPS16 & nCTEQ15\\
\hline
before reweighting & 1.56 & 2.09\\
after reweighting  & 1.02 & 1.12\\
\hline
\end{tabular}
\end{center}
\label{tab:chi2}
\end{table}
To further study the statistical properties of our results, histograms of the data residuals are shown in figure \ref{fig:residuals}. The residuals are calculated (for uncorrelated errors) as a difference between the theory value $T_i$ and corresponding data point $D_i$ normalised with the experimental uncertainty $\delta_i$. Ideally the distribution of the residuals should follow a Gaussian distribution with standard deviation of one and zero mean to which the calculated values are compared to. In addition, Gaussian fits are performed for the residuals obtained after reweighting to ease the comparison with the ideal distributions. With the original central EPPS16 and nCTEQ15 results the distributions show a behaviour diverting from the ideal Gaussian, but after reweighting a closer resemblance to that is obtained. With both nPDF sets the resulting distributions are slightly narrower than the ideal distribution but the mean is close to zero, confirming a reasonable statistical behaviour.

\begin{figure}[ptbh]
\begin{center}
\includegraphics[width=0.4\textwidth]{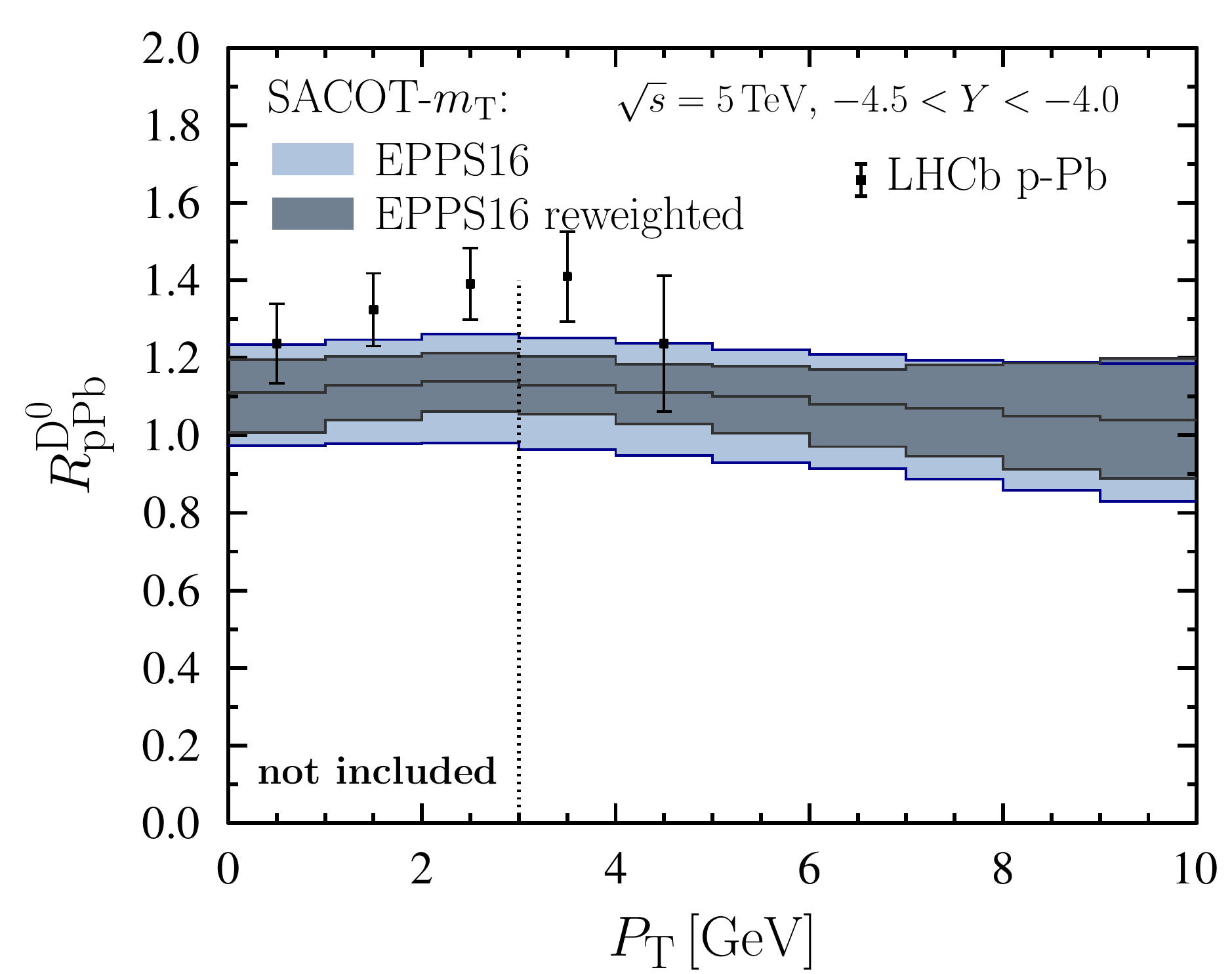}
\includegraphics[width=0.4\textwidth]{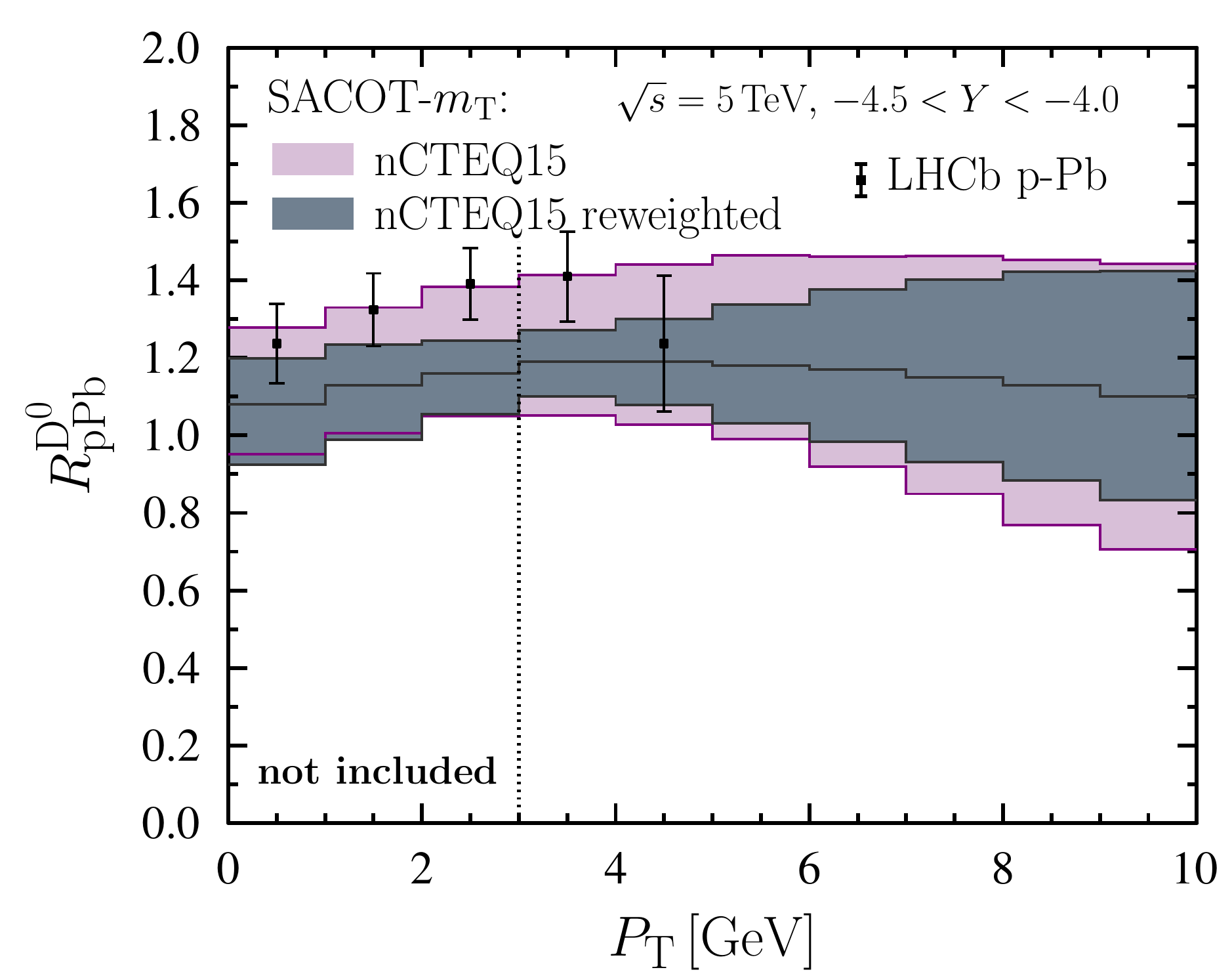}
\includegraphics[width=0.4\textwidth]{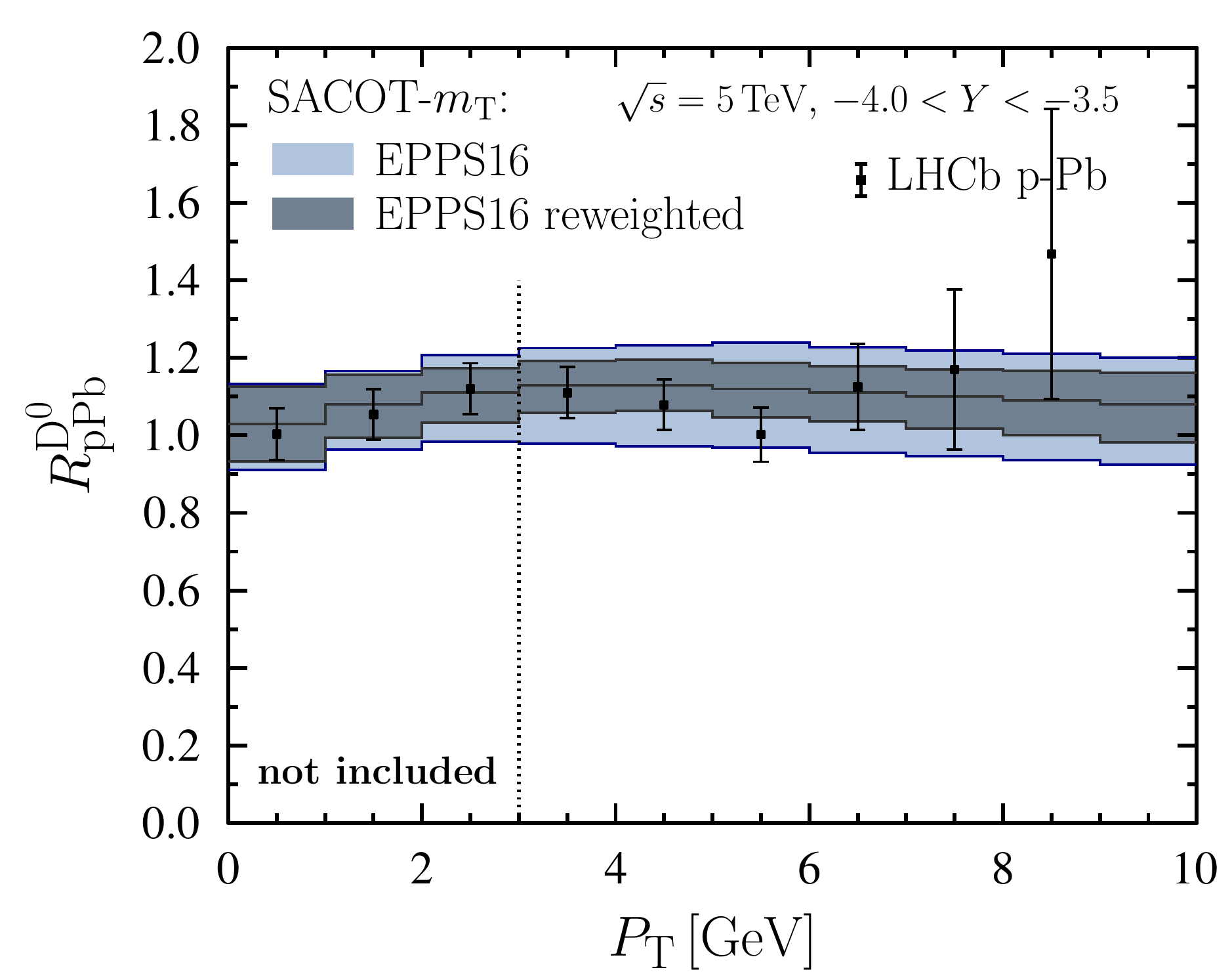}
\includegraphics[width=0.4\textwidth]{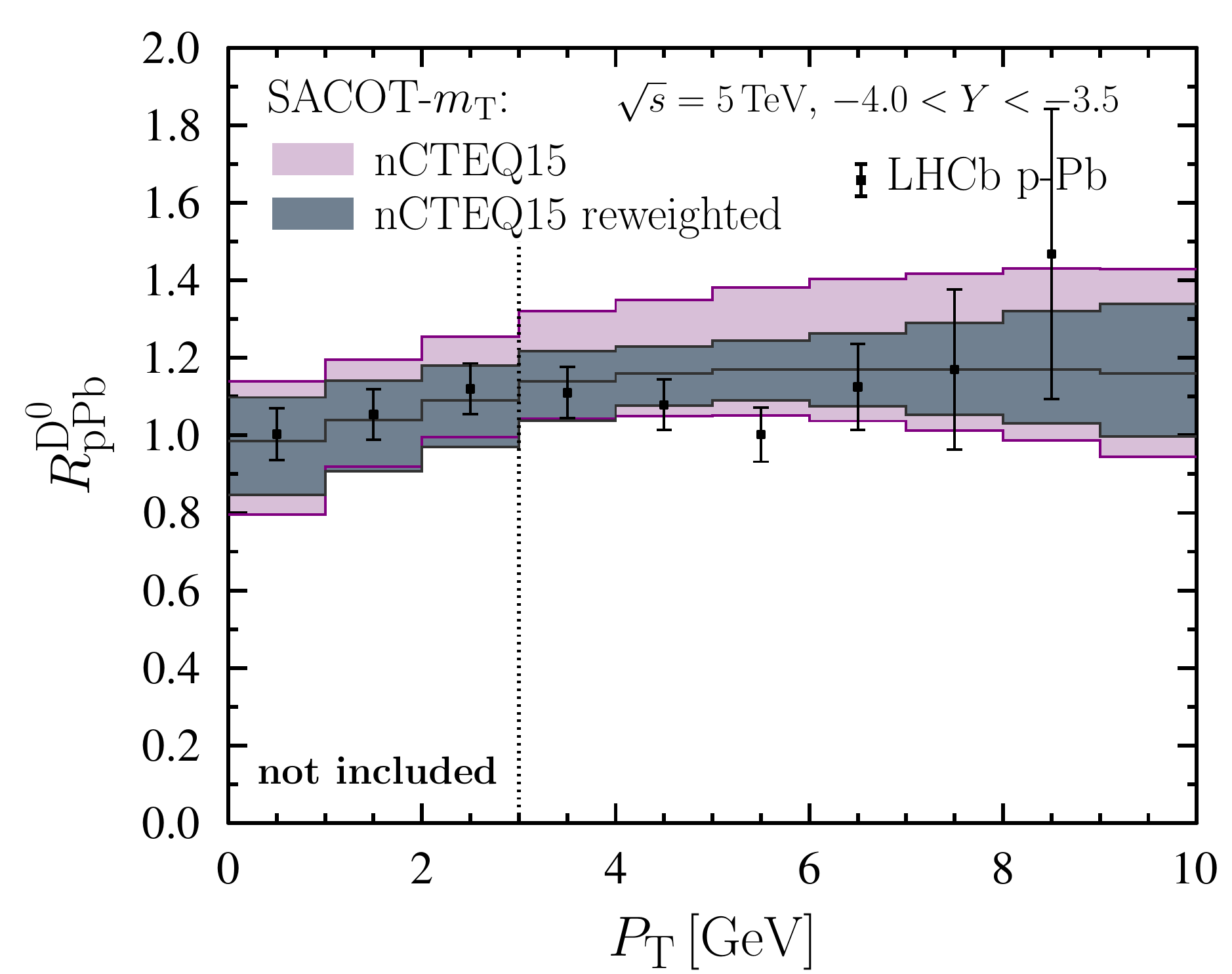}
\includegraphics[width=0.4\textwidth]{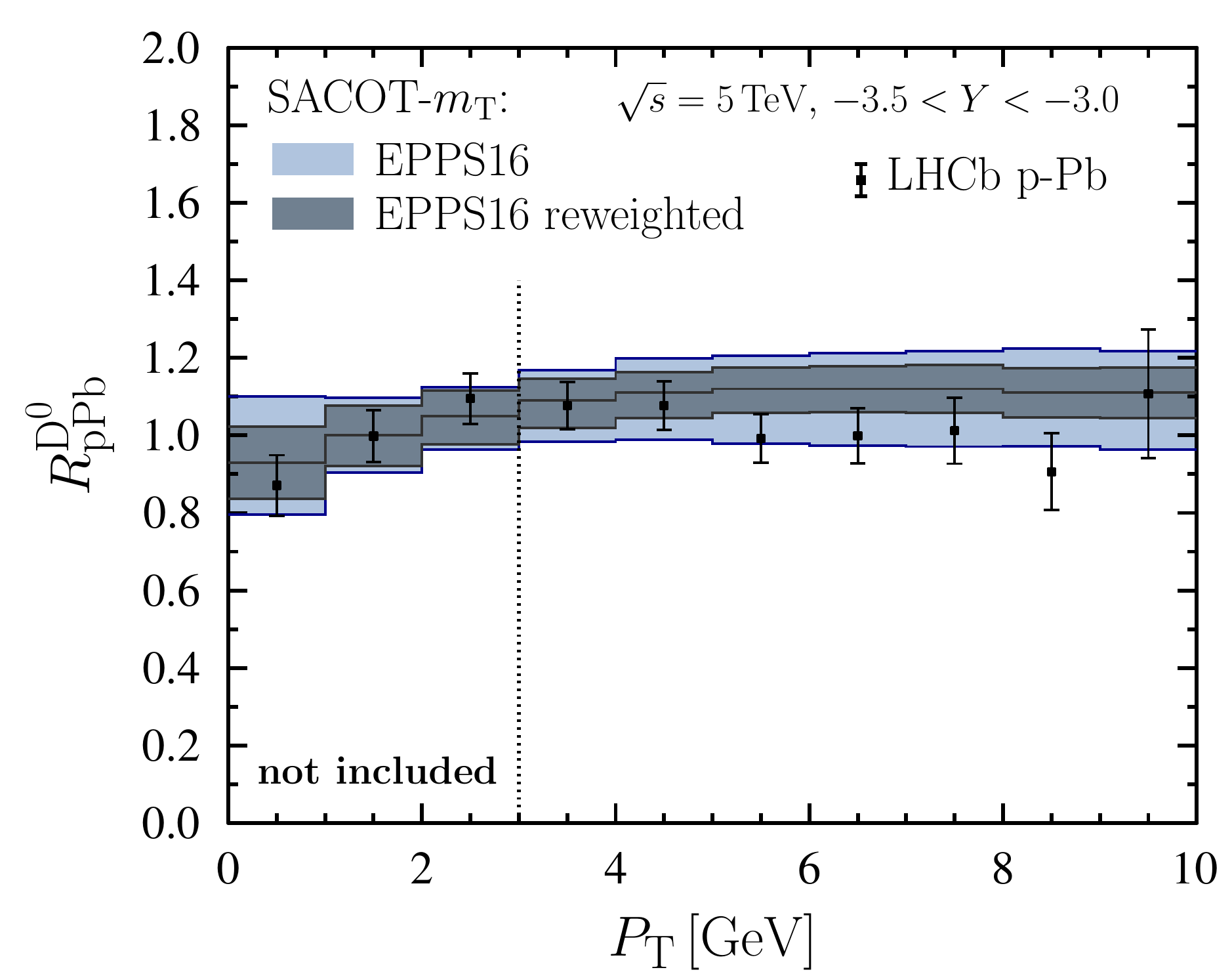}
\includegraphics[width=0.4\textwidth]{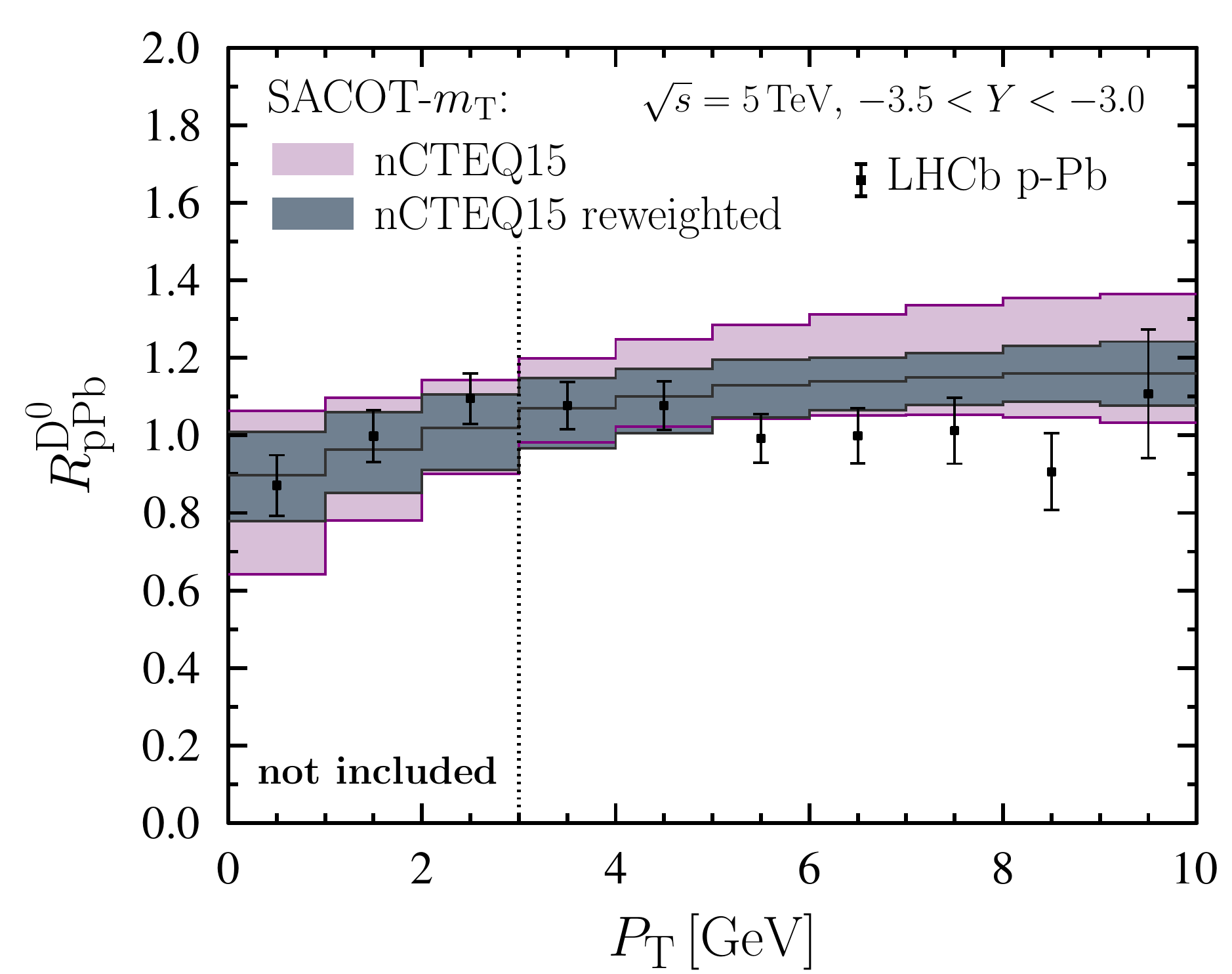}
\includegraphics[width=0.4\textwidth]{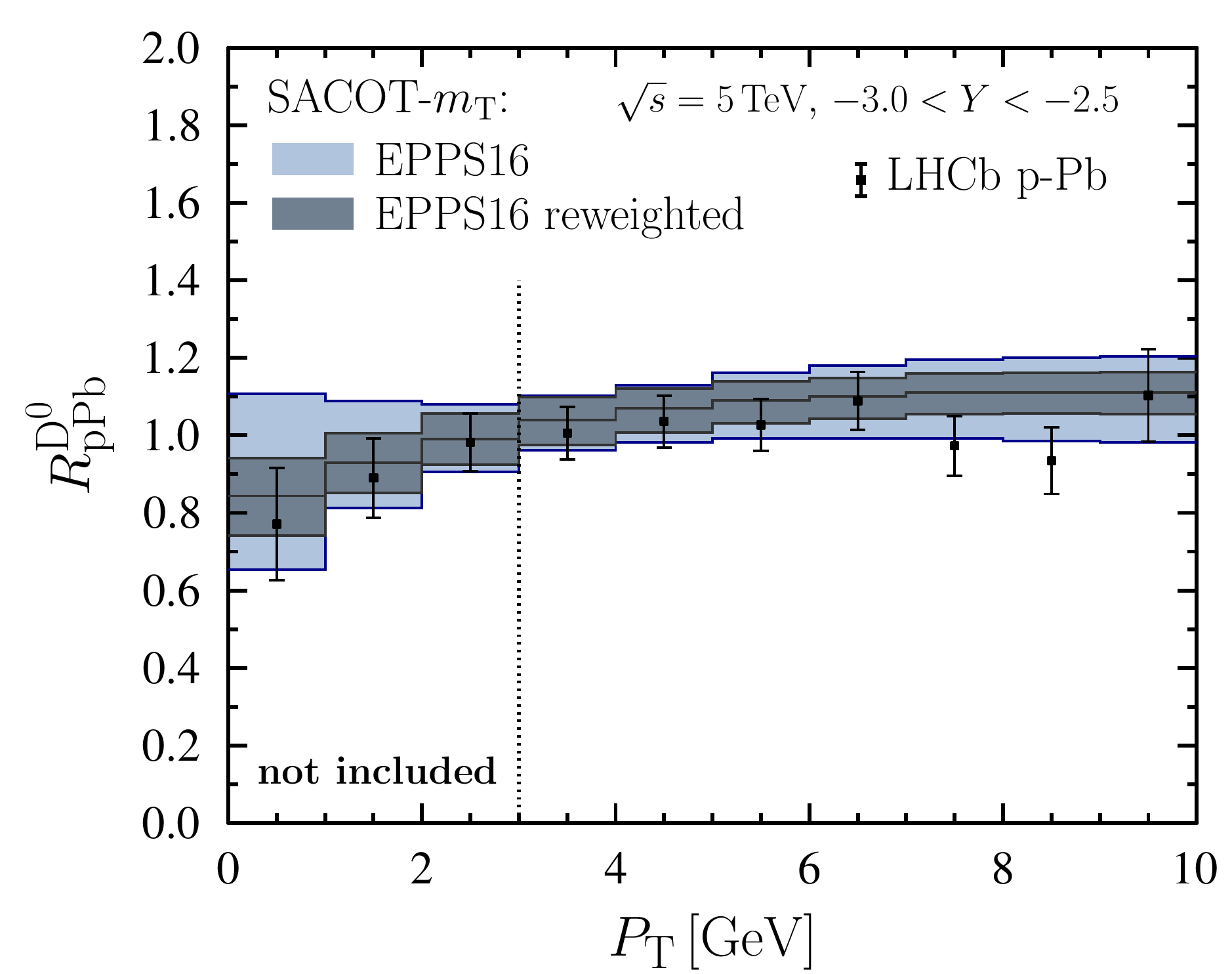}
\includegraphics[width=0.4\textwidth]{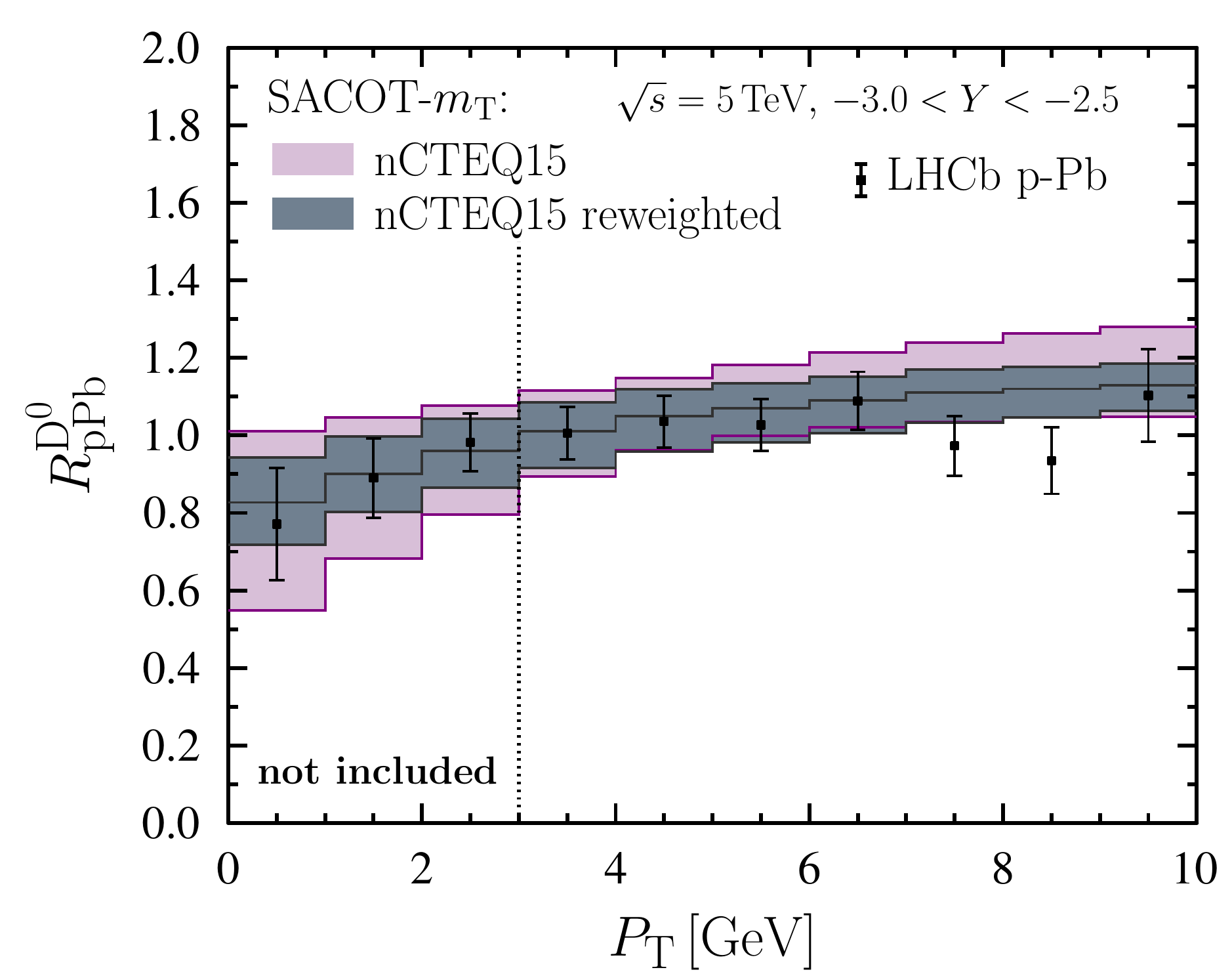}
\caption{Nuclear modification ratio for $\D0$ production at backward rapidities in p+Pb collisions at $\sqrt{s_{\mathrm{NN}}} = 5.0~\text{TeV}$ from the LHCb measurement \cite{Aaij:2017gcy} (black points with error bars) compared with the SACOT-$m_{\mathrm{T}}$ calculation using the EPPS16 (left) and nCTEQ15 (right) nPDFs with uncertainties before (light-coloured bands) and after reweighting (dark-grey bands) including the central result from the reweighted nPDFs (solid).}
\label{fig:R_pPb_backward_reweight}
\end{center}
\end{figure}

\begin{figure}[ptbh]
\begin{center}
\includegraphics[width=0.4\textwidth]{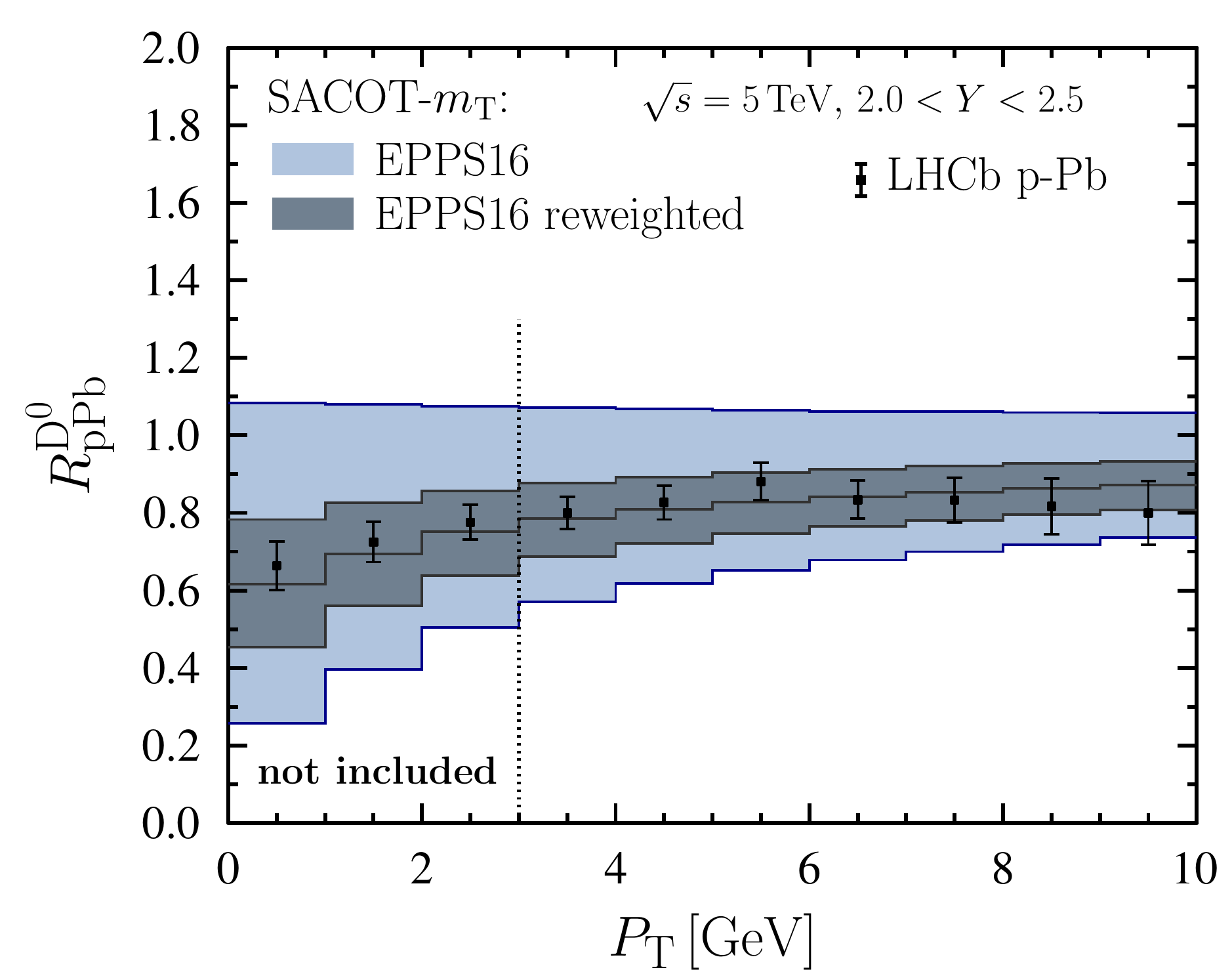}
\includegraphics[width=0.4\textwidth]{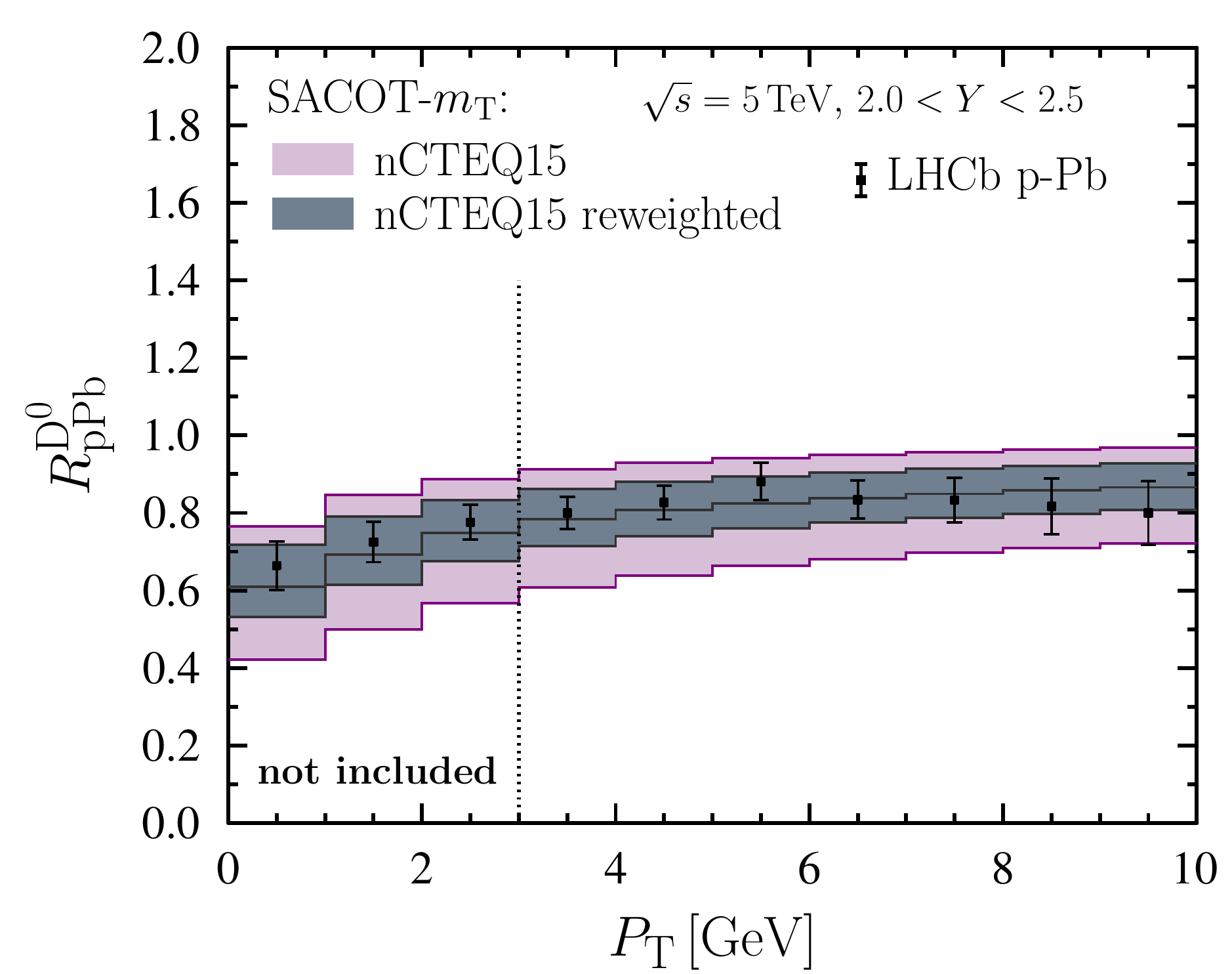}
\includegraphics[width=0.4\textwidth]{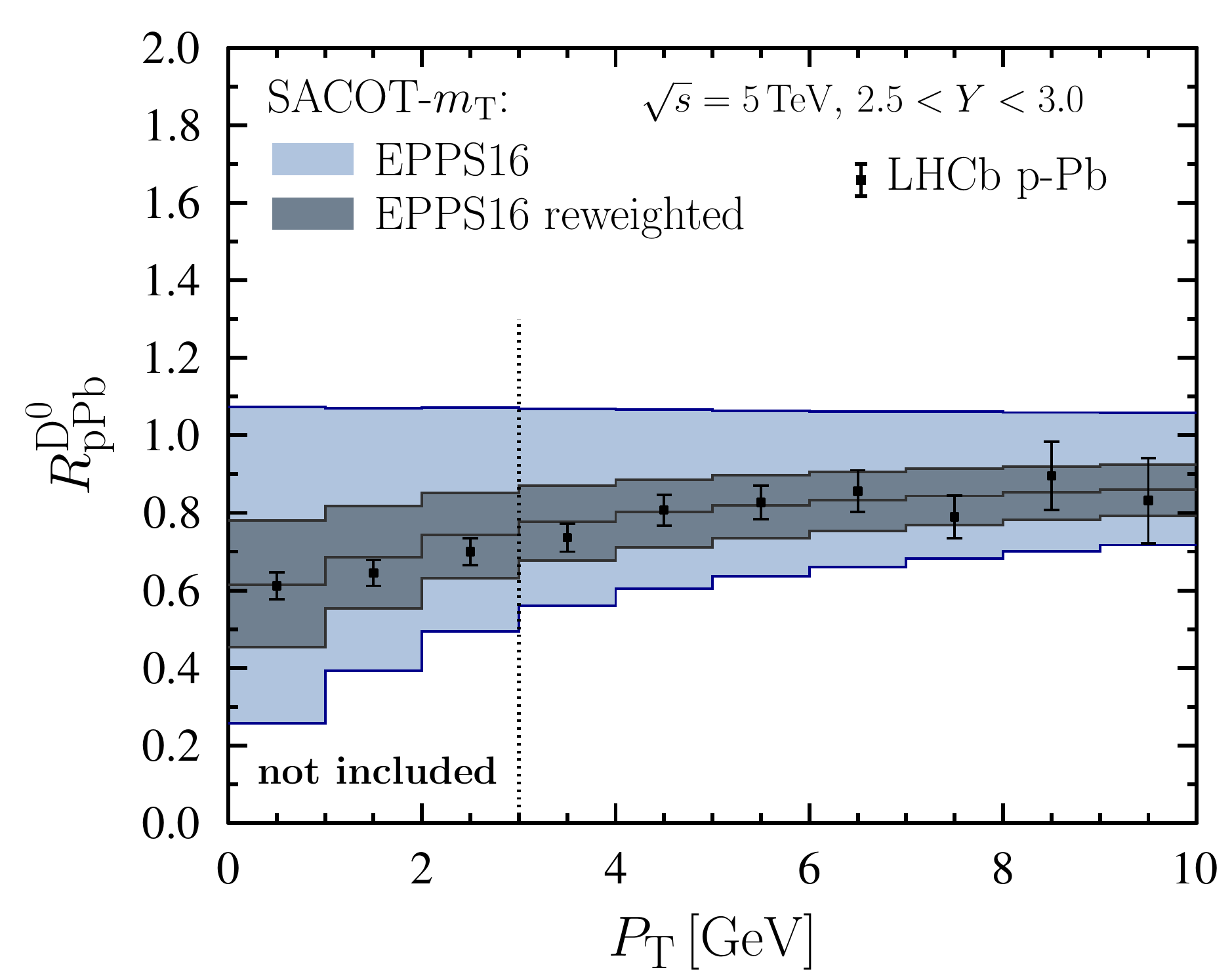}
\includegraphics[width=0.4\textwidth]{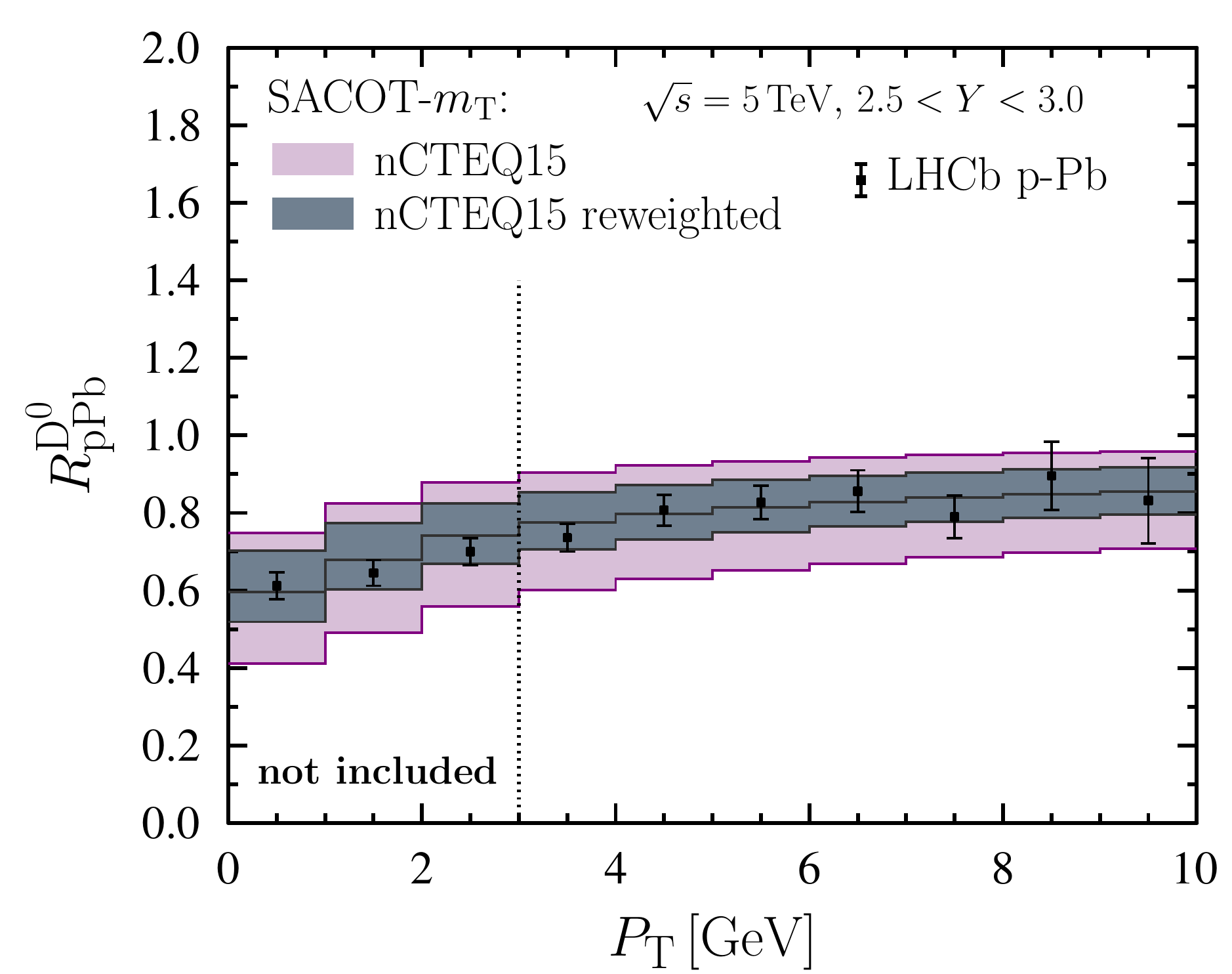}
\includegraphics[width=0.4\textwidth]{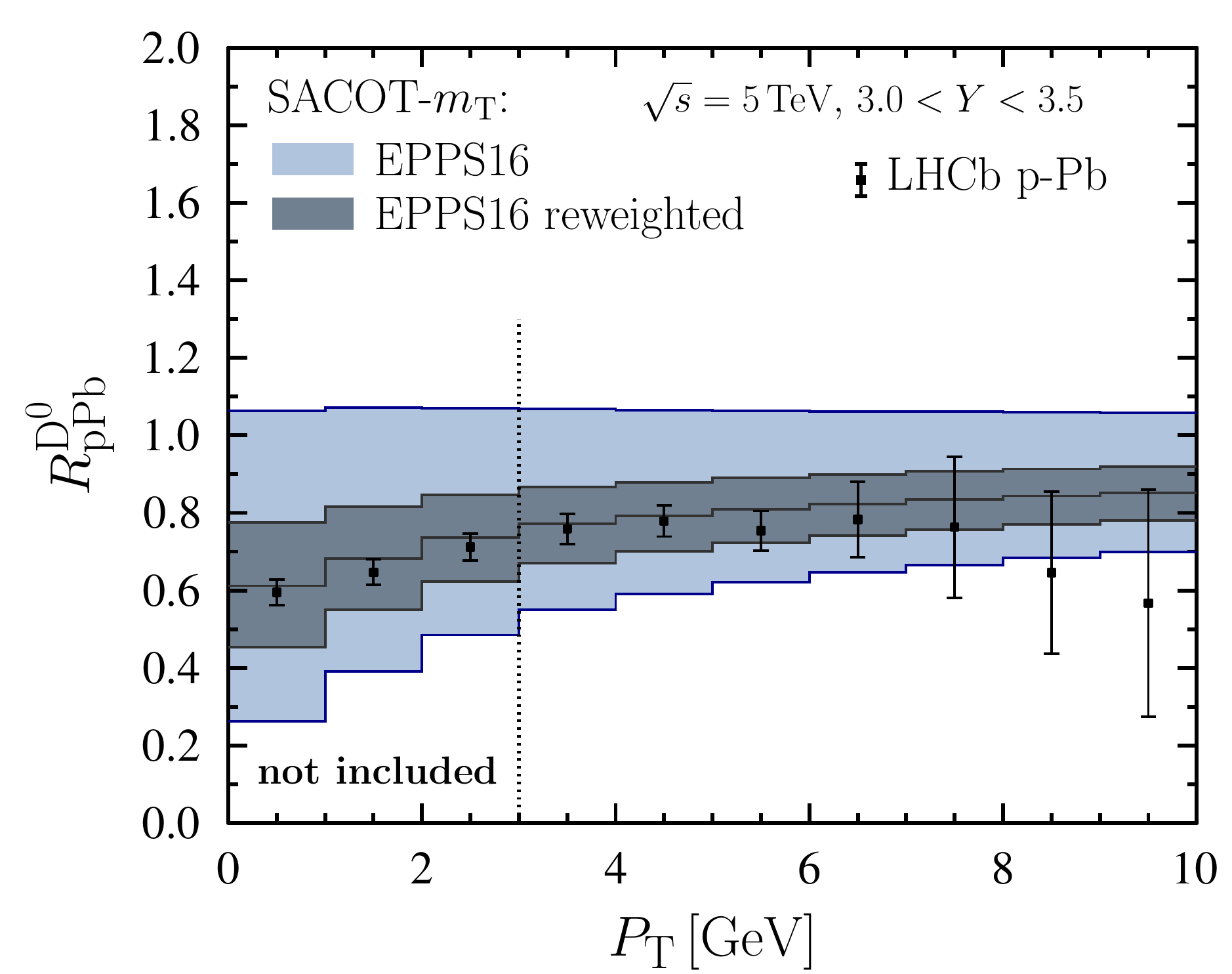}
\includegraphics[width=0.4\textwidth]{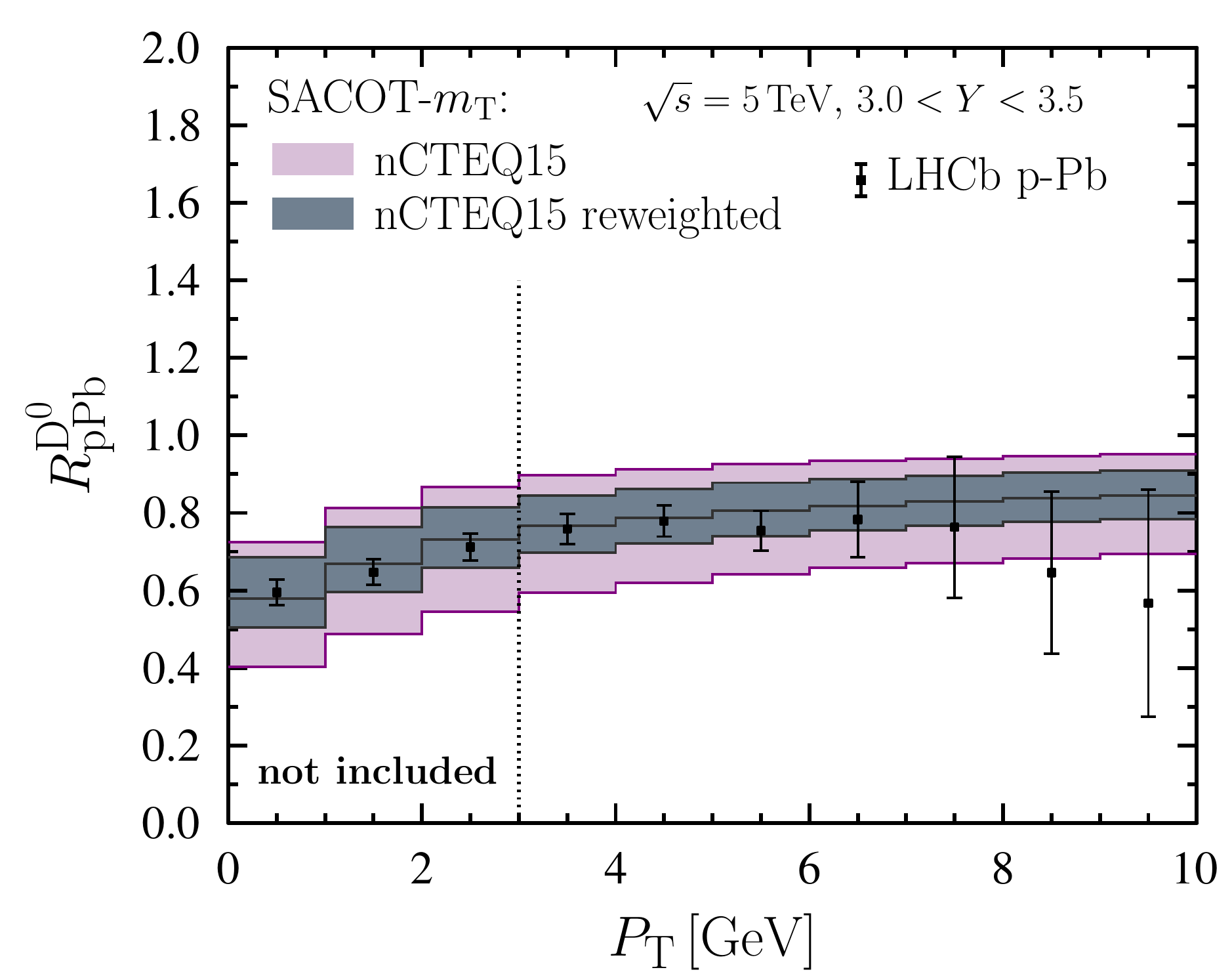}
\includegraphics[width=0.4\textwidth]{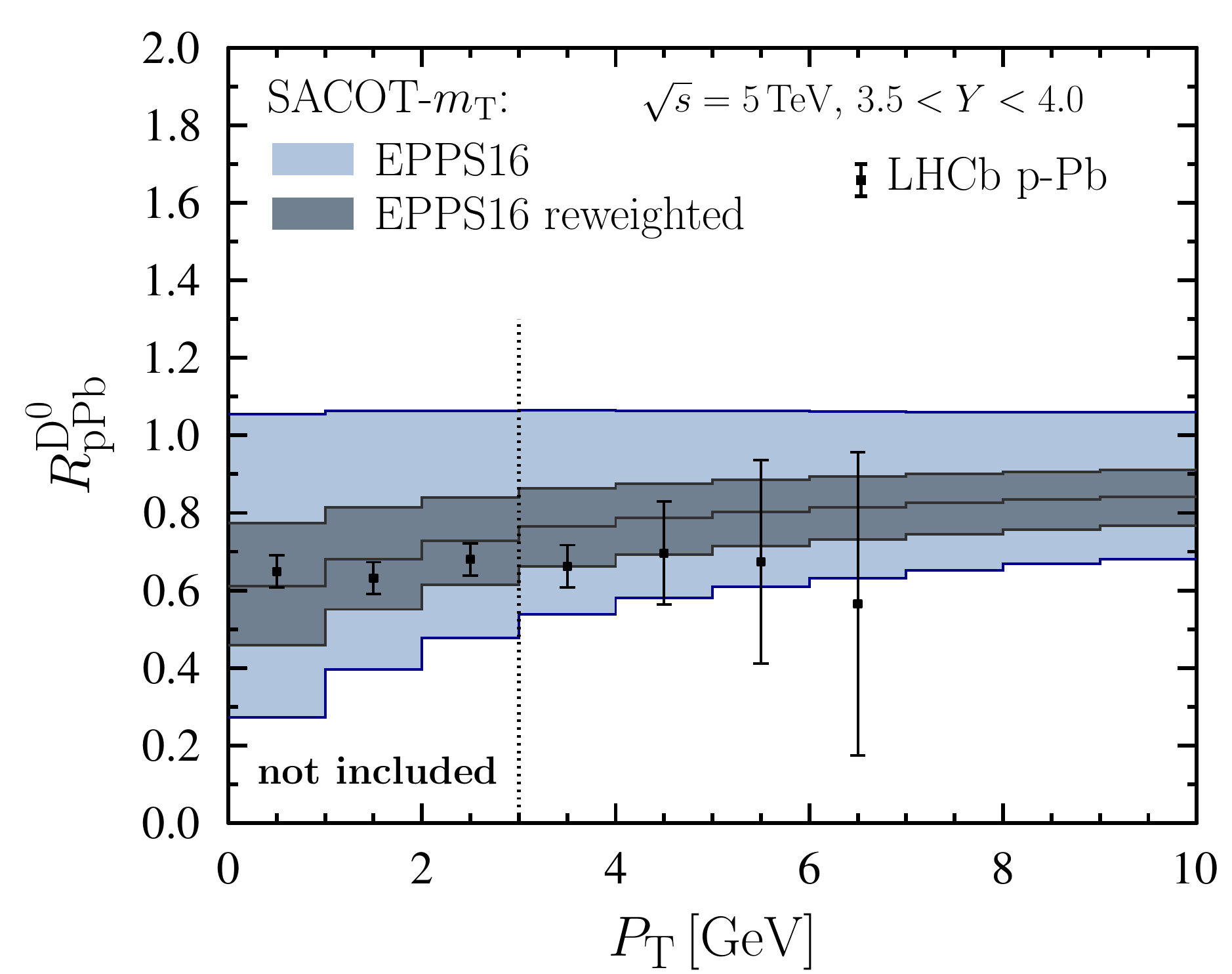}
\includegraphics[width=0.4\textwidth]{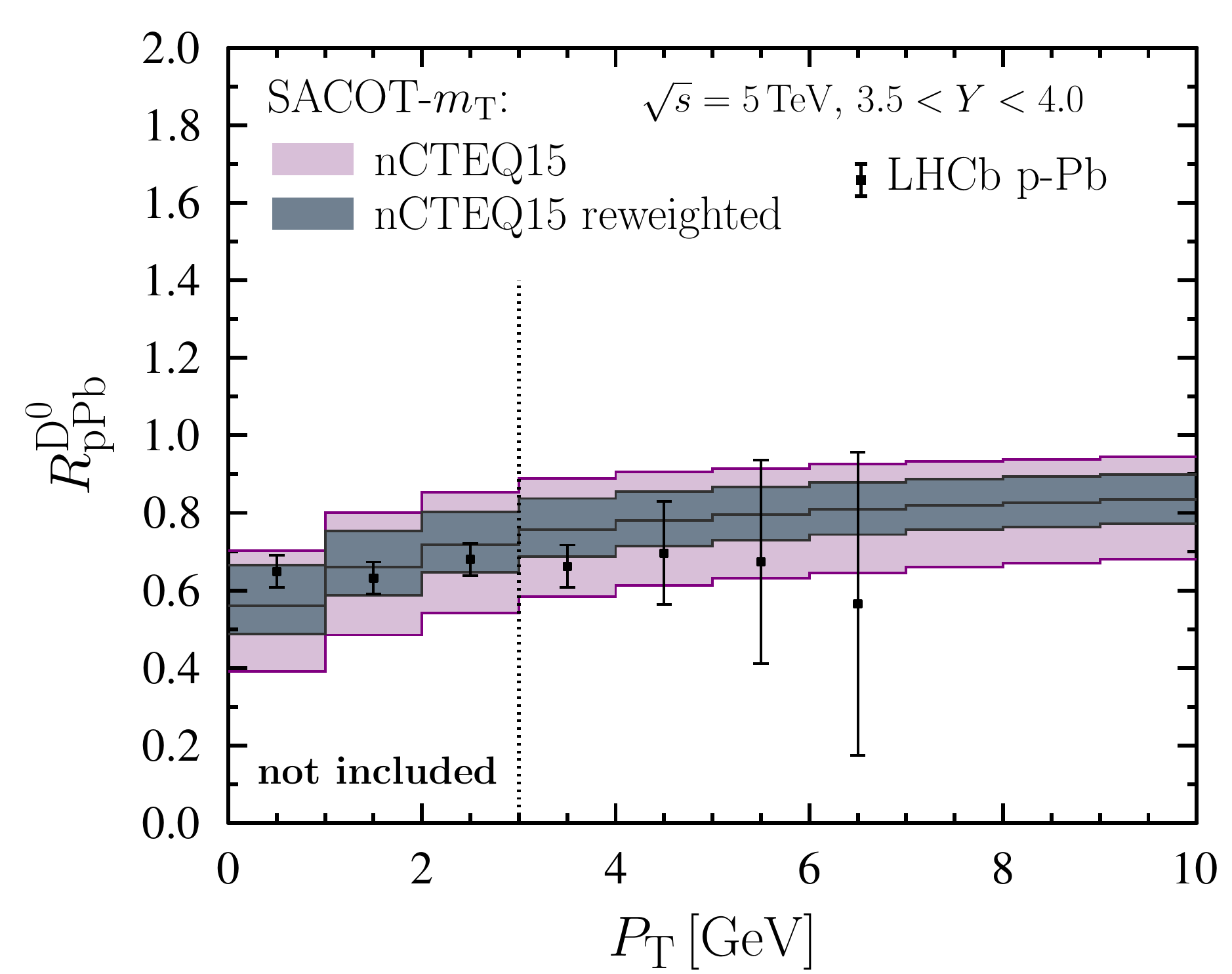}
\caption{Same as in figure \ref{fig:R_pPb_backward_reweight} but at forward rapidities.}
\label{fig:R_pPb_forward_reweight}
\end{center}
\end{figure}

The results for $R_{\mathrm{pPb}}^{\rm D^0}$ after reweighting, compared with the data and original predictions, are shown in figures \ref{fig:R_pPb_backward_reweight} and \ref{fig:R_pPb_forward_reweight}. As expected, the reweighted results are in an excellent agreement with the data across the wide rapidity range covered by the data, the only exception being the most backward bin where the data show a stronger enhancement than the reweighted PDF predictions. The new nPDF uncertainties computed from the reweighted nPDFs are significantly reduced in comparison to the original error bands. This holds especially at forward rapidities where the small-$x$ region with no previous data constraints, is probed. For the EPPS16 nPDFs an improvement of a factor of three is observed whereas for nCTEQ15 the improvement is somewhat more modest. This difference follows from a bit more rigid functional form of the nCTEQ15 parametrization which leads to smaller errors to begin with. Interestingly, even though the lowest-$\PT$ bins were not included in the analysis, the agreement remains very good also with the data points in the $\PT<3~\text{GeV}$ region. We can thus conclude that to describe these data, no physics outside collinear factorization is needed. 

\begin{figure}[ptbh]
\begin{center}
\includegraphics[width=0.99\textwidth]{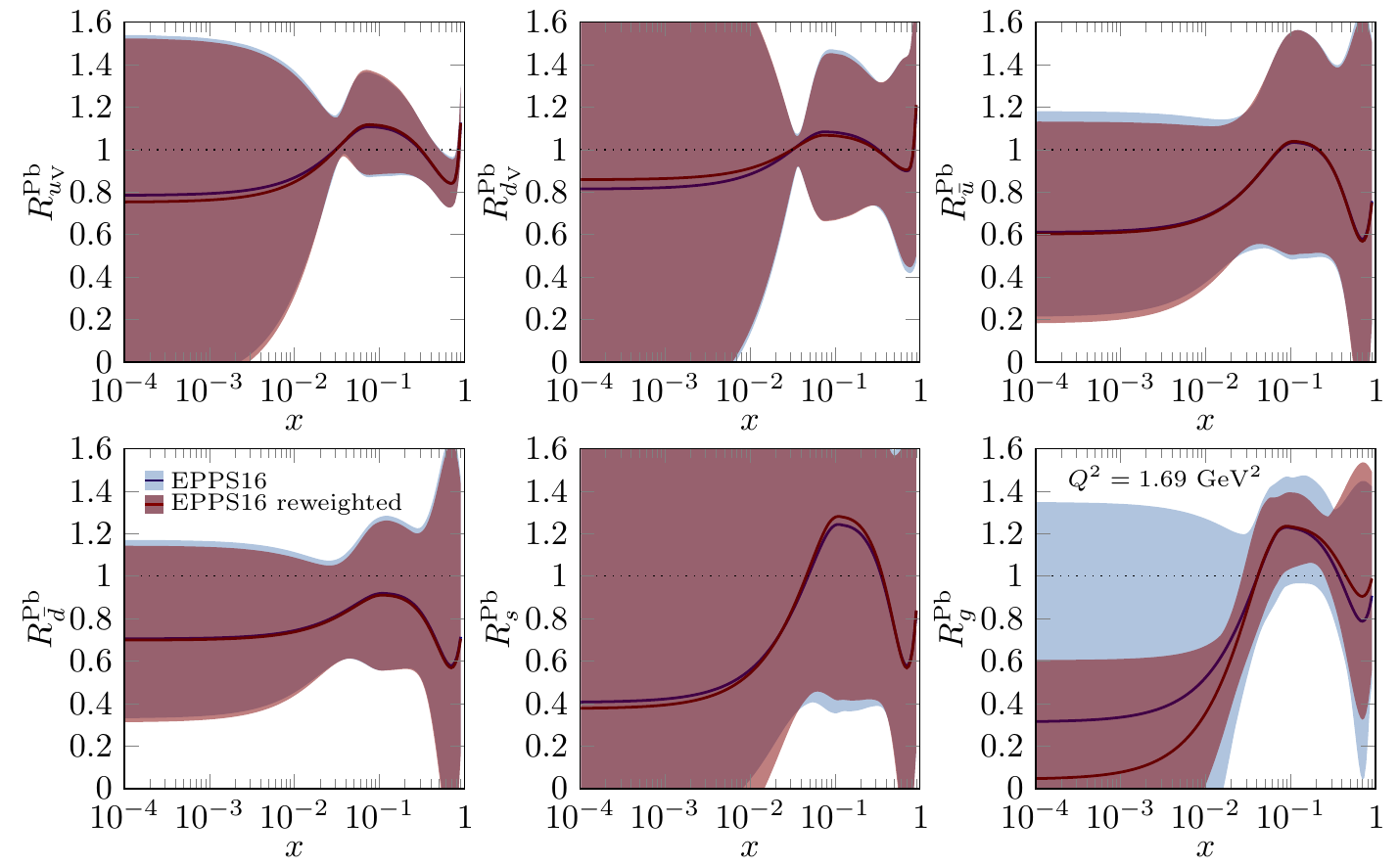}
\caption{The EPPS16 nuclear modifications for bound-proton PDFs in Pb nucleus before (blue) and after (red) reweighting with the LHCb data. The scale is $Q^2 = 1.69~\text{GeV}^2$.}
\label{fig:RPb_EPPS16_Q0}
\end{center}
\end{figure}

\begin{figure}[ptbh]
\begin{center}
\includegraphics[width=0.99\textwidth]{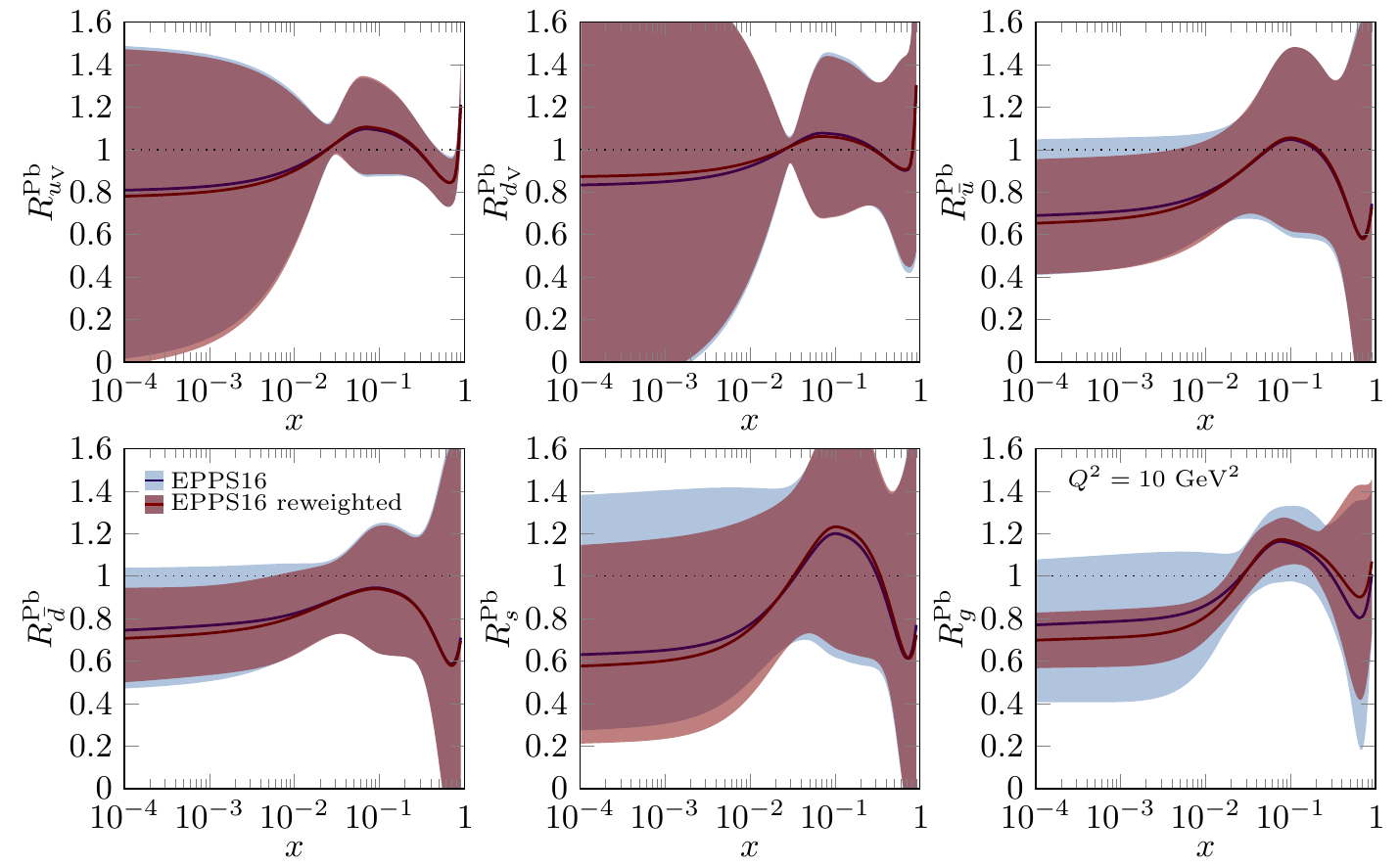}
\caption{The EPPS16 nuclear modifications for bound-proton PDFs in Pb nucleus before (blue) and after (red) reweighting with the LHCb data. The scale is $Q^2 = 10~\text{GeV}^2$.}
\label{fig:RPb_EPPS16_Q10}
\end{center}
\end{figure}

\begin{figure}[ptbh]
\begin{center}
\includegraphics[width=0.99\textwidth]{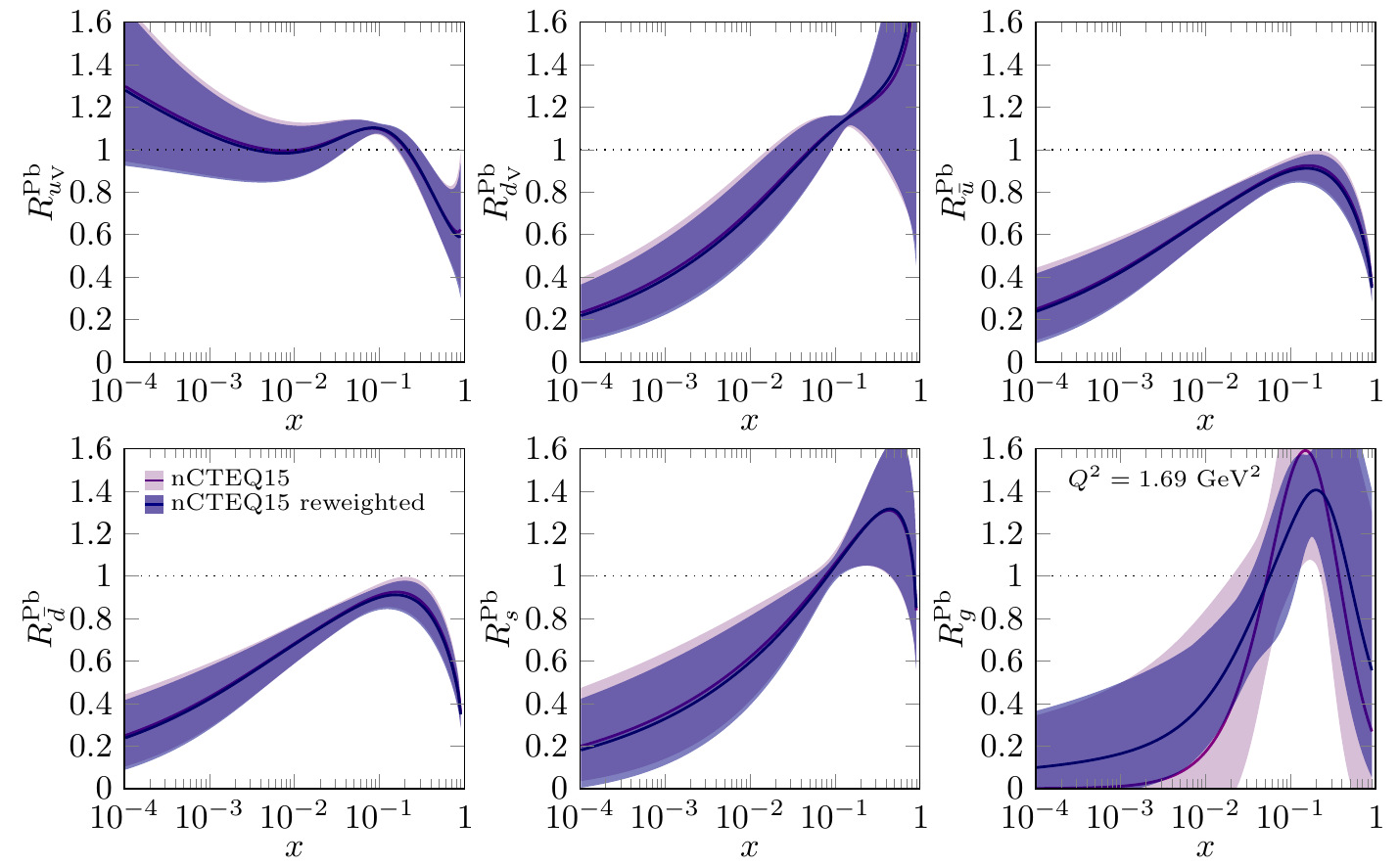}
\caption{The nCTEQ15 nuclear modifications for bound-proton PDFs in Pb nucleus before (purple) and after (blue) reweighting with the LHCb data. The scale is $Q^2 = 1.69~\text{GeV}^2$.}
\label{fig:RPb_nCTEQ15_Q0}
\end{center}
\end{figure}

\begin{figure}[ptbh]
\begin{center}
\includegraphics[width=0.99\textwidth]{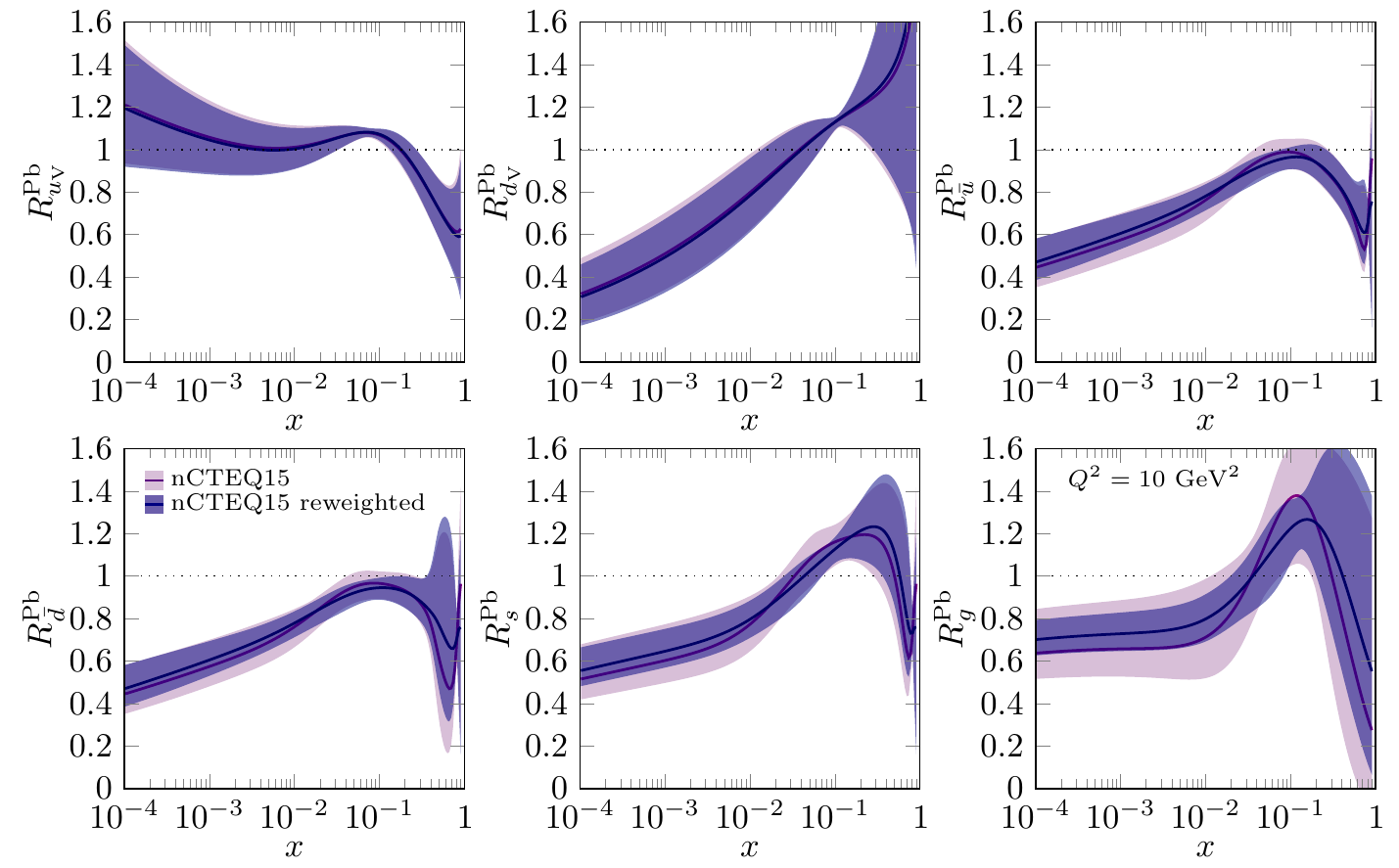}
\caption{The nCTEQ15 nuclear modifications for bound-proton PDFs in Pb nucleus before (purple) and after (blue) reweighting with the LHCb data. The scale is $Q^2 = 10~\text{GeV}^2$.}
\label{fig:RPb_nCTEQ15_Q10}
\end{center}
\end{figure}

In figures \ref{fig:RPb_EPPS16_Q0} -- \ref{fig:RPb_nCTEQ15_Q10} we finally compare the EPPS16 and nCTEQ15 nuclear modifications in bound protons, $R_i^{\rm p/Pb}(x,Q^2) = f_i^{\rm Pb}(x,Q^2)/f_i^{\rm p}(x,Q^2)$, before and after reweighting. We present the results at two different scales: the initial scale of the original analyses, $Q^2 = 1.69~\text{GeV}^2$, and a somewhat higher scale $Q^2 = 10~\text{GeV}^2$ directly probed by the considered observable when reweighting to the $\PT>3~\text{GeV}$ subset of data. The valence and sea quark distributions are shown separately for each partonic flavour. For the EPPS16 analysis these are plotted in figures \ref{fig:RPb_EPPS16_Q0} and \ref{fig:RPb_EPPS16_Q10}. The central values remain unchanged for all quark flavours but for gluons a somewhat stronger shadowing and slightly weaker EMC suppression are preferred by the data. At the parametrization scale $Q^2 = 1.69~\text{GeV}^2$ the uncertainty bands remain practically unchanged for quarks but a drastic reduction is observed for small-$x$ gluons. At $Q^2=10~\text{GeV}^2$ also the sea-quark uncertainties are slightly reduced due to the DGLAP evolution which correlates sea quarks with gluons. For gluons the strong shadowing at the initial scale is reduced to around 0.7 at $x \lesssim 0.01$ due to the evolution effects. Incidentally, the changes in the EPPS16 gluon PDFs are remarkably similar as found in ref.~\cite{Eskola:2019dui} based on the recent CMS dijet data \cite{Sirunyan:2018qel}. In addition, since the central values are only slightly modified, the good agreement with the recent $\mathrm{W}^{\pm}$ data at $\sqrt{s_{\mathrm{NN}}} = 8.16~\text{TeV}$ \cite{Sirunyan:2019dox} is expected to persist. We should also mention that the gluon errors at $Q^2 = 1.69~\text{GeV}^2$ dropping negative is of no concern. Indeed, a backward evolution by the DGLAP equations will make any gluon PDF negative at sufficiently low scales, and demanding a positive-definite gluon distribution at any arbitrary scale would be an unphysical requirement. At a deeper level, the resummation of $\log(1/x)$ terms in the DGLAP splitting functions \cite{Bonvini:2016wki} may slow down the evolution speed particularly at low $Q^2$ and thereby better retain the gluons positive. 

For nCTEQ15 the original and D-meson updated nuclear modifications are plotted in figures \ref{fig:RPb_nCTEQ15_Q0} and \ref{fig:RPb_nCTEQ15_Q10}. As was the case with EPPS16, the quark nuclear modifications remain more or less the same after reweighting with the LHCb data. The originally strong shadowing for small-$x$ gluons becomes slightly weaker after reweighting and is now rather similar to the gluon shadowing obtained with the reweighted EPPS16. The resulting uncertainties for the gluon shadowing are also on the same ballpark with EPPS16. In addition, the reweighted nCTEQ15 nuclear modifications for gluons tend to have somewhat less anti-shadowing (the bump around $x \sim 0.1$) than in the original analysis and the uncertainties are significantly reduced also in this regime. 



\subsection{Impact without the lower cut on $\PT$}

\begin{figure}[ptbh]
\begin{center}
\includegraphics[width=0.8\textwidth]{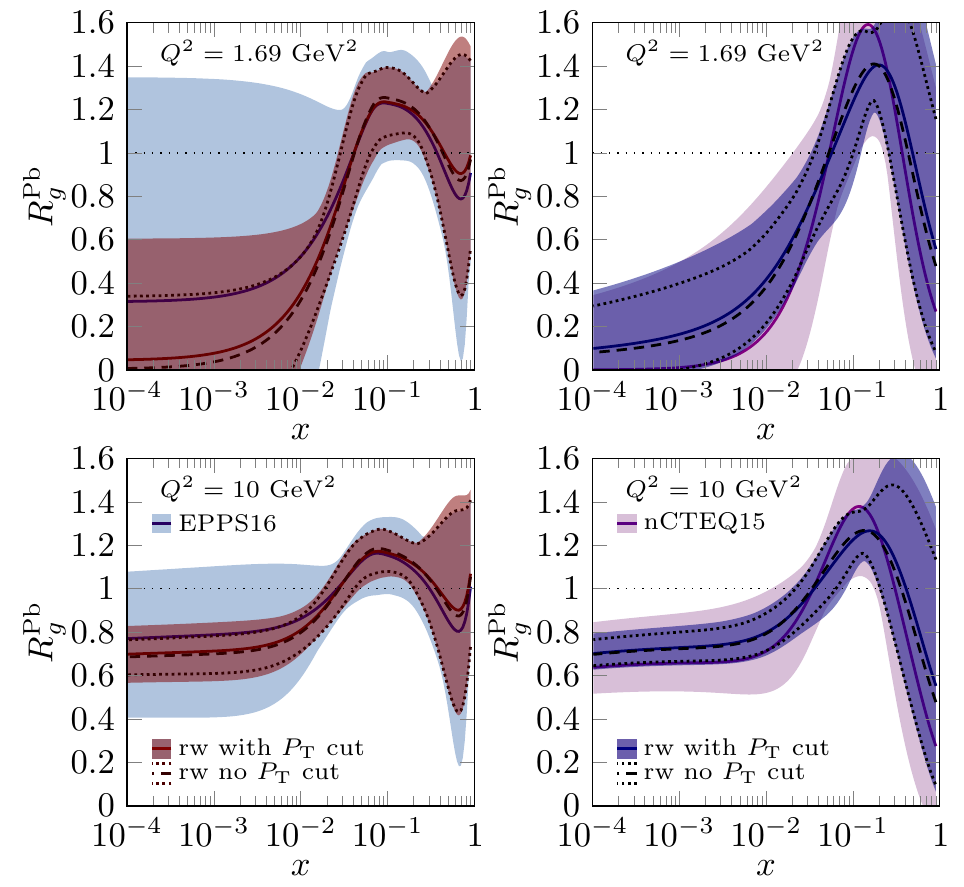}
\caption{The EPPS16 (left) and nCTEQ15 (right) nuclear modifications for bound-proton PDFs in Pb nucleus before (EPPS16 blue, nCTEQ15 purple), after reweighting with the LHCb data with $\PT > 3~\text{GeV}$ (EPPS16 red, nCTEQ15 blue), and including all data points (dotted curves). The results are shown at $Q^2 = 1.69~\text{GeV}^2$ (upper panels) and at $Q^2 = 10~\text{GeV}^2$ (lower panels).}
\label{fig:RPb_EPPS16_Q0_nocut}
\end{center}
\end{figure}

The agreement between the measured and calculated $R_{\mathrm{pPb}}^{\D0}$ was found to be very good also at $\PT < 3~\text{GeV}$ which we excluded from the reweighting due to theoretical concerns. To check how much potential constraints we threw away, we have repeated the reweighting procedure this time including all the LHCb data. The resulting gluon nPDFs at $Q^2 = 1.69~\text{GeV}^2$ and $Q^2 = 10~\text{GeV}^2$ are shown in figure \ref{fig:RPb_EPPS16_Q0_nocut} for EPPS16 and nCTEQ15. Effect for quark nPDFs was found negligible at $Q^2=1.69~\text{GeV}^2$. In both cases the reweighted central results remain practically unchanged but the uncertainties are further reduced at small $x$ in the case of EPPS16 and also at larger $x$ in the case of nCTEQ15. However, the bulk part of the uncertainty reduction still comes from the data in the ``safe region'' $\PT > 3~\text{GeV}$ such that inclusion of the $\PT < 3~\text{GeV}$ data is not critical. As we will argue next, including the lower $\PT$ data would not even increase the sensitivity to the small $x$ region significantly. 

\subsection{Sensitivity to small-$x$ region}
\label{sec:small-x}

\begin{figure}[ptbh]
\begin{center}
\includegraphics[width=0.6\textwidth]{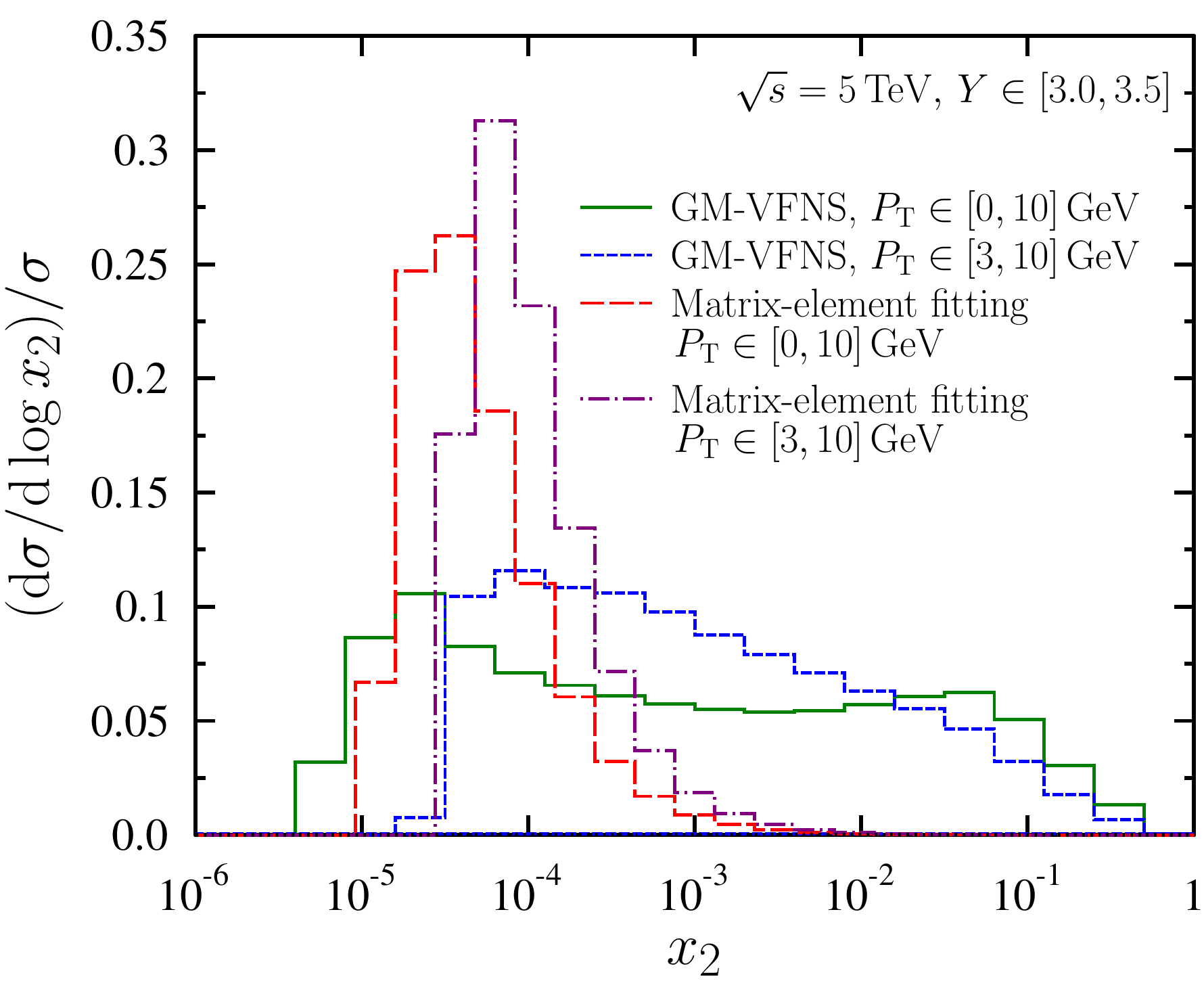}
\caption{Contributions to differential $\D0$ cross section from different values of $x_2$ at $3.0<Y<3.5$ from the GM-VFNS in $\PT$ ranges of $[0,10]~\text{GeV}$ (solid green) and $[3,10]~\text{GeV}$ (short-dashed blue) and from matrix-element fitting approach for same $\PT$ ranges (long-dashed red and dot-dashed purple).}
\label{fig:x_dist}
\end{center}
\end{figure}

The $x$ values probed by a given $\PT$ and $Y$ are often in the literature estimated with simplified leading-order kinematics, see e.g. ref.~\cite{Zenaiev:2015rfa}. To get a more complete understanding on the small-$x$ sensitivity of $\D0$ production at forward rapidities we show the contributions from different values of $x_2$ (momentum fraction in nucleus) to the D$^0$ cross section in figure~\ref{fig:x_dist}. These distributions are based on full NLO GM-VFNS calculation with EPPS16 including the convolution with fragmentation functions. The results are compared to distributions from a ``matrix-element fitting'' approach similar to the one introduced in ref.~\cite{Lansberg:2016deg} and applied in ref.~\cite{Kusina:2017gkz} to study the impact of the LHCb data on nPDFs. In the latter method the squared matrix element $|\mathcal{M}|^2$ for D-meson production is parametrized and the parameters are fitted to data from p+p collisions assuming that the only contribution is gluon-gluon initiated $2 \rightarrow 2$ scattering. The parameters used for the result in figure~\ref{fig:x_dist} are obtained from ref.~\cite{Lansberg:2016deg} but the correspondence is not guaranteed to be exact since the details of the applied two-body phase space are not explicitly defined in the reference. However, the main point here is that the assumed $x_{1,2}$ dependence which, together with PDFs, dictates the shape of the $x$ distributions is rather trivial, of the form $|\mathcal{M}|^2 \propto x_1x_2$. 
The $x$ distributions from the full NLO GM-VFNS calculation are shown for $\PT$-integrated case with and without the lower cut of $\PT > 3~\text{GeV}$. As expected, the $\D0$ meson production at forward rapidities is indeed sensitive to small-$x$ region reaching down to $10^{-5}$ in the considered $3.0<Y<3.5$ bin. However, there is still a significant contribution from larger $x$. These large-$x$ tails mainly arise from the convolutions with the fragmentation functions which smears the connection between partonic and hadronic kinematics. Also the NLO corrections contribute to the tail as discussed in ref.~\cite{Helenius:2018uul}. Maybe a bit surprisingly, the tail extends to higher values of $x$ when no lower cut on $\PT$ is applied. A very similar behaviour has been seen in the case of inclusive photon production \cite{Helenius:2014qla}. In part, this can be explained by the valence-like gluons at low scales which shift the cross section to higher $x$ region. In addition, the nuclear effects in EPPS16 are most pronounced at low scales and the shadowing further suppresses the contributions from small $x$, whereas anti-shadowing tends to increase the larger-$x$ tail. All this dilutes the extra small-$x$ constraints that could be obtained by releasing the $\PT > 3~\text{GeV}$ cut in our GM-VFNS scheme. Thus, a significant part of the reduced small-$x$ uncertainties in figure~\ref{fig:RPb_EPPS16_Q0_nocut} can be explained just by the increased statistics (24 data points more) rather than pushing to smaller $x$. These long large-$x$ tails are not visible in the distributions obtained with the matrix-element fitting approach as it assumes leading-order partonic kinematics and, in particular, a naive $|\mathcal{M}|^2 \propto x_1x_2$ behaviour of the coefficient function. Thus, in comparison to the GM-VFNS approach, the matrix-element fitting approach would overestimate the sensitivity of the LHCb data on the small-$x$ PDFs and would lead to an overly optimistic impact at small $x$ if used in a global analysis.

\subsection{Reweighting with \textsc{Powheg}}

\begin{figure}[ptbh]
\begin{center}
\includegraphics[width=0.8\textwidth]{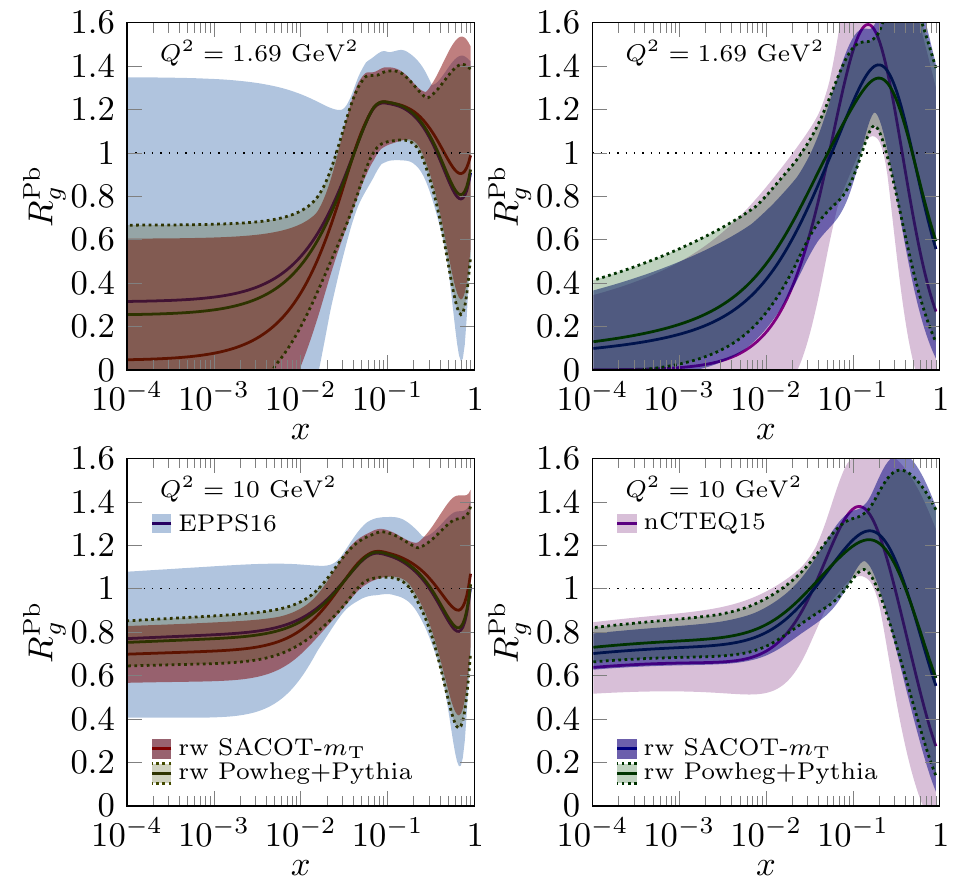}
\caption{
The EPPS16 (left) and nCTEQ15 (right) nuclear modifications for bound-proton PDFs in Pb nucleus before (EPPS16 blue, nCTEQ15 purple) and after reweighting with the LHCb data with SACOT-${m_{\mathrm{T}}}$ (EPPS16 red, nCTEQ15 blue) and with \textsc{Powheg} (green with dashed error-band limits) frameworks with a cut $\PT > 3~\text{GeV}$. The results are shown for gluons at $Q^2 = 1.69~\text{GeV}^2$ (upper panels) and at $Q^2 = 10~\text{GeV}^2$ (lower panels).}
\label{fig:RPb_EPPS16_Q0_powheg}
\end{center}
\end{figure}

To study the impact of the terms resummed in SACOT-${m_{\mathrm{T}}}$ we have performed the nPDF reweighting with the LHCb data also using the \textsc{Powheg}+\textsc{Pythia} approach introduced in section \ref{sec:Powheg}. The resulting gluon distributions are compared to the results obtained within the SACOT-${m_{\mathrm{T}}}$ scheme in figure \ref{fig:RPb_EPPS16_Q0_powheg} for EPPS16 and nCTEQ15. To avoid statistical fluctuations the cross sections with the nPDF error sets were calculated from the original events by calculating a weight for each event and each error set using the event-reweighting machinery introduced in \textsc{Powheg Box V2}. In both cases, EPPS16 and nCTEQ15, reweighting the nPDFs with the LHCb data using the \textsc{Powheg} framework leads to somewhat reduced shadowing compared to SACOT-${m_{\mathrm{T}}}$ result. This can be explained by the fact that a FFNS calculation lacks the large-$x$ contribution which is generated by gluon fragmentation in GM-VFNS as demonstrated in figure 5 in ref.~\cite{Helenius:2018uul} and discussed in section \ref{sec:small-x}. Therefore in the SACOT-${m_{\mathrm{T}}}$ scheme a stronger shadowing is required as it needs to compensate for the enhancement arising from the contribution from the anti-shadowing regime. The separation is slightly more pronounced with the EPPS16 nPDFs but in both cases the differences are within the estimated uncertainties and the reduction of the small-$x$ gluon uncertainties is similar with both theoretical setups. We can thus conclude that the constraints obtained for the nuclear PDFs are even surprisingly stable against varying theoretical approaches. We stress, however, that since the absolute cross sections are quite different, the rough agreement between the applied frameworks is due to fact that we consider data for the ratio $R_{\mathrm{pPb}}$ where many effects cancel out.

\section{Summary}
\label{sec:summary}

We have presented the first direct QCD analysis of the recent LHCb data \cite{Aaij:2017gcy} for $\D0$ meson production in p+Pb collisions and their impact on nuclear PDFs. To accomplish this we have used the Hessian reweighting method and the cross sections calculated within GM-VFNS using the recently introduced SACOT-$m_{\mathrm{T}}$ scheme at NLO \cite{Helenius:2018uul}. The advantage of the new scheme over the previous GM-VFNS implementations is that by explicitly including the heavy-quark masses in the kinematics also for processes where the $Q\overline{Q}$ pair is produced from light-flavour fragmentation, a sensible behaviour in the $\PT \rightarrow 0$ limit is always obtained. However, the description of the very low-$\PT$ regime is still somewhat arbitrary within GM-VFNS and this is one of the reasons we have concentrated mainly on the $\PT \geq 3~\text{GeV}$ region. The resulting cross sections are in a very good agreement with the single-inclusive D-meson $\PT$ spectra in the wide rapidity range covered by the LHCb measurement. We also computed predictions by a frequently used \textsc{Powheg} approach in which the heavy quarks are first produced in the partonic $2 \rightarrow 2$ and $2 \rightarrow 3$ scattering events, and then showered and hadronized with \textsc{Pythia}. This approach generally yields smaller differential cross sections than what we obtain with the GM-VFNS formalism. At very low $\PT \lesssim m_{\rm charm}$ this is hardly significant due to the large scale uncertainties and scheme dependence of the GM-VFNS calculations. At large $\PT \gtrsim 3 \, {\rm GeV}$  it is possible that the observed differences are due to the omission of contributions in which the heavy quark is produced in $2 \rightarrow 4$ processes and beyond, though within the scale uncertainties the Powheg and GM-VFNS approaches agree also there (see Fig.~11 of Ref.~\cite{Helenius:2018uul}). Thus, higher-order calculations would be needed to improve our understanding of whether this is the principal cause for the observed differences.

A very good agreement with the $R_{\mathrm{pPb}}^{\rm D^0}$ data is found with both of the considered nPDF analyses, EPPS16 and nCTEQ15, and the data are accurate enough to set significant further constraints. For quark PDFs the modifications in the central values are weak but for gluons a somewhat stronger (weaker) small-$x$ shadowing than originally in EPPS16 (nCTEQ15) is preferred by the data. The reweighting also brings the gluon shadowing in these two nPDF sets into a better mutual agreement. The main impact of the data is, however, the substantial reduction of the uncertainties for gluon nuclear modifications at $x<0.01$. In fact, these are the first data directly sensitive to small-$x$ gluons in heavy nuclei at clearly perturbative scales, and therefore provide the first unambiguous direct evidence for nuclear gluon shadowing in the context of a global analysis. The backward data seem to confirm the presence of a moderate gluon antishadowing at large $x$. We note that the effect of these data on EPPS16 are remarkably similar as recently found from dijet data at significantly higher interaction scales, though there the region $x < 0.002$ is not directly probed \cite{Eskola:2019dui}. The nPDF reweighting was repeated also with the \textsc{Powheg} setup resulting in a slightly reduced gluon shadowing but otherwize very similar results are obtained as with the SACOT-$m_{\mathrm{T}}$ scheme. It thus appears that our main results -- constraints on nuclear PDFs -- are robust against theoretical uncertainties. 

By studying how the cross section builds up from different values of nuclear $x$ we have shown that the LHCb D$^0$ data constrain nPDFs down to $x \sim 10^{-5}$ but, due to the convolution with FFs, there is still a notable contribution from the high-$x$ region. The importance of using a full QCD calculation to quantify the impact of D-meson data was also underlined. Indeed, a simplified framework can lead to an apparent increase in the sensitivity to the small-$x$ region and would therefore not provide a realistic estimation of the constraints. The good agreement between the nPDF calculation and the data down to $\PT = 0~\text{GeV}$ --- even when rejecting data points at $\PT < 3~\text{GeV}$ from the fit --- implies that the pure collinear-factorization approach is valid also in the small-$x$ region. All in all, we conclude that the LHCb D-meson data can be included in future updates of global nPDF analyses without causing conflicts with the other existing data. To more deeply test the factorization and the universality of nPDFs, data with similar $x$-reach but for a different observable would be crucial.

\acknowledgments

We wish to thank Michael Winn and Yanxi Zhang for discussions and help concerning the LHCb data. The Academy of Finland projects 297058 (K.J.E.) and 308301 (H.P. and I.H.), and the Magnus Ehrnrooth Foundation (P.P.) are acknowledged for financial support. Further support was provided by Ministerio de Ciencia e Innovaci\'on of Spain under project FPA2017-83814-P; Unidad de Excelencia Mar\'\i de Maetzu under project MDM-2016-0692; ERC-2018-ADG-835105 YoctoLHC; and Xunta de Galicia (Conseller\'\i a de Educaci\'on) and FEDER (P.P.). The Finnish IT Center for Science (CSC) is acknowledged for the computing time within the project jyy2580.

\bibliographystyle{JHEP}
\bibliography{DmesonRpPb}

\end{document}